\documentstyle[12pt]{article}
\input {epsf}
\def\beq{\begin{equation}}
\def\eeq{\end{equation}}
\def\beqn{\begin{eqnarray}}
\def\eeqn{\end{eqnarray}}

\def\ts{\textstyle}
\def\eq#1{eq.~(\ref{#1})}
\def\Eq#1{Eq.~(\ref{#1})}
\def\tr{\mathop{\rm tr}}
\def\Tr{\mathop{\rm Tr}}

\def\bra#1{\left\langle #1\right|}
\def\ket#1{\left| #1\right\rangle}
\def\VEV#1{\left\langle #1\right\rangle}

\def\p{\partial}
\def\d{{\rm d}}
\def\ee{{\rm e}^}
\def\Lm{\Lambda}
\def\hA{\hat{A}}
\def\hB{\hat{B}}
\def\lm{\lambda}
\def\tlm{\tilde{\lambda}}
\def\om{\omega}
\def\hW{\hat{W}}
\def\hg{\widehat{g}}
\def\l{\ell}

\newcommand{\hsig}[1]{%
   \widehat{\sigma}_{#1}}
\newcommand{\tK}[1]{%
   \widetilde{K}_{#1}}
\def\DVEV#1{\left\langle \!\! \left\langle #1
   \right\rangle \!\! \right\rangle}
\def\ZZ{{\sf Z}\!\!{\sf Z}}
\catcode`\@=11
\def\numberbysection{\@addtoreset{equation}{section}
\def\theequation{\arabic{section}.\arabic{equation}}}
\def\appendix{\setcounter{section}{0}
        \def\thesection{Appendix \Alph{section}}
        \def\theequation{\Alph{section}.\arabic{equation}}}
\newbox \JCCHoldBox
\newdimen \JCCLower
\newcommand \epscenterbox [2]%
{    
     \epsfxsize=#1%
     \setbox \JCCHoldBox = \hbox{\epsfbox{#2}}%
     \JCCLower = 0.5\ht\JCCHoldBox
     \advance \JCCLower by -0.3ex%
     \lower \JCCLower \box\JCCHoldBox
}
\newcommand \epscenterboxy [2]%
{    
     \epsfysize=#1%
     \setbox \JCCHoldBox = \hbox{\epsfbox{#2}}%
     \JCCLower = 0.5\ht\JCCHoldBox
     \advance \JCCLower by -0.3ex%
     \lower \JCCLower \box\JCCHoldBox
}
\begin{document}
\numberbysection
\begin{titlepage}
\begin{flushright}
 {\normalsize 
     hep-th/9703095 \\ 
     February 1997 \\}
\end{flushright}
\vfill
\begin{center}
  {\Large \bf   
   Fusion rules and macroscopic loops\\
     from discretized approach to\\ 
     two-dimensional gravity}
\footnote{
  A thesis submitted in partial fulfillment
  of the requirements for the degree of
  Doctor of Science
}
\\
\vfill
 {\bf Masahiro Anazawa}
 \footnote{e-mail: anazawa@funpth.phys.sci.osaka-u.ac.jp
 \quad JSPS research fellow} \\
\vfill
 {\it
    Department of Physics,\\
    Graduate School of Science, Osaka University,\\
    Toyonaka, Osaka, 560 Japan\\
  }
\vfill
\end{center}
\begin{abstract}
   We investigate the multi-loop correlators and the multi-point
 functions for all of the scaling operators in unitary minimal
 conformal models coupled to two-dimensional gravity from the two-matrix
 model. 
  We show that simple fusion rules for these scaling operators
 exist, and these are summarized in a compact form as fusion
 rules for loops. 
   We clarify the role of the boundary operators and discuss
 its connection to how loops touch each other.
  We derive a general formula for the n-resolvent correlators,
 and point out the structure similar to the crossing
 symmetry of underlying conformal field theory.
  We discuss the connection of the boundary conditions of 
 the loop correlators to
 the touching of loops  for the case of
 the four-loop correlators.

\end{abstract}
\vfill
\end{titlepage}
\pagenumbering{roman}
\newpage
\tableofcontents
\baselineskip 16pt
\newpage
\pagenumbering{arabic}
\begin{center}
\section{Introduction}
\label{introduction}
\end{center}

    Quantization of gravity is one of the most important issues in
 physics. The understanding of two-dimensional quantum  gravity,
 which is the
 simplest quantum gravity, has experienced  great progress through
 the study of matrix models.\footnote
 {See for example \cite{DifGZj} for review.}
   It was proposed \cite{matrix model}
 that the integral over the geometry
 of two-dimensional surface can be descretized as a sum over randomly
 triangulated surfaces 
 and such regularized two-dimensional gravity can be realized by
  hermitian matrix models.
    Feynman diagrams of the matrix models correspond to the
 dynamically triangulated surfaces and the continuum limit
 of the models then describe the theory of
 two-dimensional gravity.

    Due to the double scaling limit \cite{nonperturvative}
 the sum of the contributions from all topologys of
 two-dimensional surface
 can be treated, and thereby the matrix models have been
 drawn much attention
 as a non-perturbative definition of non-critical string
 theories.
   Following the discovery of the double scaling limit,
 many important structures of the models
 have become clear; for example, the connection to
 KdV flow \cite{Doug}, the Virasoro and $W$ constraints
   \footnote
 {The corresponding structures in continuum framework have been
 shown also in \cite{Hamada}. }
\cite{constraints,IM,GN}.
   Field theory of non-critical strings
 \cite{string field,IIKMNS}
 has been constructed based on the matrix models.

   The matrix models include infinite critical points,
 which are considered to represent 
  certain conformal matters coupled to two-dimensional gravity.
   The m-th critical point of the one-matrix model corresponds
 to the $(2m+1,2)$ minimal conformal model coupled to
 two-dimensional gravity.
   The general $(p,q)$ minimal conformal model, where 
 the central charge is 
 $c=1-\frac{6(p-q)^2}{pq}$,
 can be realized as the continuum limit of the
 discrete system where the degrees of freedom are points
 on the A(DE) Dynkin diagram \cite{Pasquier}.
  Multi-matrix chain model has been introduced as 
 a model which includes the critical points
 corresponding to the general $(p,q)$ minimal models coupled
 to gravity.
   In this model, $q$ matrices interact as a chain.
   The two-matrix model \cite{Douglas-Proc,Tada,DKK},
 which is the simplest multi-matrix chain model,
 however, turned out to include all $(p,q)$
 critical points, which was pointed out in
 \cite{Douglas-Proc,Tada} and shown explicitly in \cite{DKK}.
   We use the two-matrix model to investigate
 the unitary minimal model $(m+1,m)$
 coupled to two-dimensional gravity.

    The emergence of the infinite number of  scaling
 operators is one of the most important properties 
 of the matrix models.
    Before coupled to gravity, the minimal model has finite
 number of primary fields. Coupled to gravity, however,
 infinite number of scaling operators emerge. 
    This phenomenon can be understood as follows.
    In the Kac table we can divide the primary fields
 $\Phi_{r,s}$ into those which are inside the the minimal
 conformal grid $1\le r\le q-1,\; 1\le s \le p-1$
 and those outside, which correspond to the null states.
   Before dressed by gravity, the fields outside the minimal
 conformal grid decouple \cite{BPZ} from physical correlators.
  After gravitational dressing, 
 they cease to  decouple \cite{Kitazawa,AGBG}
 and become infinite number of  scaling operators.
   The similar phenomenon has been shown in continuum framework.
 Through the examination of the BRST cohomology of the 
 coupled system composed of  Liouville theory, the ghosts
 and the minimal matter, infinite physical states were shown
 to exist \cite{LZ,BMP}.
  These states have their counterparts in the matrix models
 as the scaling operators.  Some of the scaling operators
 do not have
 their counterparts in the BRST cohomology, which we will discuss
 later.

   In ordinary $(p,q)$ minimal conformal model
 the primary fields satisfy  certain fusion rules \cite{BPZ};
 three-point function
 $\langle \Phi_{r_1,s_1}\Phi_{r_2,s_2}\Phi_{r_3,s_3}\rangle$
 is non-vanishing only when
\beqn
  1+|r_1-r_2|\le r_3 \le \mbox{\rm mim}
 \{r_1+r_2-1,p \},\;\; r_1+r_2+r_3=\mbox{\rm odd}
 \nonumber \\
  1+|s_1-s_2|\le s_3 \le \mbox{\rm mim}
 \{s_1+s_2-1,q \},\;\; s_1+s_2+s_3=\mbox{\rm odd}
\;.
\eeqn
    It is interesting to examine how the fusion rules change
 when the matter
 couples to gravity.
 In particular, we are interested in the fusion rules
 for the gravitational descendants
 ($\sigma_j,\; j=q+1,q+2,\cdots$), 
 most of which correspond to the operators outside the
 minimal conformal grid.  Before coupled to gravity
 the corresponding fusion rules do not exist.
   Such three-point functions were examined
 from the point of view of the generalized
 KdV flow in \cite{DifK} for lower dimensional scaling operators
 in the case of $(m+1,m)$ unitary matter.
  It was shown that the gravitational primaries 
 $\sigma_j\; (j=1,\cdots,m-1)$
 satisfy fusion rules of BPZ type;
 for $j_1+j_2+j_3\le 2m-1$, 
 $\VEV{\sigma_{j_1}\sigma_{j_2}\sigma_{j_3}}$ is
 non-vanishing only when
\beq
  1+|j_1-j_2|\le j_3 \le j_1+j_2-1
\;.
\eeq
   The fusion rules were also examined in continuum framework
 \cite{AGBG}.
   As for the gravitational descendants, however, we think
 clear results have not been obtained.
    In this paper we would like to clarify the
 fusion rules for all of the scaling operators including
 the gravitational descendants in the case of unitary minimal
 model.
   This paper is based on  \cite{AII,AII2,AI,Anazawa}.
 
   Macroscopic loop correlators \cite{BDSS,AJM},
 which are the amplitudes of the surfaces with boundaries (loops)
 of fixed lengths, are the fundamental amplitudes of the
 matrix models. Although these amplitudes are hard to treat in the continuum 
 framework \cite{Nakayama}, they are defined quite naturally
 in the matrix models.
   It was shown \cite{MSS,MS}
 that these correlators have more information
 than those of local operators and that  the latter 
 correlators can be extracted
 from the former  correlators
 explicitly in the case of $c=0,1/2,1$.
   They argued there that macroscopic loops could be replaced
 by a sum of local operators in a certain situation and 
 thereby obtained the correlators of local operators from
 those of macroscopic loops.

   One of the  purposes of this paper is to generalize
 the idea in \cite{MSS}
 to the cases of the general unitary
 minimal models and to clarify the fusion rules for 
 macroscopic loops and all of the scaling operators.
   First, we derive the three- and n-loop correlators
 from the two-matrix model at the general unitary critical
 points \cite{AII2,AI}, and then derive
 the explicit forms of the 
 correlators of the scaling operators \cite{Anazawa}.
   The main conclusion is that the three-point correlators
 of all of the scaling operators satisfy certain simple fusion rules
 \cite{Anazawa}
 and the fusion rules for all of the scaling operators are summarized
 in a compact form as the fusion rules for
 three-loop correlators \cite{AII2}.

  In matrix models, there are infinite subset of the  scaling operators
 which do not have their
 counterparts in the BRST cohomology of Liouville
 theory.
  In the case of one-matrix model, Martinec, Moore and
 Seiberg \cite{MMS} argued that these operators are boundary
 operators, which correspond to the vertex operators of open
 string and couple to the boundaries of
 two-dimensional surface.
  They proved that one of them is in fact a boundary operator
 which measures the total loop length.
   We think, however, the roles of the rest of these operators 
 were not clear.
   We also clarify the role of these operators and
 its connection to the touching of loops
 in the case of general unitary models  \cite{Anazawa}.
 
   We also determine completely the forms of the
 multi-resolvent correlators, which are the Laplace
 transform of the multi-loop correlators,
   and point out that the loop correlators have
 the structures similar to those of the crossing symmetry of
 the underlying conformal field theory \cite{AI}.
   In the cases of four- and five-loop, we discuss the 
 connection of the boundary conditions of the loops to
 the touching of the loops \cite{Anazawa}.

   As another formulation of 2D gravity with matter system,
 models of string whose target spaces are 
 the Dynkin diagrams have been investigated \cite{Kostov}.
   We also comment on the connection of our results to those
 from these models.

   The paper is organized as follows.
 Sect.~2 is devoted to the review of the matrix models
 and the macroscopic loops. We limit our discussion
 to the subjects that have direct connections to
 the subsequent sections.

   In sect.~3 we derive the three-loop correlators and
 extract the three-point functions for all of the scaling operators
 through expansion of loops in terms of the local operators.
   We then show that certain simple fusion rules exist for 
  these local operators or loops.
   We also discuss the role of the boundary operators there.
 
   We derive the formulas for the multi-resolvent
 correlators in sect.~4, and give the explicit forms
 of the four- 
 and five-loop correlators.
   We point out that the structure corresponding to the
 crossing symmetry of the underlying 
 conformal field theory exists in the multi-loop
 correlators.
   We also discuss the connection of the boundary
 condition of the loops to the touching of the loops.
   
   Sect.~6 is a summary.
\newpage
\begin{center}
\section{Conformal field theory coupled to two-dimensional
 gravity (review)}
\end{center}

\subsection{Matrix models and two-dimensional gravity}

   Let us briefly review the matrix models and the connection
 to the Liouville theory,
 emphasizing on the notion of 
 the scaling operators and that of the macroscopic loops.
   We limit our discussion to the subjects which have
 direct connections to the later sections.

\subsubsection{Matrix models and random triangulation}

\begin{figure}
\begin{center}
 \qquad\qquad\epsfbox{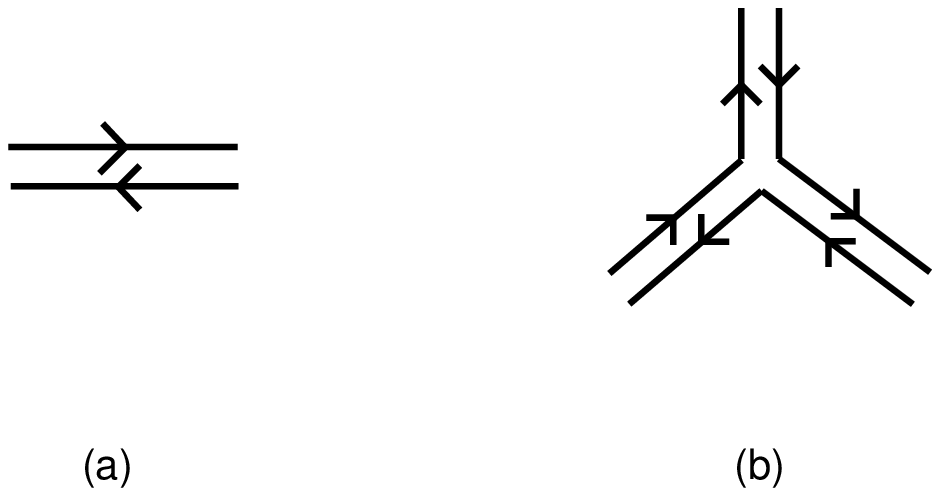}
\caption{propagator and vertex}
\label{propagator and vertex}
\end{center}
\end{figure}

 Let us consider the model defined by the path integral with
 respect to an $N\times N$ hermitian matrix $\Phi$,
\beq
\label{eq:one-matrix}
 e^Z =\int \d \Phi e^{-\frac 12 \Tr \Phi ^2
 + \frac {g}{\sqrt{N}}\Tr\Phi^3 }
\;\;,
\eeq
 where the measure is
\beq
 \d \Phi=
 \prod_i \d \Phi^i_{\ i} \prod_{i<j} \d {\rm Re}\Phi^i_{\ j}
 \d {\rm Im}\Phi^i_{\ j}
\;.
\eeq
  The propagator is
 $\VEV{\Phi^i_{\ j}\Phi^k_{\ l}}=\delta^i_l\delta^k_j$,
 and is represented by the double lines in  
 fig.~\ref{propagator and vertex}.~(a).
 The arrows connect the upper matrix indices to the 
 lower ones.
 The vertex in the action is represented by 
 fig.~\ref{propagator and vertex}~(b).
 Expanding the partition function in term of $g$,
 we find that the each Feynmam diagram represents a net on
 an orientable two-dimensional surface.
 Taking the dual of such a diagram, the vertices turn into
 triangles and the dual diagram represents a random triangulation
 of two-dimensional surface.
 Therefore, the model specified by \eq{eq:one-matrix} can be
 considered to represent a theory of random triangulation of
 2D surfaces and is expected to be a theory of 2D quantum gravity
 when we take continuum limit.

   Let us count the power of $N$ associated to each diagram.
 Changing variables $\Phi\to \Phi/\sqrt{N}$, the action becomes
 $N\Tr\left(-\frac 12 \Tr \Phi ^2 + g\Tr\Phi^3\right)$.
 From this form of action it is clear that each vertex contributes
 a factor of $N$, each propagator (edge)
 contributes a factor of $N^{-1}$, and loop (face) contributes
 a factor of $N$ due to the index summation associated.
 Each diagram has thus an overall factor
\beq
\label{eq:power of N} 
 N^{V-E+F}=N^{\chi}=N^{2-2h}\;\;,
\eeq
 where $\chi$ and $h$ are the Euler character
 and the number of genera of the surface associated
 to the diagram respectively.
 From (\ref{eq:power of N}), the partition function can be expanded as
\beq
 Z(g)=\sum_{h} N^{2-2h}Z_h(g)
\;\;\,
\eeq
 where $Z_h(g)$ represents the contribution from the surfaces of genus
 $h$.
 In the large $N$ limit, the contribution from the planar surfaces
 dominate.

\subsubsection{Continuum limit}

 When expanded in the coupling $g$, for large order $n$, $Z_0$ 
 behaves as
\beq
\label{eq:Z_0(g)}
 Z_0(g)\sim \sum_n n^{\Gamma_{str}-3}\Bigl(\frac{g}{g_*}
 \Bigr)\sim (g_*-g)^{2-\Gamma_{str}}
\;
\eeq
 and the expectation value of the number of vertices (triangles
 in the dual diagram) is given by
\beq
 \VEV{n}=\frac{\p}{\p g} \ln Z_0(g)
 \sim \frac{1}{g-g_*}
\;.
\eeq
 The partition function $Z_0(g)$ thus becomes non-analytic
 and $\VEV{n}$ diverges
 when $g$ approaches some critical value $g_*$.
 Since $\VEV{n}$ diverges as $g\to g_*$, it is expected
 that the contribution from the continuum surface with finite
 area can be obtained by rescaling the area 
 of the individual triangles to zero accordingly.
   The contribution from the continuum surface 
 is considered to correspond the non-analytic part of
 \eq{eq:Z_0(g)}.
 Therefore, the behavior of the model (\ref{eq:one-matrix})
 near the critical point is considered to represent
 two-dimensional
 quantum gravity.

\subsubsection{Multi-critical points and multi-matrix models}

 So far, we have considered the model(\ref{eq:one-matrix}),
 which consist of one kind of vertex in the action.
 As a generalization of this model,
 let us consider the the model specified by the action
\beq
 S=\frac{N}{\Lm}\Tr V(\Phi)
\eeq
 where $V(\Phi)$ is some polynomial of matrix $\Phi$.  This model
 can represent  a series of systems of consisting of 
 matter and two-dimensional gravity.
 By tuning the couplings in the potential, various critical
 points are obtained. The $m$-th critical point corresponds to
 the  $(2m+1,2)$ minimal conformal model coupled to
 two-dimensional gravity.

 As another generalization, let us consider the multi-matrix
 chain model,
\beq
\label{eq:multi matrix}
 e^Z=\int \prod^{(\alpha)} \Phi_{\alpha}\;e^{-S}
 \;,
\eeq
\beq
\label{eq:S of multi matrix}
 S=\sum_{\alpha=1}^{\nu-1} V_{\alpha}(\Phi^{(\alpha)})
 -\sum_{\alpha=1}^{\nu-2} c_{\alpha}\Phi^{(\alpha)}
 \Phi^{(\alpha+1)}
 \;.
\eeq
 Here the different matrices $\Phi^{(\alpha)}$ represent
 $\nu-1$ different matter degrees of freedom that can exist
 at the vertices.
 Note the couplings $c_{\alpha}$ 
 in the action (\ref{eq:S of multi matrix}) couple the
 matrices along a line (chain).

 It was pointed out, however, in \cite{Douglas-Proc,Tada} and
 shown explicitly in\cite{DKK}
 that it is sufficient to
 consider the two-matrix model in order to generate the most
 general critical points,
 which correspond to the $c< 1$ minimal conformal models.
 We will thus use the two-matrix model to examine the minimal
 models coupled to two-dimensional gravity in this article.

\subsection{Scaling operators in matrix models}

\subsubsection{KdV flows and scaling operators}
 
 Consider the two-matrix model with symmetric potential,
\beq
\ee Z =\int \d \hA \d \hB 
~\ee{-\frac N{\Lm} \tr \left( U(\hA)+U(\hB)-\hA\hB \right)}
\;\;,
\eeq
 where $U$ is an arbitrary polynomial of order $m$.
 In this article, we limit our discussion to the two-matrix model
 with symmetric potential
 and to the critical points which correspond to the unitary
 minimal models.
  In the case of asymmetric potential, some of the boundaries
 (loops) of the surface 
 would have fractal 
 dimensions different from the usual dimension of length.  

 Integrating out the ``angular" variables, we have
 \cite{Mehta}
\beq
 \ee Z
 =\int \d \lm_{i}\d \tlm_i \Delta(\lm)\Delta(\tlm)
 ~\ee{-\frac N{\Lm} \sum_i\left( U(\lm_i)+U(\tlm_i)
 -\lm_i\tlm_i \right)} 
 \;\;.
\eeq
 Here $\Delta(\lm)$ is the Vandermonde determinant, $\lm$ and $\tlm$
 represent the eigenvalues of the matrices $\hA$ and $\hB$
 respectively.

 We introduce the orthogonal polynomials
 $|j\rangle =\xi_j(\lm)$ and
 $\langle k| =\xi_k(\tlm)$ by the orthonormality relation
\beqn
 \VEV{j|k}&=&\int\d \mu ~\xi_j(\tlm)\xi_k(\lm)=\delta_{jk}
  \;\;,\nonumber \\
 &&\;\;
 \d \mu \equiv \d \lm \d \tlm
 ~\ee{-\frac N{\Lm} \left( U(\lm)+U(\tlm)-\lm\tlm \right)}
\;.
\eeqn
We define matrices $A$ and $P$ by
\beqn
A_{nm}&=&\VEV{n|\lm|m}
\label{eq:matrix-A}
 \\
 P_{nm}&=&
 \bra{n} {\frac{\p}{\p\lm}}\ket{m}
\label{eq:matrix-P}
\; .
\eeqn
 It is obvious from the definition (\ref{eq:matrix-A}) and
 (\ref{eq:matrix-P}), that $A$ and $P$ obey the Heisenberg
 commutation relations
\beq
\label{eq:Heisenberg}
[P,A]= 1  
\;.
\eeq
 Now the important fact is that the operators $P$ and $A$
 have non-zero matrix elements $P_{ij}$ and $A_{ij}$
 only if $|i-j|$ is sufficiently small.
 Since the bounds are independent of $N$, in the limit
 $N\to \infty$, $P$ and $A$ become differential operators
 (in $x$, the cosmological constant) of finite order.
   The continuum scaling limit of the two-matrix model is
 abstracted to the mathematical problem of finding solution
 to \eq{eq:Heisenberg}.
 Let us consider the $(m+1,m)$ critical point which
 corresponds to the $(m+1,m)$ minimal model.
 After suitable renormalization,  $A$ is
 given by 
\beq
\label{eq:A-diff}
 A=D^{m} + u(x)_{m-2}D^{m-2} +u(x)_{m-3}D^{m-3}
 +\cdots +u(x)_{0} \;,\;\; D=\p_{x}\;.
\eeq
 (By a change of basis of the form $A\to f^{-1}(x)A f(x)$,
 the coefficient of $D^{m-1}$ may be always be set to zero.)
 and $P$ is given by \cite{Doug}
\beq
\label{eq:P-diff}
 P=(L^{m+1})_{+} \;.
\eeq
  Here $L\equiv A^{1/m}$ is a pseudodifferential operator
 satisfying
\beq
 A=L^{m}\;,\;\;
 L=D+a_1 D^{-1}+a_2 D^{-2}+\cdots \;,
\eeq
 and $(L^{\alpha})_{+}$ denotes the nonnegative (differential
 operator) part of $L^{\alpha}$.
 Substituting eqs. (\ref{eq:A-diff}) and (\ref{eq:P-diff})
 into \eq{eq:Heisenberg}, we find that 
 differential equations for $u(x)_i$.  These equations determine
 $u(x)_i$
 up to $m-1$ integration constants $t_i\;(i=1,\cdots,m-1)$.
  The dependence of $L$ on the constants $t_i$ is given by
 the first $m-1$ generalized KdV flows:
\beq
 \frac{\p}{\p t_i} L=\left[ (L^i)_+,\; L\right]
\;\;.
\eeq 
  In general, the perturbation from the $(m+1,m)$ point is represented
 by the independent flows in term of commuting operators 
 $\{L^i |i=1,2,\cdots\;\ne 0\;(mod\; m)\}$:
\beqn
 \label{eq:KdV flow}
 \frac{\p}{\p t_i} A &=&\left[ (L^i)_+,\; A\right]
 =-\left[ (L^i)_-,\; A\right]
 \;\;, \nonumber \\
 \frac{\p}{\p t_i} P &=&\left[ (L^i)_+,\; P\right]
 =-\left[ (L^i)_-,\; P\right]
\;\;.
\eeqn
  The infinite number of directions of the perturbation
 correspond to flows along RG trajectories between various
 critical theories, identified with the $(p,q)$ minimal model
 coupled to two-dimensional gravity.
   Since the perturbation is realized by adding an infinite
 number of relevant matter operators dressed by gravity to
 the original critical action,
 the correlation functions of the scaling operators
 are defined by
 the following relation:
\beq
 \VEV{ \sigma_{j_1}\sigma_{j_2}\cdots\sigma_{j_n} }
 =\frac{\p}{\p t_{j_1}}\frac{\p}{\p t_{j_2}}\cdots
 \frac{\p}{\p t_{j_n}} \log Z
 \;.
\eeq

\subsubsection{Correlators from KdV flow}

   The correlation functions of the scaling operators
 $\sigma_j$
 on the sphere were calculated in \cite{DifK} for lower
 $j$ from the point of view of  the KdV flow.
 
  Let us define the following infinite number of commuting 
 operators on the sphere:
\beq
\frac 1n Q_n=\sum_{j=0}^{\infty}
 \left({n-j-1\atop j-1}\right) \frac{(-\frac 12 u)^j}{j}
 D^{n-2j}
\eeq
with
\beq
 \left[Q_n,Q_k\right]=0
\;\;.
\eeq
 The operators $A$ and $P$ on the sphere are given by
\beqn
\label{eq:A,P on the sphere}
 A&=&(Q_m)_+=L^{m}=D^{m}-\frac 12 m u D^{m-2}+\cdots
 \;, \nonumber \\
 P&=&(Q_{m+1})_+\;,
\eeqn
 where $u$ is the two-point function of dressed identity
 operator.
   Substituting eqs. (\ref{eq:A,P on the sphere}) into
 \eq{eq:Heisenberg}, we have
\beq
 \frac u2=\left( \frac x{m+1}\right)^{1/m}
\;.
\eeq
 
 The correlation functions of $\sigma_j$ for lower $j$
 can be calculated from the KdV flow.
   For example,
 from eqs. (\ref{eq:KdV flow}), we obtain the following expression for
 the one-point function:
\beq
 \frac{\p}{\p t_n}u=-2({\rm Res}\;L^n)^{'}
 \;,
\eeq
 where ${\rm Res}\;L^n$ is the coefficient of $D^{-1}$
 in $L^{n}$ and $L^{n}\;(n\le 2m-3)$ can be replaced with
 $Q_n$ due to the relation
\beq
 Q_{n}-L^{n}=\frac nm c_n D^{n-2m}
 +{\cal O}(D^{n-2m-2})
 \;.
\eeq
 The two-point functions can be calculated from 
\beq
 \frac{\p}{\p t_n}\frac{\p}{\p t_k}u=
 2({\rm Res}\;[ L^n_{-},L^k_{+} ])^{'}
 \;.
\eeq
 As for explicit results of the correlators,
 we mention these in sec.~\ref{sec:three-point functions}. 

\subsubsection{Connection to continuum theory}
  The $m$-th critical point of one-matrix model corresponds to
 the $(2m+1,2)$ minimal model coupled to two-dimensional gravity.
   The scaling operators are naively expected to
 correspond to
 the following operators in the $(2m+1,2)$ minimal model coupled
 to Liouville theory in the continuum framework:
\beq
 \sigma_{j}\leftrightarrow
 \int e^{\alpha_j \varphi} \Phi_{1,m-1-j}
 \;,\;\; j=0,\cdots,m-2,
\eeq
 where $\alpha_{j}=\frac 12 \gamma(m-j)$, $\Phi_{r,s}$
 are the primary fields of the corresponding
 conformal field theory.
 
   This correspondence fails however.
 In \cite{MSS}, it was argued that the discrepancies were
 due to contact terms which arise when two operators are
 at coincident points.  They showed  explicitly
 the correct correspondence
 by the analytic redefinition of coupling constants
 $t_j$
\beq
 t_n=C^i_n \hat{t}_i+C^{ij}_{n}\hat{t}_i\hat{t}_j+\cdots
\;,
\eeq
  mainly for the case of one-matrix model.
   The original frame of operators $\sigma_{j}$ and couplings
 $t_j$ is referred to as the KdV frame and the new frame of 
 operators $\hsig{j}$ and $\hat{t}_j$ is referred to as
 conformal field frame.
  The wave function of $\hsig{j}$ 
 is proportional to the modified Bessel function
 $K_{\frac jq}(2\sqrt{\mu}\l)$ so that it satisfies the 
 (minisuperspace) Wheeler-deWitt equation
\beq
 \left[ -\left(\l \frac{\p}{\p \l}\right)^2
 +4\mu\l^2 +\Bigl(\frac jq\Bigr)^2 \right]
 \Psi_j(\l) =0
\;,
\eeq
 which is a desirable property.

  The BRST cohomology of the coupled system of Liouville theory,
  ghosts and  the $(p,q)$ minimal matter was examined
 in \cite{LZ,BMP}.
   It turned out that the BRST cohomology is spanned by
 infinite operators of the form
\beq
\label{eq:ops-Liouville}
  \tilde{{\cal O}}_j~\ee{\alpha_j \varphi},
 \;\;\; \frac{\alpha_j}{\gamma}=
 \frac{p+q-j}{2q}
 \;\;\; j\ge 1,\;\ne 0\;({\rm mod}\;p),
 \;\ne 0\;({\rm mod}\;q)
\;,
\eeq
 where $\varphi$ is the Liouville field and
 $\gamma=(\sqrt{25-c}-\sqrt{1-c})/\sqrt{12}$.
  The operators $\tilde{{\cal O}}_j$ are made of ghosts,
 matter and derivatives of $\varphi$.
   On the other hand, the scaling operators $\sigma_j$
 of matrix model at the $(p,q)$ critical point scale
 like Liouville operators of the form
\beq
\label{eq:ops-matrix}
  \tilde{{\cal O}}_j~\ee{\alpha_j \varphi},
 \;\;\; \frac{\alpha_j}{\gamma}=
 \frac{p+q-j}{2q}
 \;\;\; j\ge 1,\;\ne 0\;({\rm mod}\;q)
\;.
\eeq
 Apart from the discrepancy of the operators with
 $j=0\;({\rm mod}\;p)$, the two calculations are in
 remarkable agreement.
   It was argued in \cite{MMS} that the scaling operators
 with  $j=0\;({\rm mod}\;p)$ are boundary operators, which
 couple to the boundaries of two-dimensional surface
 and correspond to
 the vertex operators of open string.

\subsection{Macroscopic loops}

\subsubsection{Macroscopic loops in two-matrix model}
\label{sec:macroscopic loops}

 In the two-matrix model, the operators
\beq
 \Tr\hA^{n_1}\;,\;\;\Tr\hB^{n_2}
\eeq
 create holes with boundaries of lattice lengths
 $n_1$ and $n_2$ respectively.
    The correlation functions of $\Tr\hA^{n_i}$ or
 $\Tr\hA^{n_i}$ are expected to
 become those of macroscopic loops
 in the limit 
 $a n_i\to \l_i$ with $\l_i$ finite,
 when the unit lattice length $a$ approaches zero.

  It is convenient to consider first the correlators of the
 resolvents
\beq
 \hW^+(p_i)=\Tr\frac 1{p_i-\hA}\;,
 \;\;
 \hW^-(p_i)=\Tr\frac 1{p_i-\hB}\;,
\eeq
 where $p_i$ is a parameter corresponding to 
 the bare boundary cosmological constant of each
 loop.
  Due to the formal expansion
\beq
 \hW^+(p_i)=\sum_{n=0}^{\infty}
 \frac{\Tr \hA^n}{p_i^{n+1}}
 \;,
\eeq
 the resolvents include the contributions from loops of 
 any length.
   The correlators become singular when $p_i$ approach
 some critical value $p_*$.
   Since the contributions from loops of finite
 continuum length
 corresponds to those of infinite lattice length, 
 continuum loop correlators are defined as the inverse-Laplace
 image of non-analytic part of the resolvent correlators
 with respect to $\zeta_i=(p_i-p_*)/a$.

\subsubsection{`Classical' solutions to Heisenberg relation}
\label{sec:classical functions}

 In later sections we will use extensively the `classical' solutions
 to the `classical' Heisenberg relation.
  Let us explain these in this subsection
 \ref{sec:classical functions}. 
 
   Since we would like to examine the correlators
 on the sphere,
 we are interested only in the planar limit ( large $N$ limit ).
  The Heisenberg relation (\ref{eq:Heisenberg}) turns into
\beq
\label{eq:Heisenberg-2}
 [P,A]=\frac{\Lm}{N}
\;,
\eeq
 after rescaling $P\to \frac{N}{\Lm}P$.
 From \eq{eq:Heisenberg-2}, we see that
 $\frac{\Lm}{N}$ plays the role of Planck constant.
   It is thus expected that the corresponding `classical'
 functions would be much easier to handle than the operators
 $A$ and $P$ in the large $N$ limit. 

 At this point it is useful to change notation for the indices
 of the matrix elements:
\beq
 A_k(n)=A_{n-k,n} \quad,\quad P_k(n)=P_{n-k,n}
 \nonumber \\
\;.
\eeq
 Here $n$ represents the position of the matrix element on the
 diagonal, and $k$ is its deviation from it.
  Then the action of the operators $A$ and $P$ on the
 orthogonal polynomial basis is described by
\beq
 A\ket n=\sum_{k=-1}^{m-1}\ket {n-k}A_k(n)
 \;\;,\;\;
 P\ket n=\sum_{k=-1}^{(m-1)^2}\ket {n-k}P_k(n)
\;\;.
\eeq
 The `classical' functions are defined by
\beqn
 A(\om,s)&=&\sum_{k=-1}^{m-1}\ee{k\om}A_k(n)
 \;\,
 \nonumber \\
 P(\om,s)&=&\sum_{k=1}^{(m-1)^{2}}\ee{k\om}P_k(n)
 \;,
\eeqn
 where $s$ is the continuous variable
\beq
s=\frac nN {\Lm}
\eeq
and $\om$ is its conjugate coordinate.

  The equation of motion 
\beq
 \frac{\Lm}{N}P_{ij}=\VEV{i|U'(\lm)|j}
 -A^{T}_{ij}
\;,
\eeq
  which is obtained by doing an integration by parts,
 reads
\beq
\label{eq:eq of motion -2}
P(\om,s)=U'\Bigl( A(\om,s-\frac{\Lm}{N} \frac \p{\p\om})\Bigr)
   \cdot 1
   -A (-\om+\frac{\Lm}{N} \frac \p{\p s},s)\cdot 1
 \;,
\eeq
 when expressed in term of the classical functions.
 In the planar limit, \eq{eq:eq of motion -2} reads
\beq
\label{eq:eq of motion -3}
P(\om,s)=U'\Bigl( A(\om,s)\Bigr)-A(-\om,s) 
\;,
\eeq 
 and the Heisenberg commutation relation (\ref{eq:Heisenberg-2})
 is replaced by the Poisson bracket
\beq
\label{eq:Poisson bracket}
\{ P(\om,s),A(\om,s)\}\equiv
 {\p P\over\p s}{\p A\over\p\om}-{\p P\over\p\om}
 {\p A\over\p s}=-1 
\;.
\eeq
 Note that in the large $N$ limit the classical functions
 $A$ and $P$ depend on $\Lm$ only through $s$, which is
 easily seen from eqs.~(\ref{eq:eq of motion -2}) and
 (\ref{eq:eq of motion -3}).

 Let us find the solution to the classical Heisenberg relation
 \eq{eq:Poisson bracket} near the (m+1,m) critical point,
 which corresponds to the $(m+1,m)$ unitary minimal
 conformal model.

 At the critical point, one expects the following singular
 behavior of $A$ and $P$:
\beq
\label{eq:scaling law}
A(z)-A_{*}\sim (1-z)^m 
\;,\;\; 
P(z)-P_{*}\sim (1-z)^{m+1}
\;.
\eeq
 From the scaling laws (\ref{eq:scaling law}),
 the solution to the Heisenberg relation is given by
 \cite{DKK}
\beqn
\label{eq:classical functions}
A(z,s)-A_{*}&=&
  2\eta^m\cosh m\theta
 \nonumber \\
P(z,s)-P_{*}&=&
  2\eta^{m+1}\cosh (m+1)\theta
 \nonumber \\
s-\Lm_*&=&
  (m+1)\eta^{2m}
\;.
\eeqn
 Here  $P_*$, $A_*$ and $\Lm_*$ denote the critical values of
 the corresponding quantities and the parametrization
\beq\label{2mat-ae}
\om=2\eta \cosh \theta
\eeq
is used.
  We will use the classical functions (\ref{eq:classical functions})
 extensively  to calculate the loop correlators
 in later sections.

\subsubsection{Loops in semi-classical Liouville theory}
\label{sec:semi-classical}

   When we discuss the loop correlators, it turns out to be very
 helpful to consider these correlators semi-classically in 
 Liouville theory. Let us explain these  \cite{MSS} briefly
 in this subsection \ref{sec:semi-classical}.
   In the continuum framework, two-dimensional gravity part of the coupled
 system can be described by Liouville theory based on the action,
\beqn
 S_{L}[\varphi;\hg]
 &=&\frac 1{8\pi}\int_{\Sigma}\d^2\xi\sqrt{\hg}
 \hg^{ab}\p_a\varphi\p_b\varphi
 \nonumber \\
 &+&\frac Q{8\pi}\left(\int_{\Sigma}\d^2\xi
 \sqrt{\hg}~\widehat{R}\varphi
 +\oint_{\p\Sigma}\d \widehat s ~\widehat k \varphi \right)
 \nonumber \\
 &+&\frac{\mu}{8\pi\gamma^2}\int_{\Sigma}\d^2\xi
 \sqrt{\hg}\;\ee{\gamma}\varphi
 +\frac{\rho}{4\pi\gamma^2}\oint_{\p \Sigma}
 \d \widehat{s}\; \ee{\gamma\varphi/2} 
\;,
\eeqn
 where $\hg_{ab}$ is a reference metric and 
 $\hg_{ab}~\ee{\gamma\varphi}$ is 
 a physical metric, $\widehat{R}$ and $\widehat{k}$ are
 respectively 
 the curvature and the extrinsic curvature of the boundary with
 respect to the reference metric $\hg_{ab}$.
  We denote by $\mu$ and $\rho$
  bulk and boundary cosmological constants respectively.
 Classically, $Q=2/\gamma$ where $\gamma$ is the Liouville
 coupling constant.
   Let us consider the correlation function
\beq
 \VEV{\prod_i \ee{\alpha_i\varphi(z_i)}}
 =\int {\cal D}\varphi~ \ee{-S_{L}}
 \prod_i \ee{\alpha_i\varphi(z_i)}
\;.
\eeq
 We obtain the classical equation of motion:
\beq
\label{eq:eq of Liouville} 
 \frac 1{4\pi}\Delta\varphi
 -\frac{\mu}{8\gamma}\;\ee{\gamma\varphi}
 +\sum_i \alpha_i\delta^{(2)}(z-z_i) =0
\;.
\eeq
 Since the curvature of the physical metric is 
\beq
 R[\ee{\gamma\varphi}\hg]=-\ee{-\gamma\varphi}
 \Delta(\gamma\varphi)
\;,
\eeq
 \eq{eq:eq of Liouville} describes a surface with constant
 negative curvature and
 the inserted  operators $\ee{\alpha_i\varphi(z_i)}$
 play the role of the sources of curvature.
  Note that in the absence of a boundary a solution exists
 only when
\beq
 X=\sum_{i}\alpha_i +\frac Q2 (2h-2)
\eeq
 is positive, where $h$ is the number of handles.
 The nature of the surface and hence the nature of associated
 quantum states depend crucially on the sign of $X$.
   When there are boundaries, a classical solution always
 exists. Let us restrict our attention to the case with a single
 boundary and discuss whether the boundary can be replaced
 by local sources of curvature.
  In this case the nature of the surface depends crucially
 on the sign of
\beq
 Y=X+\frac 12 Q=\sum_i \alpha_i -\frac 12 Q \chi
\;.
\eeq
 
\underline{Case 1}: Fixed $\mu,\;Y>0$.
 When the loop is shrunk to a point, there exists a classical
 solution with constant negative curvature. A small loop
 behaves like a local source of curvature $Q/2$.

\underline{Case 2}: Fixed $\mu,\;Y<0$.
 When the loop is shrunk to a point, there is no classical
 solution with constant negative curvature.
  We can understand this case better if we constrain the
 area of the surface to be $A$.

\underline{Case 2-1}: Fixed $A \gg \l^2,\;Y<0$.
  The classical solution has positive constant curvature
 and the small-$\l$ limit is smooth and the loop becomes 
 as a puncture with curvature $Q/2$.

\underline{Case 2-2}: Fixed $A \ll \l^2,\;Y<0$.
 The classical solution has negative constant curvature
 and the loop cannot be thought of as a local disturbance.

 So far we have discussed classically.
 Semi-classically, $Y$ must be modified by
\beq
 Y=X+\alpha_{\rm mim}
\;,
\eeq
 where $\alpha_{\rm mim}$ is the curvature associated with
 the
 lowest dimension operator ${\cal O}_{\rm min}$ ( the dressed
 identity operator, in the case of the unitary minimal matter)
 because
 this is the maximum curvature that can be localized in a
 point in the quantum theory. 
  Similar observations follow in semi-classical discussion.
 In case 1 and case 2-1 the loop can be replaced by a sum of local
 operators and the contribution to the amplitude give rise to
 non-analytic terms in $\mu$.
 In case 2-2 the loop cannot be replaced by a
 sum of local operators and the contribution to the amplitude 
 give rise to analytic terms in $\mu$; the loop length
 $\l$ plays the role of ultraviolet cutoff.

\newpage
\begin{center}
\section{Three-loop correlators and fusion rules}
\label{sec:3-loop}
\end{center}

     In this section we consider the loop correlators
  in the unitary minimal models $(m+1,m)$ coupled to
 two dimensional
  gravity and the physical information we can extract from these.
    As shown in the case of one-matrix model \cite{MSS},
 the loop correlators 
 are expected to have much more information than those of local
 operators.
     We calculate the three-loop correlators in the
  systems stated above,
  from the two-matrix model
  with symmetric potential, at the $(m+1,m)$ critical points
  and show that simple fusion rules exist for the
 loop correlators and for all of the scaling operators.

\subsection{Formula for n-resolvent correlator}
\label{sec:formula for n-resolvent}

    Consider the connected part of the n-point correlators of the
 resolvents, which we introduced in sec.~\ref{sec:macroscopic loops}:
\beq
 \hat{W}^{+}(p_{i})\equiv \Tr\frac{1}{p_i -\hA}
 \qquad,\qquad
 \hat{W}^{-}(p_{j})\equiv \Tr\frac{1}{p_j -\hB}
 \qquad .
\label{eq:hW+-}
\eeq

   First, let us show briefly the formula for the n-resolvent correlators
  we obtained in \cite{AI}\footnote{The formula for the multi-loop
 amplitudes in the case of the general one-matrix model was
 derived in \cite{AJM}.}.
     The explicit derivation of the formula will
  be shown in section ~\ref{sec:n-resolvent} later. 
     At the $(m+1,m)$ critical point, we obtained the following formula for 
  the n-resolvent correlator:
\beqn
 \left(\frac{N}{\Lambda}\right)^{n-2} \DVEV{
 \prod_{i=1}^{n} \Tr \frac{1}{p_{i} - \hA} } 
  = \prod_{i=1}^{n} \left( - \frac{\p}{\p(a \zeta_{i})} \right)
 R^{(n)}(\zeta_{i},\Lm_{i} ) |_{\Lambda_{i} = \Lambda} \;\;\;.
\label{eq:n-resolvent}
\eeqn
    Here we denote by $\left\langle\!\VEV{\cdots}\!\right\rangle$
  the connected part of the
  averaging with respect to the matrix integrations and  $R^{(n)}$
  is some function of $\zeta_{i}$ and $\Lm_{i}$ 
  through $z_{i}^{*}$ and their derivatives with respect to the bare
  cosmological constant $\Lm_{i}$.
  Note that we introduced independent cosmological constants
 $\Lm_{i}$ for each loop for the convenience of the calculation.
  We put $\Lm_{i}=\Lm$ at the end of the calculation
 in \eq{eq:n-resolvent}.
  The function  $z_{i}^{*}$ of $\zeta_{i}$ and $\Lm_{i}$
 is parametrized as follows,
\beq
  z_{i}^{*} = \exp (2\eta_{i} \cosh \theta_{i})
  \;\;\; ,
\label{eq:zi*}
\eeq
where
\beq
 p_{i}- p_{*} \equiv a \zeta_{i}
 = A(z_{i}^{*}; \Lm_{i}) - A_{*}
 = a M_{i} \cosh m\theta_i \quad,\quad
 \eta_{i}=(a M_{i}/2)^{1/m} \quad ,
\label{eq:pi}
\eeq
\beq
 \Lm_{i}-\Lm_*=-(m+1) \eta_{i}^{2m}=-a^2 \mu_{i}\;, \;\;
 \left(\frac{M_{i}}{2}\right)^2 = \frac{\mu_{i}}{m+1}\;\;\;.
\label{eq:Lmi}
\eeq
 where $ p_{*}$ and $A_{*} $ represent  the critical values of
 $p_{i}$  and  $A(z;\Lm)$  respectively.
 We denote by $\zeta_i$ and $\mu_i$
 the renormalized boundary and
 bulk cosmological constants  for the corresponding loop
 respectively.

     The origin of  the parametrization  \eq{eq:zi*}  comes from
 the planar solution to the Heisenberg algebra 
 (\ref{eq:classical functions}).
     In fact, the function $A(z,\Lm)$ in \eq{eq:pi} represents
  the solution  at the $(m+1,m)$ critical point.

  The function $R^{(n)}$ is easily written down for lower $n$.
  For $n=2,3$, we have
\beqn
 \frac{\p}{\p \Lm}
 R^{(2)}  &=&  \sum_{i=1}^{2} 
 \frac{\p z^*_i}{\p \Lm_i}
 \prod_{j (\neq i)}^{2} (z_{i}^{*}-z_{j}^{*})^{-1}
 \;\;,
\label{eq:LmR2} \\
 R^{(3)} &=& \sum_{i=1}^{3} 
 \frac{\p z^*_i}{\p \Lm_i}
 \prod_{j (\neq i)}^{3} (z_{i}^{*} -z_{j}^{*})^{-1}
 \;\;\; .
\label{eq:R3}
\eeqn
  For $n=4, 5$, the correlators can be written compactly
  using graphs as  introduced below:
\beqn
 R^{(4)}  
 &=&
 \sum \; \frac{ \p}{\p\Lambda_{i_1} }
 \Biggl\{
 \epscenterboxy{2.5cm}{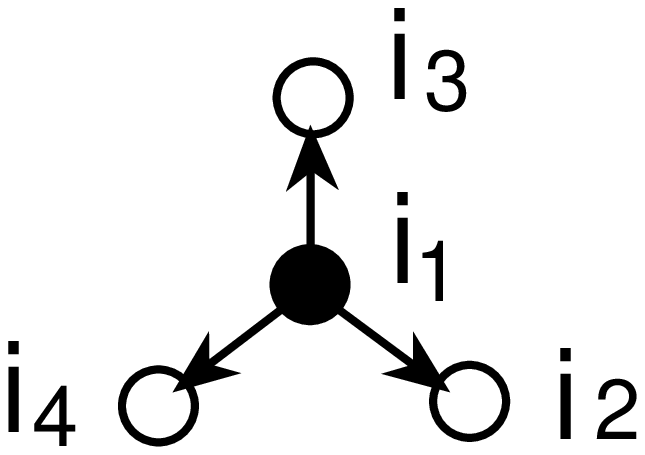}
 \Biggr\} \nonumber \\
 &+& 
 \sum \; \Biggl\{
 \epscenterboxy{1.8cm}{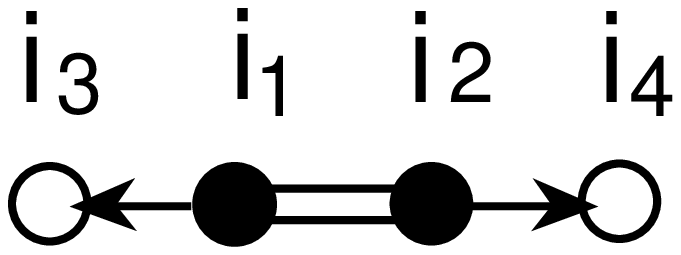}
 \Biggr\} 
 \quad  ,
\label{eq:R4}
\eeqn
\beqn
 R^{(5)}  
 &=&
 \sum \; \Bigl(\frac{ \p}{\p\Lambda_{i_1} }\Bigr)^2
 \Biggl\{
 \epscenterboxy{2.7cm}{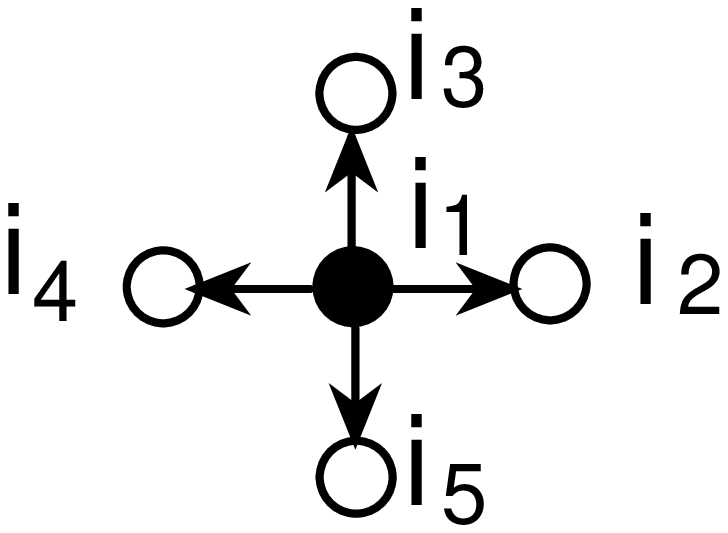}
 \Biggr\} \nonumber \\
 &+& 
 \sum \; \Bigl(\frac{ \p}{\p\Lambda_{i_1} }\Bigr)
 \Biggl\{
 \epscenterboxy{2.5cm}{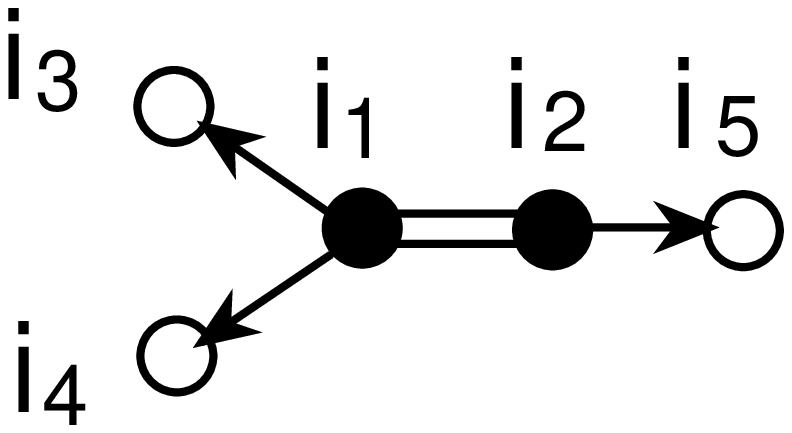}
 \Biggr\} \nonumber \\
 &+& 
 \sum \; \Biggl\{
 \epscenterboxy{1.8cm}{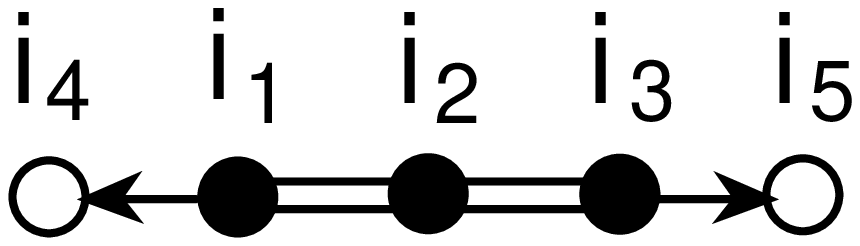}
 \Biggr\} 
 \quad.
\label{eq:R5}
\eeqn
In these figures a double line linking circle $i$ and circle $j$,
  a single line
  having an  arrow from circle $i$ to circle $j$  and a solid circle $i$ 
 represent
$(z_{i}^{*} -z_{j}^{*})^{-2}$, 
$(z_{i}^{*} -z_{j}^{*})^{-1}$ and  
$\frac{\p z^*_i}{\p \Lambda_i}$ respectively.
 The summations are over all possible graphs that
 have the same topology specified.
Each graph appears just for once in the summation.
Note that the links  to the external circles
 are not double lines but   the single ones with arrows and that the
 internal circles are solid circles.

For general $n$, the function $R^{(n)}$ is expressed in
 the same way.
The rule is as follows.
First, we consider all possible graphs which have $n$ circles
 and 
$n-1$ links in the same way as
  in the case $n=5$. 
Second, if  the internal solid circle $i$ has  $l_{i}$ links in
 each graph, the graph is operated by
$\prod_i \Bigl (\frac{\p}{\p \Lambda_i} \Bigl)^{l_{i-2} }$.
Then the summation over  all graphs gives the expression
 for  $R^{(n)}$.

\subsection{Three-loop correlators}

\subsubsection{Derivation of three-resolvent correlators}

   As a example, let us calculate explicitly the three-loop
 correlator, which we examined in \cite{AII2},
 in order to understand how we got the formula 
 from the classical solution to Heisenberg relation
 at the $(m+1,m)$ critical
 point.

   In the second quantized free fermion formalizm ( see, for example, 
 \cite{BDSS,IM} ), 
 the connected part of the correlator consisting of
 the product of
 arbitrary analytic functions $f^{(i)}(\hA)$
 in two-matrix model can be expressed as
\beqn
\lefteqn{
 \DVEV{ \prod_{i=1}^{n}\Tr f^{(i)}(\hA) }
 }
 \nonumber \\
 &&=
 \langle\!\langle N|\prod_{i=1}^{n}
 :\int \d\mu_{i} \Psi^{\dagger}(\tlm_i) f^{(i)}(\lm_i)
 \Psi(\lm_i) : |N \rangle\!\rangle
 \nonumber \\
 &&=
 \langle\!\langle N|
 \prod_{i=1}^{n} :a^{\dagger}_{k_i} a_{l_i} :
 |N \rangle\!\rangle
 \prod_{i=1}^{n} \int\d\mu_i
 \xi_{k_i}(\tlm_i) f^{(i)} (\lm_i) \xi_{k_i}(\lm_i)
 \nonumber \\
 &&=
 \langle\!\langle N|
 \prod_{i=1}^{n} :a^{\dagger}_{k_i} a_{l_i} :
 |N \rangle\!\rangle
 \prod_{i=1}^{n} \langle k_i|f^{(i)}(A)|l_i \rangle
 \;\; ,
\label{eq:free-fermi}
\eeqn
where
\beq
 \Psi(\lm)=\sum_k a_k \; \xi_k(\lm) 
 \; ,
\label{eq:2ndop}
\eeq
\beq
 \Psi^{\dagger}(\tlm)=\; \sum_k a^{\dagger}_k \xi_k(\tlm) 
\label{eq:2ndop-dagger}
\eeq
 are second quantized free fermion fields
 constructed from the orthogonal polynomials $\xi_k$ and 
 $\left.\left| N \right\rangle\!\right\rangle$ is
 a state corresponding to the filled fermi sea,
\beq
 a_k \left.\left| N \right\rangle\!\right\rangle =0
 \;\; \mbox{for} \;\; k\geq N
\label{eq:fermi-sea-1}
\eeq
\beq
 a^{\dagger}_k \left.\left| N \right\rangle\!\right\rangle =0
 \;\; \mbox{for} \;\; k\leq N-1
 \; .
\label{eq:fermi-sea-2}
\eeq
   The normal ordering $:\cdots :$ is with respect to
 $\left.\left| N \right\rangle\!\right\rangle$.
    The connected three-point
  correlator consisting of arbitrary analytic functions
  $f(\hA)$, $g(\hA)$, and $h(\hA)$
  can thus be expressed as
\beqn
 \DVEV{ \Tr f(\hA)\;\Tr g(\hA)\; \Tr h(\hA)  }
 &=& \sum_{i=0}^{N-1} \sum_{k=N}^{\infty} \sum_{l=N}^{\infty}
 \;[f(A)]_{ik} \;[g(A)]_{kl} \;[h(A)]_{li} \nonumber \\
 &-& \sum_{i=0}^{N-1} \sum_{k=N}^{\infty} \sum_{l=0}^{N-1}
 [f(A)]_{ik} \;[h(A)]_{kl} \;[g(A)]_{li}
 \; ,
\label{eq:fgh-1}
\eeqn
where
$
 [f(A)]_{ik} \equiv \langle i| f(A) |k \rangle
$.

    Because we are interested in the case of large $N$ limit only,
  it is convenient to use
  the `classical' function introduced
 sect.~\ref{sec:classical functions}.
    Note that the `classical' function depends on the
  bare cosmological constant $\Lambda$ only
  through $s$ when we take $N\to\infty$ limit.
It is, therefore, legitimate
to introduce
$
\displaystyle
 A(z,s) \equiv \lim _{N\to \infty }A(z,s,\Lambda)
:$
\beq
 A(z,s,\Lambda) = A(z,s) + O(1/N)
\quad .
\label{eq:A(z)}
\eeq

    Since the matrix elements $A_{ij}$ only near the diagonal
 are not zero,
 the matrix element $[f(A)](i)_k \equiv [f(A)]_{i-k,i}$ 
 can be replaced with the coefficient of $z^k$ for
 the `classical' 
 function $A(z,s=\Lambda)$,
\beq
 [f(A)](N)_k =\frac 1{2\pi i} \oint \frac{\d z}{z^{k+1}} f\left(
 A(z,s=\Lambda)\right) + O(1/N)
\quad ,
\label{eq:f(A)-1}
\eeq
 at large $N$ limit. 

In the right-hand side of \eq{eq:fgh-1},
the leading terms in $1/N$ of the first term and
those of the second term
get cancelled. We have to consider
the
next leading
terms. For any integer $\epsilon$, we obtain
\beqn
 [f(A)](N+\epsilon)_k &=& \frac 1{2\pi i} \oint\frac{\d z}{z^{k+1}} f\left(
 A(z,s=\Lambda)\right) \nonumber\\
 &+& \frac{\Lambda\epsilon}N \frac 1{2\pi i} \oint\frac{\d z}{z^{k+1}} \left.
 \frac{\p A(z,s)}{\p s} \right|_{s=\Lambda}
 \frac{\p f\left(A(z,\Lambda)\right)} {\p A} \nonumber \\
 &+&(\mbox{the part independent of $\epsilon$}) +O(1/N^2)
 \quad .
\label{eq:f(A)-2}
\eeqn
 The part independent of $\epsilon$ comes from the terms $O(1/N)$ in
 \eq{eq:f(A)-1}. The second term is
 responsible for the computation in what follows.
 Using \eq{eq:f(A)-2}
 and considering the terms $1/N$ in \eq{eq:fgh-1}, we obtain
\beqn
 \lefteqn{\!\!\!\!
 \DVEV{ \Tr f(\hA)\; \Tr g(\hA)\; \Tr h(\hA) }
  =
 \frac{\Lm}N \sum_{\delta_1=0}^{\infty}\sum_{\delta_2=0}^{\infty}
 \sum_{\delta=0}^{\infty }
 }
 \nonumber \\
 &&
 \left\{ (\delta_2-\delta_1)\;\frac{\p}{\p s}[f(A)](N)_{\delta_1+\delta_2+1}\;
 [g(A)](N)_{\delta-\delta_2}\; [h(A)](N)_{-\delta-\delta_1-1} \right.
 \nonumber \\
 &&+  (\delta+\delta_2+1)\; [f(A)](N)_{\delta_1+\delta_2+1}\; \frac{\p}{\p s}
 [g (A)](N)_{\delta-\delta_2}\; [h(A)](N)_{-\delta-\delta_1-1} \nonumber \\
 &&+ \left. (\delta-\delta_1)\; [f(A)](N)_{\delta_1+\delta_2+1}\;
 [g(A)](N)_{\delta-\delta_2}\; \frac{\p}{\p s}[h(A)](N)_{-\delta-\delta_1-1}
 \; \right\} \nonumber \\
 &&+ O(1/N^2)
 \quad .
\label{eq:fgh-2}
\eeqn
  Here we have used the fact that at large $N$, the original summations in
 \eq{eq:fgh-1}
  can be safely replaced with the triple summations from zero
  to infinity.

    The summations can be carried out  after putting this equation 
  into the form of contour integrals using \eq{eq:f(A)-1}.
  The three-point function \eq{eq:fgh-1} in the planar
  limit can thus
  be expressed in terms of the `classical' function in the
  form of contour integrals,
\beqn
 \lefteqn{\!\!\!\!
 \DVEV{ \Tr f(\hA)\; \Tr g(\hA)\; \Tr h(\hA) } 
 =
 \frac{\Lm}{N} \frac{1}{(2\pi i)^3} \oint_{|z_1|>|z_2|>|z_3|}
 \d z_1\d z_1\d z_3
 } \nonumber \\
 &&
 \left\{\; \frac{z_1}{(z_1-z_2)^2(z_1-z_3)^2} \;
 \frac{\p}{\p \Lm} \Bigl[f\left(A(z_1)\right) \Bigr]
 g\left(A(z_2)\right) h\left(A(z_3)\right) \right.
 \nonumber \\
 &&+\;
 \frac {z_2}{(z_2-z_1)^2(z_2-z_3)^2} \; 
 f\left(A(z_1)\right)
 \frac{\p}{\p \Lm} \Bigl[\left(A(z_2)\right) \Bigr]
 h\left(A(z_3)\right)
 \nonumber \\
 &&+ \left. \;
 \frac {z_3}{(z_3-z_1)^2(z_3-z_2)^2} \;
 f\left(A(z_1)\right) g\left(A(z_2)\right)
 \frac{\p}{\p \Lm} \Bigl[h\left(A(z_3)\right)\Bigr]
 \; \right\}
 \quad ,
\label{eq:fgh-3}
\eeqn
where we set $A(z) \equiv A(z, s = \Lambda)$.
 The condition for the contour paths $|z_1|>|z_2|>|z_3|$
 follows from the condition
 that makes the infinite summations converge.

   From the above expression (\eq{eq:fgh-3}), the three-point resolvent in
the planar limit is expressed as
\beqn
 \lefteqn{\!\!\!\!
 \frac N{\Lm_*} \DVEV{ \Tr\frac 1{p_1-\hA}\; \Tr\frac 1{p_2-\hA}\;
 \Tr\frac 1{p_3-\hA} }
 =
 \frac 1{(2\pi i)^3} \oint_{|z_1|>|z_2|>|z_3|}
 \d z_1\d z_2\d z_3
 } 
 \nonumber \\
 &&
 \left\{\; \frac {z_1}{(z_1-z_2)^2(z_1-z_3)^2} \;
 \frac{\p}{\p\Lm}\Biggl[\frac{1}{p_1-A(z_1)} \Biggr]
 \frac{1}{p_2-A(z_2)} \frac{1}{p_3-A(z_3)} \right. \nonumber \\
 &&+\;
 \frac {z_2}{(z_2-z_1)^2(z_2-z_3)^2}
 \frac{1}{p_1-A(z)}
 \frac{\p}{\p\Lm} \Biggl[ \frac{1}{p_2-A(z_2)} \Biggr]
 \frac{1}{p_3-A(z_3)} \nonumber \\
 &&+ \left. \;
 \frac {z_3}{(z_3-z_1)^2(z_3-z_2)^2} \;
 \frac{1}{p_1-A(z_1)} \frac{1}{p_2-A(z_2)}
 \frac{\p}{\p\Lm} \Biggl[\frac{1}{p_3-A(z_3)}\Biggr]
 \;\right\}
 ,
\label{eq:p1p2p3-1}
\eeqn
 where the contour of $z_1$ encircles that of $z_2$
 and similarly the contour of $z_2$ encircles that of $z_3$.
  We pick up only the contributions form the poles
 at $z_{i}=z_{i}^*$ of the parts 
 $1/[p_i-A(z_i)]$.  The contributions from the poles
 at  $z_{i}=z_{j}^*\;(i\ne j)$ give rise to terms which
 do not have the inverse Laplace image.
  We thus discard these terms.

    We obtain the following rather simple expression for the
 three-resolvent correlator,
\beqn
 \lefteqn{\!\!\!\!
 \frac{N}{\Lm_*} \DVEV{
 \Tr\frac 1{p_1-\hA}\; \Tr\frac 1{p_2-\hA}\;
 \Tr\frac 1{p_3-\hA}  }
 }
 \nonumber \\
 &=&
 -\frac{1}{a^3} \left\{
 \frac{\p}{\p\Lm_1}
 \left[ \frac{1}{(z_1^*-z_2^*)^2(z_1^*-z_3^*)^2}
 \frac{\p z_1^*}{\p \zeta_1}\right]
 \frac{\p z_2^*}{\p \zeta_2}
 \frac{\p z_3^*}{\p \zeta_3} 
 +
 (1\leftrightarrow 2)
 +
 (1\leftrightarrow 3)
 \right\}
 \nonumber \\
 &=&
 -\frac{1}{a^3}
 \frac{\p}{\p\zeta_1} \frac{\p}{\p\zeta_2}
 \frac{\p}{\p\zeta_3}
 \left. \left\{
 \frac{1}{(z_1^*-z_2^*)(z_1^*-z_3^*)}
 \frac{\p z_1^*}{\p \Lm_1}
 +
 (1\leftrightarrow 2)
 +
 (1\leftrightarrow 3)
 \right\} \right|_{\Lm_i=\Lm}
 .
 \nonumber \\
\label{eq:p1p2p3-2}
\eeqn
 Here, the poles $z_i^*$ are determined through the
 classical solution to the Heisenberg relation and are
 parametrized as \eq{eq:zi*}.

   \Eq{eq:p1p2p3-2}  agrees with the set of \eq{eq:n-resolvent} and
 \eq{eq:R3}, the formula for
 three-resolvent correlator.
  We have shown how we can get the formula
 for $n$-resolvent correlators for the case of $n=3$ explicitly.

\subsubsection{Three-loop correlators in terms of loop lengths}

    Next, let us consider to how to get the
 expression for three-loop
 correlators in terms of the loop lengths,
 performing inverse Laplace
 transformation with respect to the renormalized boundary cosmological
 constants $\zeta_{i}$.
   We will show, later, that much physical information can be extracted 
 from the three-loop correlator.
 The generalization to higher loop will be discussed
 in sect.~4.

    First, we will show we can put \eq{eq:p1p2p3-2} into
 a form in which
 the correlator is expressed as a sum of the product of three
 factors each of which depends only on $\zeta_i$ corresponding
 to individual loop.
   In order to show this, first note that \eq{eq:p1p2p3-2} can be written as
\beqn
 \frac N\Lm_* \Bigl\langle \Tr\frac 1{p_1-\hA} &\Tr&\frac 1{p_2-\hA}\;
 \Tr\frac 1{p_3-\hA} \Bigr\rangle _c \nonumber \\
 &=&\frac 1{a^3 2^2 m(m+1)}\left(\frac{aM}2\right)^{-2-1/m}
\frac{\p}{\p\zeta_1}
 \frac{\p}{\p\zeta_2}\frac{\p}{\p\zeta_3} F(\theta_1,\theta_2,\theta_3)
 \;,
 \nonumber \\
 &&
\label{eq:p1p2p3-3}
\eeqn
where
\beqn
 F(\theta_1,\theta_2,\theta_3)
 &=&
 \frac{1}{(\cosh\theta_1-\cosh\theta_2)(\cosh\theta_1-\cosh\theta_3)}
 \;\frac{\sinh (m-1)\theta_1}{\sinh m\theta_1} \nonumber\\
 &+& \quad (1\leftrightarrow 2) \quad + \quad (1\leftrightarrow 3)
 \quad .
\label{eq:F123-1}
\eeqn
 Here the following identity is crucial;
\beqn
 \lefteqn{\!\!\!
 \frac{1}{\cosh\alpha-\cosh\beta}
 \left(\;\frac{\sinh (n-k)\alpha}{\sinh n\alpha}
 -\frac{\sinh (n-k)\beta}{\sinh n\beta} \;\right) \mbox{                 }
 } \nonumber\\
 \mbox{  }
 &&=-2\sum_{j=1}^{n-k}\sum_{i=1}^{k}
 \;\frac{\sinh (n-j-i+1)\alpha}{\sinh n\alpha}
 \;\frac{\sinh (n-j-k+i)\beta}{\sinh n\beta}
 \quad .
\label{eq:cosha}
\eeqn
    Making use of the above identity twice, 
  we find that \eq{eq:F123-1}
is written as a triple sum where the summand factorizes
into three factors associated with individual loops:
\beqn
 \lefteqn{\!\!\!\!\!
 F(\theta_1,\theta_2,\theta_3) } \nonumber \\
 &=&
 4\sum_{k=1}^{m-1}\sum_{j=1}^{m-k}\sum_{i=1}^{k}
 \;\frac{\sinh (m-k)\theta_1}{\sinh m\theta_1}
 \;\frac{\sinh (m-j-i+1)\theta_2}{\sinh m\theta_2}
 \;\frac{\sinh (m-k-j+i)\theta_3}{\sinh m\theta_3}
 \; .
 \nonumber\\
\label{eq:F123-2}
\eeqn

   Here, we should specify the definition of continuum
 amplitudes at large $N$.
  Since the leading term of
 $\left\langle\!\! \VEV{\prod_{i=1}^{n}\;\hat{W}
 (p_i) }\!\!\right\rangle$ is of order of $a^{-n}\kappa^{n-2}$,
 where $\kappa \equiv a^{-2-\frac{1}{m}}(\Lm_*/N)$,
 we should renormalize to obtain continuum quantities.
 The renormalized resolvent is defined as
\beq
 \hat{W}^{\mbox{\it ren}}(\zeta_i)
 = \frac{a}{\kappa}\;\hat{W}(p_i)
 =\frac{a}{\kappa}\Tr\frac{1}{p_i-\hA}
 \; ,
\label{eq:Wren}
\eeq
 and the renormalized expectation is defined as
\beq
 \VEV{\cdots\cdots}^{\mbox{\it ren}}
 = \kappa^2 \left\langle\!\VEV{\cdots\cdots}\!
 \right\rangle
\; .
\label{eq:VEVren}
\eeq
 We will omit the superscript {\it ren} from now on.
 The continuum three-loop correlator
 $\VEV{w^+(\l_1) w^+(\l_2) w^+(\l_3)}$
 is defined by the inverse
 Laplace image of the continuum resolvent correlator,
 that is,
\beqn
\lefteqn{
 \VEV{\hW^+(\zeta_1) \hW^+(\zeta_2) \hW^+(\zeta_3) }
 }
 \nonumber \\
 &&= \int^{\infty}_{0} d \ell_{1}
      \int^{\infty}_{0} d \ell_{2}
      \int^{\infty}_{0} d \ell_{3}
      e^{- \zeta_1 \ell_1} e^{- \zeta_2 \ell_2}
      e^{- \zeta_2 \ell_3}
      \VEV{w^+(\l_1) w^+(\l_2) w^+(\l_3)}
 \;\;\;  \nonumber \\
 &&\equiv {\cal L}
 \left[\VEV{w^+(\l_1) w^+(\l_2) w^+(\l_3)}\right]
 \quad ,
\label{eq:Laplace}
\eeqn
 where $w^{\pm}(\l)$ is an operator which makes  hole with finite
 boundary (loop) length $\l$.
 Due to the following formula for the inverse Laplace image
\beqn
{\cal L}^{-1}\Biggl[\frac{\partial}{\partial \zeta}
         \frac{\sinh k \theta}{\sinh m \theta}\Biggr]
= - \frac{M \ell}{\pi} \sin \frac{k \pi}{m}~  K_{\frac{k}{m}} (M \ell)
\quad ,
\label{eq:inverse-Laplace}
\eeqn
 where $K_{\nu}(z)$ is the modified Bessel function,
 we can obtain the continuum three-loop amplitude in a rather compact
 form:
\beqn
 \lefteqn{
 \VEV{w^+(\l_1) w^+(\l_2) w^+(\l_3)}
 } \nonumber \\
 &&= -\frac{1}{m(m+1)} \left(\frac M2\right)^{-2-\frac 1m}
 \nonumber \\
 && \;\;
 \left(\frac M2\right)^{3} \l_1 \l_2 \l_3
 \sum_{k=1}^{m-1}\sum_{j=1}^{m-k}\sum_{i=1}^{k} \;
 \tK{\frac{m-k}{m}}(M \ell_1)\;
 \tK{\frac{m-j-i+1}{m}}(M \ell_2)\;
 \tK{\frac{m-k-j+i}{m}}(M \ell_3)
 \; ,
 \nonumber \\
 \quad
\label{eq:w+w+w+-1}
\eeqn
 where we introduced a notation,
\beq
 \tK{p}(M\l)\equiv \frac{\sin\pi |p|}{\pi/2} \;
 K_{p}(M\l)
 \quad .
\label{eq:tK}
\eeq
   The expression for the summation 
 in \eq{eq:w+w+w+-1} looks asymmetric with respect to
 the loop indices.
 By elementary algebras, we can convert it
 into a form which have explicit symmetry with respect to
 the interchange of loops:
\beqn
 \lefteqn{
 \VEV{w^+(\l_1) w^+(\l_2) w^+(\l_3)}
 } \nonumber \\
 &&= -\frac{1}{m(m+1)} \left(\frac M2\right)^{-2-\frac 1m}
 \nonumber \\
 && \;\;
 \left(\frac M2\right)^{3} \l_1 \l_2 \l_3
 \sum_{(k_1-1,k_2-1,k_3-1) \atop \in {\cal D}_{3}^{(m)} } \;
 \tK{1-\frac{k_1}{m}}(M \ell_1)\;
 \tK{1-\frac{k_2}{m}}(M \ell_2)\;
 \tK{1-\frac{k_3}{m}}(M \ell_3)
 \;\; ,
 \nonumber \\
 \quad
\label{eq:w+w+w+-2}
\eeqn

  Here  we have denoted by ${\cal D}_{3}^{(m)}$
\beqn
 {\cal D}_3^{(m)}
 &=& \Biggl\{ (a_1,a_2,a_3) \Bigm|
      \sum^3_{i (\not= j)} a_i  - a_j \ge 0 
      \;\mbox{for}\; j=1\sim 3\;,
 \nonumber \\
 &&\quad
       \sum^3_{i=1}a_i = {\rm even}\le 2(m-2)\;,
       a_i=0,1,2,\cdots\;  
       \Biggr\}
\;.
\label{eq:D3(m)}
\eeqn

   Eqs.~(\ref{eq:w+w+w+-2}) and (\ref{eq:D3(m)}) give the
 final expression for the three loop
 correlator at the $(m+1,m)$ critical point.
   It is interesting that the selection rule obtained
 in eqs.~(\ref{eq:w+w+w+-2}) and (\ref{eq:D3(m)})
 agrees  exactly  with the fusion rules for
 the diagonal primary fields
 in the Kac table
 of underlying conformal field theory of the unitary minimal
 model $(m+1,m)$ \cite{BPZ}.
   In fact, the fusion rules for the diagonal primary fields  read as
\beqn
 \langle \phi_{i i}~ \phi_{j j}~ \phi_{k k}\rangle
\not= 0 \quad,
\eeqn
if and only if
$i+j \geq k+1$ and two other permutations and $~i+j+k
{}~(=$ odd) $\leq 2m-1$ hold.
This set of rules  is nothing but ${\cal D}_{3}^{(m)}$.

\subsubsection{Boundary conditions of loops}
\label{sec:boundary conditions}

    A similar expression to \eq{eq:w+w+w+-2} was obtained
 in \cite{Kostov},
 where loop correlators were examined for closed string with
 one-dimensional discrete target space, that is, the degrees
 of freedom for 
 matter part are labeled by the points of Dynkin diagram.
    The matter degrees of freedom are labeled also by
 the discrete momentum $p$ instead of the discrete 
 target space coordinate $x$.
    They examined the loop correlators 
 treating  the boundary condition of each loop $\l_i$ to be 
 specified by a single momentum $p_i$. 
    Thus, it follows directly that the three-loop correlator 
 which is specified by three momenta $p_1, p_2$ and $p_3$ 
 and loop lengths,
 is proportional to the expectation value for
 wave functions for matter part, that is,
\beq
 \tilde{C}_{p_1 p_2 p_3}=\sum_{x} S^{x}_{(p_1)} S^{x}_{(p_2)}
 S^{x}_{(p_3)}  / S^{x}
 \;,
\label{eq:Cp1p2p3}
\eeq
 where $S^{x}_{(p_i)}$ and $S^{x}$ are the  wave function
 of a point particle moving on the discrete target space
 with  momentum $p_i$ and that for ground state respectively,
\beqn
 && S^{(x)}_{p}=\langle x| p\rangle
 =\left(\frac 2{h-1}\right)^{1/2}\;
 \sin \pi p x 
 \\
 &&\qquad\qquad\qquad
 x=1,2,\cdots,h-1
 \nonumber \\ 
 &&\qquad\qquad\qquad
 p=1/h,2/h,\cdots,(h-1)/h \;.
 \nonumber
\label{eq:Spx}
\eeqn
 For example, for the $A_{h-1}$ Dynkin diagram,
 each momentum takes discrete values of
 $\frac{1}h,\cdots,\frac{h-1}h$ and
 $\tilde{C}_{p_1 p_2 p_3}$ is nonvanishing
 only when $h p_i$ satisfy the equivalent rule we found in
 \eq{eq:w+w+w+-2} and \eq{eq:D3(m)}, that is,
\beq
 (hp_1-1, hp_2-1, hp_3-1) \in {\cal D}_{3}^{(h)}
 \;.
\label{eq:hp}
\eeq
   
     The similarity between the three-loop correlator
  in the case of the closed strings in discrete
  target space and that in the two-matrix model we found
 in eqs. (\ref{eq:w+w+w+-2}) and (\ref{eq:D3(m)})
  indicates that the each terms in the sum in
 \eq{eq:w+w+w+-2} represents the amplitude with
 the the loops specified by the momentum $\frac{k_i}m$.
   It appears that we can decompose the loop operator
 in the two-matrix into parts each of which specified
 by a momentum $\frac km$:
\beq
 ``w^{+}(\l)\sim\sum_{k=1}^{m-1}\; c^{+}_k\;
 w_{k}(\l)"
 \;\; ,
\label{eq:w+-wk}
\eeq
  
%

   From the selection rules in \eq{eq:D3(m)}, we can deduce the fusion
 rules for the gravitational descendants as well as for the
 gravitational primaries. In other words, some fusion rules
 are satisfied among all of the scaling operators including the
 gravitational descendants as well.
   We suggested first that the selection rules in the three-loop
 correlator correspond to those for the gravitational
 primaries by examining the limit of small loop length 
 \cite{AII}. In \cite{Anazawa}, we obtained the fusion rules for
 all of the scaling operators from the three-loop correlator.
   We will discuss these issues in detail in later subsections
 after we examine the two-loop amplitude and
 an expansion of the loop operator in terms of
 local operators.

\subsection{Expansion of loop operators}

   In \cite{MSS,MS} it was proposed that a loop operator
 can be replaced
 by a sum of local operators if 
 the loop correlator does not diverge
 when the loop shrink to a point.
  This was discussed explicitly
 in the case of the one-matrix model and $c=1$ case.
   We apply this idea to the general minimal models coupled
 to gravity and discuss the correlation functions for
 the scaling operators.
   In order to derive the form of the expansion of the loop
 in terms of the local operators, let us first consider the
 two- and one-loop amplitudes.

\subsubsection{Two-loop correlators from three-loop correlators}

   Let us derive two- and one-loop correlators
 from the three loop correlator \eq{eq:w+w+w+-2}.
  As we shrink one of the three loops,
 the three-loop correlator
 should approach the derivative of the two-loop correlator
 with respect to the  cosmological constant.
   Consider shrinking  the third loop $M\l_3$ in \eq{eq:w+w+w+-2}.
 Since, for $M\l \ll 1$, we have
\beqn
 \frac{M\l}{2} \tK{1-\frac{k}{m}}(m\l)
 &&=\frac{M\l}{2} \; \left\{I_{-1+\frac{k}{m}}(M\l)
 -I_{1-\frac{k}{m}}(M\l) \right\}
 \nonumber \\
 &&\approx \frac{M\l}2 \; I_{-1+\frac{k}{m}}(M\l)
 \nonumber \\
 &&\approx \left(\frac{M\l}2\right)^{\frac{k}m}
 \; \frac{1}{\Gamma(\frac{k}m) }
 \;\; ,
\label{eq:tKsim}
\eeqn
 the leading contribution in the summation in \eq{eq:w+w+w+-2}
 comes from the part of $k_3=1$ and  we have 
\beqn
 \lefteqn{
 \VEV{w^+(\l_1) w^+(\l_2) w^+(\l_3)}
 } \nonumber \\
 &&\approx
 -\frac{1}{m(m+1)} \frac{ \l_3^{\frac{1}m } }
 {\Gamma(1/m)}  
 \; \l_1 \l_2
 \sum_{k=1}^{m-1} \;
 \tK{1-\frac{k}{m}}(M \ell_1)\;
 \tK{1-\frac{k}{m}}(M \ell_2)\;
 \;\; ,
\label{eq:w+w+w+sim}
\eeqn
 for $M\l_3 \ll 1$.
  This should be proportional to the derivative of two-loop
 correlator with respect to $\mu$.
   In fact, by the explicit calculation similar to 
 the case of  three-loop,
 one can obtain \cite{DKK,AII}
\beqn
 \lefteqn{
 \frac{\p}{\p \mu}
 \VEV{w^+(\l_1) w^{\pm}(\l_2)}
 } \nonumber \\
 &&=
 -\frac{1}{m(m+1)}
 \; \l_1 \l_2
 \sum_{k=1}^{m-1} \; (\pm)^{k-1} \;
 \tK{1-\frac{k}{m}}(M \ell_1)\;
 \tK{1-\frac{k}{m}}(M \ell_2)\;
 \;\; .
\label{eq:muw+w+-}
\eeqn
  It is clear that \eq{eq:w+w+w+sim} and \eq{eq:muw+w+-}
 are consistent.
 
  Due to a relation
 $\frac{\p}{\p \mu}=\frac{2}{m+1} \frac{\p}{M\p M}$
 and a formula of a integral (for $\alpha\ne\beta$),
\beq
 \int^{z} dz z K_{\nu} (\alpha z) K_{\nu} (\beta z)
 = \frac{-z}{\beta^2 - \alpha^2}
  \Bigl\{
        \beta K_{\nu} (\alpha z) K_{\nu - 1} (\beta z)
      - \alpha K_{\nu - 1} (\alpha z) K_{\nu} (\beta z)
  \Bigr\}
  ,
\label{eq:KnuKnu}
\eeq
 one can obtain the two-loop correlator (for $\l_1\ne\l_2$):
\beqn
 \lefteqn{
 \VEV{w^+(\l_1) w^{\pm}(\l_2)}
 } \nonumber \\
 &&=
 \frac{1}{m}
 \frac{M}{2} \frac{\l_1 \l_2}{\l_1^2 -\l_2^2} \;
 \sum_{k=1}^{m-1} \; (\pm)^{k-1}
 \left\{
 \l_1 \tK{\frac{k}{m}}(M \ell_1)\;
 \tK{1-\frac{k}{m}}(M \ell_2)\;
 \right. \nonumber \\
 &&\qquad\qquad\qquad\qquad\qquad\qquad -
 \left.
 \l_2 \tK{1-\frac{k}{m}}(M \ell_1)\;
 \tK{\frac{k}{m}}(M \ell_2)\;
 \right\}
 \nonumber \\
 &&=
 \frac{1}{m}
 \frac{M}{2} \frac{\l_1 \l_2}{\l_1 +(\pm)^m \l_2} \;
 \sum_{k=1}^{m-1} \; (\pm)^{k-1}
 \tK{\frac{k}{m}}(M \ell_1)\;
 \tK{1-\frac{k}{m}}(M \ell_2)\;
 \;\; .
\label{eq:w+w+-}
\eeqn

\subsubsection{One-loop amplitudes from two-loop correlators}

   Shrinking $\l_2$ in \eq{eq:w+w+-} as well, one should have 
 the derivative of the one-loop amplitude with respect to
 the cosmological constant.
   For $M\l_2 \ll 1$, we have
\beq
 \VEV{w^+(\l_1) w^{\pm}(\l_2)}
 \approx
 \frac{1}{m}
 \frac{ \l_2^{\frac{1}m } } {\Gamma(1/m)}  
 \; \left(\frac M2 \right)^{\frac{1}m}
 \tK{\frac{1}{m}}(M \l_1)
 \;\; .
\label{eq:w+w+-sim}
\eeq
  In fact by explicit calculation, one can obtain
\beq
 \frac{\p}{\p \mu}
 \VEV{w^{\pm}(\l_1)}
 =
 \frac{1}{m}
 \; \left(\frac M2 \right)^{\frac{1}m}
 \tK{\frac{1}{m}}(M \l_1)
 \;\; .
\label{eq:nuw+-}
\eeq
   Note that this amplitude is nothing but the wave function
 of the dressed identity operator.
  Performing the integral with respect to $\mu$, one obtain the one-loop
 amplitude:
\beq
 \VEV{w^{\pm}(\l_1)}
 = 
 - \left(1+\frac 1m\right) \; \l_{1}^{-1} \;
 \frac{\sin \pi/m}{\pi/2}
 \; \left(\frac M2\right)^{1+\frac{1}{m}}
 K_{1+\frac{1}{m}}(M \l_1)
 \;\; .
\label{eq:w+-}
\eeq
   Note that $\l_1 \VEV{w^{\pm}(\l_1)}$ is the wave function
 of a boundary operator  $\hsig{1+m}$ \cite{MMS} which couples to
 the boundary of two-dimensional surfaces (i.e. loops),
\beq
 \l_1 \VEV{w^{\pm}(\l_1)}
 = \left(1+\frac 1m \right)
 \left(\frac M2\right)^{1+\frac 1m}
 \; \tK{1+\frac 1m}(M\l_1)
 \;\; .
\label{eq:l1w+-}
\eeq

\subsubsection{Expansion of loops in local operators}

  In \cite{MSS}, in the case of one-matrix, it was argued
 that the loop operator can be expanded in terms of local
 operators inside the loop correlators, that is,
 the loop can be replaced with the infinite combination of
 local operators,
  except some special cases.
   Whether this replacement can be done safely or not is
 connected with whether the
 corresponding classical solution  exits or not in the limit
 of small length of corresponding loop.
   This claim is quite natural because, in the one-matrix model,
 all of the scaling operators are expressed in term of
 one matrix $\Phi$ as 
\beq
 \sigma_{j}=\Tr(1-\Phi)^{j+1/2}=\sum_{n}a_{n}(j) n^{-1}
 \Tr \Phi^{n} 
\;.
\eeq 
   On the other hand, in the two-matrix model, since there are two kinds of
 matrix $\hA$ and $\hB$, there can exist many kinds of microscopic
 loops, $\Tr(\hA^{n_1}\hB^{n_2}\hA^{n_3}\cdots )$. 
   In the case of the two-matrix model, thus, the direct connection of
 the scaling operators  to the loops $\Tr\hA^{\l/a}$ or 
 $\Tr\hB^{\l'/a}$ is not clear.
  At first sight, the expansion of loops in terms of local
 operators is not legitimate.
   We think, however, this expansion is possible
 by the following reason.
   When one of the loops on two-dimensional surface shrunk
 to a microscopic loop, 
 the loop represents local deformation of the surface.
  The microscopic loop can be considered
 to be replaced by the insertions of local operators.
  As we will see later,
 the loop correlators except one-loop case
 are continuous when the length of one of
 the loops approaches zero, so that we expect that
 a macroscopic loop can also be replaced by a sum
 of local operators.

   In the following, we find the form of the expansion of 
 the loop operator.
 First, let us represent the two-loop correlator in term of
 off shell
 renormalizable wave function \cite{MSS,MS},
\beq
 (E\sinh \pi E)^{1/2}\; K_{iE}(M\l)
\;\; .
\label{eq:off-shell}
\eeq
 We show that the two-loop correlators are expressed as
\beqn
 \lefteqn{
 \VEV{w^+(\l_1) w^{\pm}(\l_2)}
 =
 \sum_{k=1}^{m-1} \frac 1{2m} (\pm)^{k-1}
 \left( \frac{\sin \pi \frac km}{\pi/2} \right)^{2} \;
 } \nonumber \\ 
 && \qquad\qquad
 \int_{0}^{\infty}\d E
 \frac{E\sinh \pi E}{\cosh \pi E -\cos \pi \frac km}
 \; K_{iE}(M\l_1) \; K_{iE}(M\l_2)
\label{eq:w+w+-off}
\eeqn
From \eq{eq:muw+w+-}, it is reasonable to assume 
\beq
 ``w^{\pm}(\l)= \sum _{k=1}^{m-1} (\pm)^{k-1}
 \frac{\sin \pi \frac km}{\pi/2} \;
 w_{k}(\l)"
\label{eq:w+-sum}
\eeq
 in the case of two- and three-loop correlators, 
 where we have introduced loop operators
 $w_{k}(\l)$ which represent loops with 
 some distinct matter boundary condition 
 (see sect.~\ref{sec:boundary conditions}).
    From \eq{eq:muw+w+-}, we have
\beq
 \VEV{w^+(\l_1) w^{\pm}(\l_2)}
 =\sum_{k=1}^{m-1} (\pm)^{k-1}
 \left( \frac{\sin \pi \frac km}{\pi/2} \right)^{2} \;
 \VEV{w_{k}(\l_1) w_{k}(\l_2)}
\label{eq:w+w+-sum}
\eeq
and 
\beq
 \frac{\p}{\p M}
 \VEV{w_{k}(\l_1) w_{k}(\l_2)}
 =
 -\frac 1{m} \frac M2 \l_1 \l_2 \; 
 K_{1-\frac km}(M\l_1)\;K_{1-\frac km}(M\l_2)
 \; .
\label{eq:Mwkwk}
\eeq
   Making use of a formula 
\beq
 K_{\nu}(z)\;K_{\nu}(w)
 =\frac 12 \int_{0}^{\infty} \frac{\d t}{t}\;
 K_{\nu}\left(\frac{zw}{t}\right)\;
 \exp\left(-\frac t2 -\frac{z^2+w^2}{2t}\right)
\label{eq:KnuzKnuw}
\eeq
 and replacing $t$ with $tM^2$, we have
\beqn
 \lefteqn{ 
 M \l_1 \l_2 \; K_{1-\frac km}(M\l_1)\;
 K_{1-\frac km}(M\l_2)
 } \nonumber \\
 &&=
 \frac 12 \int_{0}^{\infty} \frac{\d t}{t}\;
 K_{1-\frac km} \left(\frac{\l_1\l_2}{t}\right)\;
 \exp\left(-\frac {tM^2}{2} -\frac{\l_1^2+\l_2^2}{2t}\right)
 \;\; .
\label{eq:int-1}
\eeqn
  Carrying out the integral with respect to $M$, and from
 \eq{eq:Mwkwk}, we have
\beqn
 \lefteqn{
 \VEV{w_{k}(\l_1) w_{k}(\l_2)}
 =
 \frac 1{4m} 
 } \nonumber \\
 &&
 \int_{0}^{\infty} \frac{\d t}{t}\;
 \frac{\l_1\l_2}{t}
 K_{1-\frac km} \left(\frac{\l_1\l_2}{t}\right)\;
 \exp\left(-\frac {tM^2}{2} -\frac{\l_1^2+\l_2^2}{2t}\right)
 \; .
\label{eq:int-2}
\eeqn
 Due to a formula,
\beq
 z K_{1-|p|}(z)
 =
 \int_{0}^{\infty}\d E
 \frac{E\sinh \pi E}{\cosh \pi E -\cos \pi p}
 \; K_{iE}(z)
 \;\; , 
\label{eq:zK-int}
\eeq
 the right hand side of \eq{eq:int-2} turns into
\beqn
\lefteqn{
 \frac 1{4m}
 \int_{0}^{\infty} \frac{\d t}{t}
 } \nonumber \\
 &&
 \int_{0}^{\infty}\d E
 \frac{E\sinh \pi E}{\cosh \pi E -\cos \pi p}
 \; K_{iE} \left( \frac{\l_1 \l_2}{t} \right)
 \exp\left(-\frac {tM^2}{2} -\frac{\l_1^2+\l_2^2}{2t}\right)
 \; .
\label{eq:int-3}
\eeqn
 Using a formula \eq{eq:KnuzKnuw} again, \eq{eq:int-3} turns out to be
\beq
 \frac 1{2m}
 \int_{0}^{\infty}\d E
 \frac{E\sinh \pi E}{\cosh \pi E -\cos \pi \frac km}
 \; K_{iE}(M\l_1) \; K_{iE}(M\l_2)
 \; .
\label{eq:int-4}
\eeq
 Putting \eq{eq:int-4} and \eq{eq:w+w+-sum} together,
 we have proved \eq{eq:w+w+-off}. 

   Let us go back to \eq{eq:w+w+-off} and perform the
 $E$-integral.
 The integral can be carried out  by deforming the contour.
 The residues for poles
\beq
 E=\pm i (\frac km +2n)
 \;\; , \;\; n=0,\pm 1, \pm 2, \cdots
 \;\; ,
\eeq
 contribute to the integral and, after all, we obtain the
 following expansion for the two-loop correlators (for
 $\l_1 < \l_2$) 
\beqn
 \lefteqn{
 \VEV{w_{k}(\l_1) w_{k}(\l_2)}
 } \nonumber \\
 &&=
 \frac 1m \sum_{n=-\infty}^{\infty} \;
 \Bigl(\frac km +2n\Bigr)\;
 \left( \frac{\sin \pi \frac km}{\pi/2} \right)^{-1} \;
 I_{|\frac km +2n|}(M\l_1)
 \; K_{\frac km +2n}(M\l_2)
 \nonumber \\
\label{eq:wkwk-exp}
\eeqn
 and
\beqn
 \lefteqn{
 \VEV{w^+(\l_1) w^{\pm}(\l_2)}
 } \nonumber \\
 &&=
 \frac 1m \sum_{k=1}^{m-1} \sum_{n=-\infty}^{\infty} \;
 (\pm)^{k-1}
 \Bigl|\frac km +2n\Bigr|\; I_{|\frac km +2n|}(M\l_1)
 \; \tK{\frac km +2n}(M\l_2)
 \; .
\label{eq:w+w+--exp}
\eeqn


    Since the two-loop correlators \eq{eq:w+w+-} or \eq{eq:w+w+--exp}
 do not diverge when
 the length of one of the
 loops approaches zero, we expect that one of the loops can be replaced
 by an infinite combination of local operators.
   From the consideration of the minisuperspace
 Wheeler-deWitt equation
 and the scaling behavior, we expect that
 the wave function of the scaling operator
 $\hsig{j}$ is proportional to
 $M^{j/m}K_{j/m}(M\l)$.
  \Eq{eq:nuw+-} and \eq{eq:l1w+-} indicate that the following 
 normalization of the wave function 
\beqn
 \VEV{\hsig{j}\; w^{+}(\l)}
 &=&
 \frac jm \left(\frac M2\right)^{\frac jm}
 \;\frac{\sin \frac jm \pi}{\pi/2}
 \;K_{\frac jm}(M\l) 
 \nonumber \\
 &=&\frac jm \left(\frac M2\right)^{\frac jm}
 \;\tK{\frac jm}(M\l) 
 \;\; , \;\;  j\geq 1 \; \ne 0 \;(\mbox{mod}\; m)
 \;\;.
\label{eq:wave-function+}
\eeqn
  would be reasonable.
   Note that
 the normalization factor
 $\sin \frac jm \pi$ in \eq{eq:wave-function+} is consistent
 because there are no scaling operators
 $\hsig{j}$ for $j=0$ (mod $m$) in the matrix model.
   Comparing \eq{eq:wave-function+} with \eq{eq:w+w+--exp},
 we expect the following expansions of the loop operators
 in term of the local operators:
\beq
``\; w^{\pm}(\l)=\frac 1m \sum_{k=1}^{m-1}
 \sum_{n=-\infty}^{\infty} (\pm)^{k-1}
 \left(\frac M2\right)^{-|\frac km +2n|}
 \; I_{|\frac km +2n|}(M\l)\;\hsig{|k+2mn|}\; "
 \; .
\label{eq:w+--exp}
\eeq
   These expansions are the generalizations of those in the case of
 the one-matrix model \cite{MSS} to the case of arbitrary unitary
 minimal model coupled to two-dimensional gravity.

   Since the loop correlators are symmetric under the interchange
 of two kinds of loops, that is,
 $\VEV{w^+(\l_1)\;w^+(\l_2)}=\VEV{w^-(\l_1)\;w^-(\l_2)}$,
 the wave functions of the scaling operators with respect to
 loop $w^-(\l)$ are read as
\beq
 \VEV{\hsig{j}\;w^-(\l)}=(-1)^{j-1} \VEV{\hsig{j}\;w^+(\l)}
 \;\; .
\label{eq:wave-function-}
\eeq
 The wave functions with respect to the loop $w_{k}(\l)$ are
\beq
 \langle \hsig{|k+2mn|} w_{k'}(\l) \rangle
 =\delta_{kk'}\left(\frac km +2n\right)
 \left(\frac M2\right)^{|\frac km +2n|}
 K_{|\frac km +2n|}(M\l) 
 \;\; .
\label{eq:wave-function-k}
\eeq

\subsection{Relation to the multi-matrix model}

  Let us comment on the relation between loops in the two-matrix model
 and those in the multi-matrix chain models.
 The lowest critical point of the $(m-1)$-matrix chain model
 represents also the $(m+1,m)$ minimal model,
 which corresponds 
 to $A_{m-1}$ Dynkin diagram, coupled to
 two-dimensional gravity.
 From the observation,
\beqn
 |x=1\rangle &=&\sum_{k=1}^{m-1} 
 \ts{ |p=\frac km \rangle \left(\frac 2{m-1}\right)^{1/2}\;
 \sin\pi\frac km }
 \;\; ,
\label{eq:x=1} \\
 |x=m-1\rangle &=&\sum_{k=1}^{m-1} 
 \ts{ |p=\ts{\frac km} \rangle \left(\frac 2{m-1}\right)^{1/2}\;
 \sin\pi\frac km (m-1) }
 \nonumber \\
 &=&\sum_{k=1}^{m-1} 
 \ts{ |p=\frac km \rangle \left(\frac 2{m-1}\right)^{1/2}\;
 (-1)^{k-1} \sin\pi\frac km }
 \;\; ,
\label{eq:x=m-1}
\eeqn 
 where $\VEV{x|p}$ is the wave function introduced in \eq{eq:Spx},
 we think $w^+(\l)$ and $w^-(\l)$ should correspond
 $|x=1\rangle$ and $|x=m-1\rangle$ respectively.
 The loop operator $w^{(x)}(\l)$ created by the $x$-th matrix $\hA^{(x)}$
 of the $(m-1)$-matrix chain model, thus, corresponds to
\beq
 |x\rangle =\sum_{k=1}^{m-1} 
 \ts{ |p=\frac km \rangle \left(\frac 2{m-1}\right)^{1/2}\;
 \sin\pi\frac km x }
 \;\; ,
\label{eq:x}
\eeq
 and would be represented accordingly as
\beq
 w^{(x)}(\l)= \sum _{k=1}^{m-1}
 \frac{\sin \pi \frac km x}{\pi/2} \;
 w_{k}(\l)
 \;\; ,\;\;
 x=1,\cdots,m-1
 \;\; .
\eeq
  We think this relation is valid
 at least for loop correlators
 with less than four loops.
  Using the relation, we can construct the loop correlators
 of the multi-matrix models
 from those of the two-matrix model.

\subsection{Three-point functions and fusion rules}
 \label{sec:three-point functions}

\subsubsection{One- and two-point functions}

 Let us consider the correlators of the scaling operators.
 We can extract these from loop correlators
 due to the relation \eq{eq:w+--exp}.

 Since the one-loop amplitude diverges when the loop length
 approaches to zero, this amplitude include the contribution
 which is not represented by the local operators.
  Putting the one-loop amplitude into
\beqn
 \lefteqn{
 \VEV{w^+(\l)}=-\left(\frac M2\right)^{2+\frac 1m}
 \left( \tK{2+\frac 1m}(M\l)-\tK{\frac 1m}(M\l) \right)
 }
 \nonumber \\
 &&=
 \left(\frac M2\right)^{2+\frac 1m}
 \left( I_{2+\frac 1m}(M\l)-I_{-2-\frac 1m}(M\l)
 -I_{\frac 1m}(M\l)+I_{-\frac 1m}(M\l) \right)
 \; ,
 \nonumber \\
\label{eq:w+I}
\eeqn
 and  extracting the parts proportional to $I_{\nu}$ ($\nu>0$),
 which parts would be considered as the contributions from
 local operators, we can obtain the one-point functions of
 the scaling operators:
\beqn
 \VEV{\hsig{1}}&=&-m \left(\frac M2\right)^{2+\frac 2m}
 \;\; ,
\label{eq:sig(1)} \\
 \VEV{\hsig{1+ 2m}}
 &=&m \left(\frac M2\right)^{4+\frac 4m} 
 \;\; ,
\label{eq:sig(1+2m)} \\
 \VEV{\hsig{j}}&=&0 \;\; ,\;\; j\ne 1,1+2m
 \;\; .
\label{eq:sig(j)}
\eeqn

 Let us turn to the two-point functions.
 Substituting \eq{eq:w+--exp} into \eq{eq:wave-function+},
 we obtain the two-point
 functions:
\beq
 \VEV{\hsig{i} \hsig{j} }
 =\delta_{ij}\;j\; \left(\frac M2\right)^{2j/m }
 \;\; ,
 \;\; i,j\ne 0\; (\mbox{mod}\;m)
 \;\; .
\label{eq:sig(i)sig(j)}
\eeq
  Note that we obtain diagonal two-point functions.

\subsubsection{Three-point functions}

 As for three-point functions, using the formula,
\beqn
 z K_{1-|p|}(z)
 &=&
 \int_{0}^{\infty}\d E
 \frac{E\sinh \pi E}{\cosh \pi E -\cos \pi p}
 \; K_{iE}(z)
 \nonumber \\
 &=&
 \pi\sum_{n=-\infty}^{\infty}
 \frac{|p+2n|}{\sin \pi|p+2n|} I_{|p+2n|}(z)
 \;\; , 
\label{eq:zK-exp}
\eeqn
 we first expand the three-loop correlator \eq{eq:w+w+w+-2}  as
\beqn
 \lefteqn{ \!\!\!
 \VEV{w^+(\l_1)w^+(\l_2)w^+(\l_3)}
 =
 \frac{-1}{m(m+1)}\left(\frac M2\right)^{-2-\frac 1m}\;
 \sum_{{\cal D}_3} \sum_{n_1=-\infty}^{\infty}
 \sum_{n_2=-\infty}^{\infty} \sum_{n_3-\infty}^{\infty}
 } \nonumber \\
 &&
 \ts{
 \left(\frac{k_1}{m}+2n_1\right) \left(\frac{k_2}{m}+2n_2\right)
 \left(\frac{k_3}{m}+2n_3\right) }
 I_{|\frac{k_1}m+2n_1|}(M\l_1) I_{|\frac{k_2}m+2n_2|}(M\l_2)
 I_{|\frac{k_3}m+2n_3|}(M\l_3)
 \; .
 \nonumber \\
\label{eq:w+w+w+-exp}
\eeqn
 Comparing \eq{eq:w+w+w+-exp} with \eq{eq:w+--exp},
 we can extract
 the three-point functions \cite{Anazawa}:
\beqn
 \lefteqn{
 \VEV{\hsig{|k_1+2mn_1|}\hsig{|k_2+2mn_2|}\hsig{|k_3+2mn_3|} }
 } \nonumber \\
 &&= C_{k_1k_2k_3} \frac{-1}{m(m+1)}
 \prod_{i=1}^{3}(k_i+2mn_i)
 \left(\frac M2\right)^{-2-\frac 1m 
 +\sum_{i=1}^{3}\frac 1m |k_i+2mn_i| }
 \; ,
\label{eq:sigsigsig}
\eeqn
 where
\beq
 C_{k_1k_2k_3}=\left\{
  \begin{array}{rc}
  1 \;\; ,& \;\; (k_1-1,k_2-1,k_3-1)\in{\cal D}_3^{(m)}\\
  0 \;\; ,& \;\;  \mbox{otherwise}
  \end{array}\right.
 \;\; .
\label{eq:Ck1k2k3}
\eeq
   For $n_i=0$, \eq{eq:sigsigsig} is nothing but
 the correlator of the gravitational primaries.
   For the gravitational primaries,
  \eq{eq:sig(i)sig(j)} and \eq{eq:sigsigsig} agree with
 the correlators obtained in
 \cite{DifK} from the generalized KdV flow up to a factor $-2$.
  Note that we obtain, here, the correlators of the gravitational
 descendants as well.
 
  In \cite{AGBG}, the fusion rules for the gravitational
 primaries were examined in continuum framework.
    Note that we have found here the fusion rules
 for the gravitational descendants
 as well as for the gravitational primaries.
  These fusion rules are similar to those for
 the gravitational primaries
 due to the factor $C_{k_1k_2k_3}$ in \eq{eq:sigsigsig}.
   Introducing the equivalence classes
 $[\hsig{k}]$ by
 the equivalence relation
\beq
 \hsig{k}\sim \hsig{|k+2mn|}\;\;,\;\;
 n\in \ZZ
 \;\;,
\label{eq:sigksim}
\eeq
 we can consider the fusion rules in \eq{eq:sigsigsig}
 as fusion rules
 among $[\hsig{k}]$ ($k=1,\cdots,m-1$).
  Note, here, that  the class $[\hsig{k}]$ does not
 correspond to
 the set which consist of the gravitational primary
 ${ \cal O}_k$ and its
 gravitational descendants
 $\sigma_l\left({ \cal O}_k\right)\;,l=1,2,\cdots$
 in \cite{DifK} introduced from the viewpoint of KdV flow.
 The three-loop correlator \eq{eq:w+w+w+-2} represents
 the fusion rules for all of the scaling operators including
 the gravitational descendants in a compact form.

\subsection{Further on the fusion rules}

    Let us examine the fusion rules in \eq{eq:sigsigsig}
 further and consider the relation of the scaling operators
 to the primary fields in the corresponding
 conformal field theory.

  In the $(p,q)$ minimal conformal model,
 the primary field $\Phi_{rs}$ has the conformal dimension
\beq
 \Delta_{r,s}=\frac{(pr-qs)^2-(p-q)^2}{4pq}
 \;\; ,
\label{eq:Delta-rs}
\eeq
  where $r$ and $s$ are positive integers. Since 
\beq
 \Delta_{r,s}=\Delta_{r+q,s+p}=\Delta_{q-r,p-s}
 \;\; ,
\label{eq:Delta-rs-2}
\eeq
 the corresponding primary fields can be regarded as the same
 one. The integers $r$ and $s$ can thus be restricted
 in the range
\beq
 \left\{
 \begin{array}{l}
 1\le r \le q-1 \\
 1\le s \;\ne 0 \;(\mbox{mod}\; p) \\
 pr < qs 
 \end{array}\right.
\label{eq:rs-range}
\eeq
 (see fig.~\ref{fig:pic1} ).
   In fig.~\ref{fig:pic1},
 the primary fields in the region $(\!(2)\!)$ or
 $(\!(2)\!)'$ are the secondary fields of those in the region
 $(\!(1)\!)$.  In general, the fields in the region $(\!(n+1)\!)$
 or $(\!(n+1)\!)'$ are the secondaries
 of the fields in $(\!(n)\!)$ or $(\!(n)\!)'$.
   Since the secondary fields correspond to null vectors,
 those fields  decouple. 
  One can thus construct consistent conformal field theory
  which include only the primary fields in the region $(\!(1)\!)$
 (i.e. inside the minimal table), that is,
  the minimal model $(p,q)$ \cite{BPZ}.
    Coupled to Liouville theory, however, the fields outside the
  the minimal table fail to decouple \cite{Kitazawa} and 
 infinite physical states emerge accordingly.
 These states are considered to correspond to
 the primaries outside 
 the minimal table.
   This correspondence is implied by the
 BRST cohomology  \cite{LZ,BMP} of the coupled system.

   Denoting the gravitational dimension of the dressed operator for
 $\Phi_{r,s}$ by  $\Delta^{G}_{r,s}=1-\frac{\alpha_{r,s}}{\gamma}$,
  in the minimal model coupled to Liouville theory, the following
 relation was shown \cite{LZ,BMP},
\beq
 \frac{\alpha_{r,s}}{\gamma}
 =\frac{p+q-|pr-qs|}{2q}
 \;\; ,
\label{eq:alpha-rs}
\eeq
 where $r$ and $s$ take the values in the range \eq{eq:rs-range}.
    On the other hand, in the matrix model, the corresponding relation 
  for the scaling operator $\hsig{j}$ is
\beq
 \frac{\alpha_{j}}{\gamma}
 =\frac{p+q-j}{2q}
 \;\; .
\label{eq:alpha-j}
\eeq
 From \eq{eq:alpha-rs} and \eq{eq:alpha-j}, we should take as
\beq
 j=|pr-qs| \;\; , \;\; j=1,2,\cdots\;\ne 0\;(\mbox{mod}\;q)
 \;\; ,
\label{eq:j=}
\eeq
 for $\hsig{j}$.

   Consider now  the relation of $\hsig{|k+2mn|}$
 to the primary field $\Phi_{r,s}$ of the unitary $(m+1,m)$
 minimal model.
   Let us first compare the two sets 
\beq
 S_{k}=\Bigl\{|k+2nm|\;\bigm|\;n\in\ZZ \;\Bigr\}
 \;\; ,
\label{eq:Sk}
\eeq
 and
\beqn
 \Bigl\{\;|pr-qs|\;\Bigr\}&=&\Bigl\{\;|(m+1)r-ms|\;\Bigr\}
 \nonumber \\
 &=& \Bigl\{\;r'+(s-r-1)m\;\Bigr\}
 \;\; ,
\label{eq:|pr-qs|}
\eeqn
 where $r$ and $s$ are positive integers in the range
\beq
 \left\{
 \begin{array}{l}
   1\le r\le m-1  \\ 
   1\le s  \\
   r+1\le s
 \end{array} \right.
\label{eq:rs-range-2}
\eeq
 and $r'\equiv m-r$.
 Note that we include $s=0$ (mod $m+1$) here.
 Dissolving the set $S_k$ into  two sets as
\beq
 S_{k}=S^+_{k}\oplus S^-_{k}
 \;\; ,
\label{eq:Sk-2}
\eeq
where
\beqn
 S^+_{k}&=&\Bigl\{k+2nm\;\bigm|\;n=0,1,2,\cdots\;\Bigr\}
 \;\; , \nonumber \\
 S^-_{k}&=&\Bigl\{(m-k)+(2n'+1)m\;\bigm|\;n'=0,1,2,
 \cdots\;\Bigr\}
 \;\; ,
\label{eq:S+S-}
\eeqn
 and comparing \eq{eq:S+S-} and \eq{eq:|pr-qs|},
 we can express  the sets $S^+_k$ and $S^-_k$
 in terms of $|(m+1)r-ms|$ as
\beqn
 \!\!\!
 S^+_k&=&\Bigl\{\;|(m+1)r-ms|\;\bigm|\;r'=k,\;s-r=2n+1,
 \;n=0,1,\cdots\; \Bigr\}
 \; ,
 \nonumber \\
 \!\!\!
 S^-_k&=&\Bigl\{\;|(m+1)r-ms|\;\bigm|\;r=k,\;s-r=2n'+2,
 \;n'=0,1,\cdots\;\Bigr\}
 \; .
\label{eq:S+S--2}
\eeqn
 From \eq{eq:S+S--2}, the following correspondence is obtained:
\beqn
 \hsig{|k+2mn|}\;( n\ge 0) &\leftrightarrow&
 \Phi_{m-k,\;r+2n+1} \; ( n\ge 0)
 \nonumber \\
 \hsig{|k+2m(-1-n')|}\;( n'\ge 0 ) &\leftrightarrow&
 \Phi_{k,\;r+2n'+2} \; ( n'\ge 0)
 \;\;,
\label{eq:sigPhi}
\eeqn
 where $s\ne 0\;(\mbox{mod}\;m+1)$.

 In \cite{MMS}, it was suggested that the scaling operators
 $\hsig{j},\;j=0\;(\mbox{mod}\;m+1)$, should be identified
 as the boundary operators which couple to the boundaries of
 two-dimensional surface.
 These scaling operators do not have their
 counterparts in the BRST cohomology of the system coupled to
 Liouville theory.
   From
\beq
 |(m+1)r-ms|=(s-r)(m+1) -s =0\;(\mbox{mod}\;m+1)
 \;\;,
\eeq
 $s=0\;(\mbox{mod}\;m+1)$ in the range \eq{eq:rs-range-2} should
 correspond to the boundary operators.

\begin{figure}
 \begin{center}
\setlength{\unitlength}{0.7mm}
\begin{picture}(300,100)(-10,-10)
 \put(0,0){\thicklines\vector(1,0){170} }
 \put(0,0){\thicklines\vector(0,1){80} }
 \put(0,40){\thicklines\line(1,0){170} }
 \multiput(50,0)(50,0){3}{\thicklines\line(0,1){45} }
 \multiput(0,0)(100,0){2}{\line(5,4){50} }
 \put(50,40){\line(5,-4){50} }
 \multiput(40,30)(20,0){6}{\circle*{2}}
 \multiput(30,10)(20,0){7}{\circle*{2}}
 \multiput(30,20)(10,0){13}{\thicklines\circle {2.5}}
 \multiput(50,30)(20,0){6}{\thicklines\circle {2.5}}
 \multiput(50,30)(20,0){6}{\circle*{1}}
 \multiput(20,10)(20,0){7}{\thicklines\circle {2.5}}
 \multiput(20,10)(20,0){7}{\circle*{1}}
 \put(-10,-10){\makebox(8,8)[bl]{0}}
 \put(-10,65){\makebox(8,8)[bl]{r}}
 \put(-10,40){\makebox(8,8)[bl]{4}}
 \put(170,-10){\makebox(8,8)[b]{s}}
 \put(50,-10){\makebox(8,8)[b]{5}}
 \put(100,-10){\makebox(8,8)[b]{10}}
 \put(150,-10){\makebox(8,8)[b]{15}}
 \put(170,28){\makebox(10,30)[bl]
    {$\gets\;\hsig{l}\bigl({\cal O}_1\bigr)$} } 
 \put(170,18){\makebox(10,30)[bl]
    {$\gets\;\hsig{l}\bigl({\cal O}_2\bigr)$} } 
 \put(170,8){\makebox(10,30)[bl]
    {$\gets\;\hsig{l}\bigl({\cal O}_3\bigr)$} } 
  \put(20,50){\vector(0,-1){37}}
  \put(30,50){\vector(0,-1){27}}
  \put(40,50){\vector(0,-1){17}}
  \put(16,55){\makebox(10,10)[b]
    {${\cal O}_3$} } 
  \put(26,55){\makebox(10,10)[b]
    {${\cal O}_2$} } 
  \put(36,55){\makebox(10,10)[b]
    {${\cal O}_1$} } 
 \put(65,80){\circle*{2}}
 \put(73,78){\makebox(10,30)[bl]
    {: $\;\hsig{|1+2mn|}$} } 
 \put(65,70){\thicklines\circle{2.5}}
 \put(73,68){\makebox(10,30)[bl]
    {: $\;\hsig{|2+2mn|}$} } 
 \put(65,60){\circle*{1}}
 \put(65,60){\thicklines\circle{2.5}}
 \put(73,58){\makebox(10,30)[bl]
    {: $\;\hsig{|3+2mn|}$} } 
\end{picture}
 \end{center}
\caption{the scaling operators $\hsig{|k+2mn|}$
 in the $(5,4)$ minimal model coupled to 2D gravity}
\label{fig:pic2}
\end{figure}
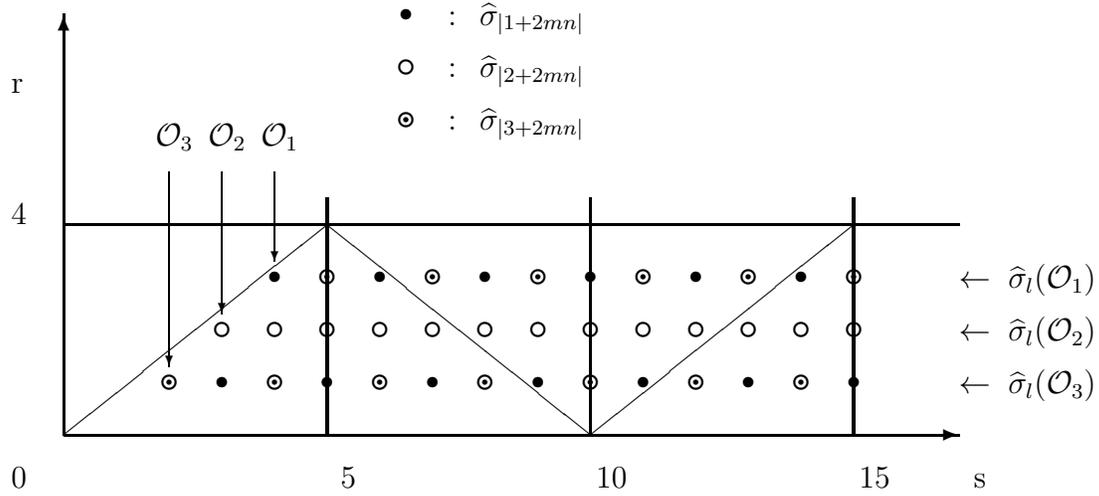

   As an example, we depicted the scaling operators
 on the r-s plane for the case of $m=4$ in 
 fig.~\ref{fig:pic2}.

   Concerning the scaling operators inside the minimal table,
 the fusion rules involving non-diagonal operators dose not
 agree with the
 fusion rules of the unitary minimal model;
\beqn
 \lefteqn{ \!\!\!\!
 \VEV{\Phi_{r_1,\;s_1}\Phi_{r_2,\;s_2}\Phi_{r_3,\;s_3}}
 \ne 0 \;\;{\mbox{iff}}
 } \nonumber \\
 &&\!\!\!\! \!\!\!\! \!\!\!\! \!\!\!
 (r_1-1,r_2-1,r_3-1)\in{\cal D}_3^{(m)}
 \;\;\mbox{and}\;\;(s_1-1,s_2-1,s_3-1)\in{\cal D}_3^{(m+1)}
 .
\label{eq:KPZ-fusion}
\eeqn
  For example, in the Ising model ($m=3$) the three point
 function for the energy operators vanish,
\beq
 \VEV{\Phi_{1,3}\Phi_{1,3}\Phi_{1,3}}=0
 \;.
\eeq
 Coupled to gravity, however,
 the corresponding three point function does not vanish.

\begin{figure}
 \begin{center}
\setlength{\unitlength}{0.7mm}
\begin{picture}(300,100)(-10,-10)
 \put(0,0){\thicklines\vector(1,0){170} }
 \put(0,0){\thicklines\vector(0,1){80} }
 \put(0,40){\thicklines\line(1,0){170} }
 \multiput(50,0)(50,0){3}{\thicklines\line(0,1){50} }
 \multiput(0,0)(100,0){2}{\line(5,4){50} }
 \put(50,40){\line(5,-4){50} }
 \put(-10,-10){\makebox(8,8)[bl]{0}}
 \put(-10,65){\makebox(8,8)[bl]{r}}
 \put(-10,40){\makebox(8,8)[bl]{q}}
 \put(170,-10){\makebox(8,8)[b]{s}}
 \put(50,-10){\makebox(8,8)[b]{p}}
 \put(100,-10){\makebox(8,8)[b]{2p}}
 \put(150,-10){\makebox(8,8)[b]{3p}}
 \put(30,10){\makebox(10,10)[bl]{$(\!(1)\!)$}}
 \put(60,10){\makebox(10,10)[bl]{$(\!(2)\!)$}}
 \put(130,10){\makebox(10,10)[bl]{$(\!(3)\!)$}}
 \put(80,25){\makebox(10,10)[bl]{$(\!(2)\!)'$}}
 \put(110,25){\makebox(10,10)[bl]{$(\!(3)\!)'$}}
\end{picture}
 \end{center}
\caption{the range of $(r,s)$ }
\label{fig:pic1}
\end{figure}
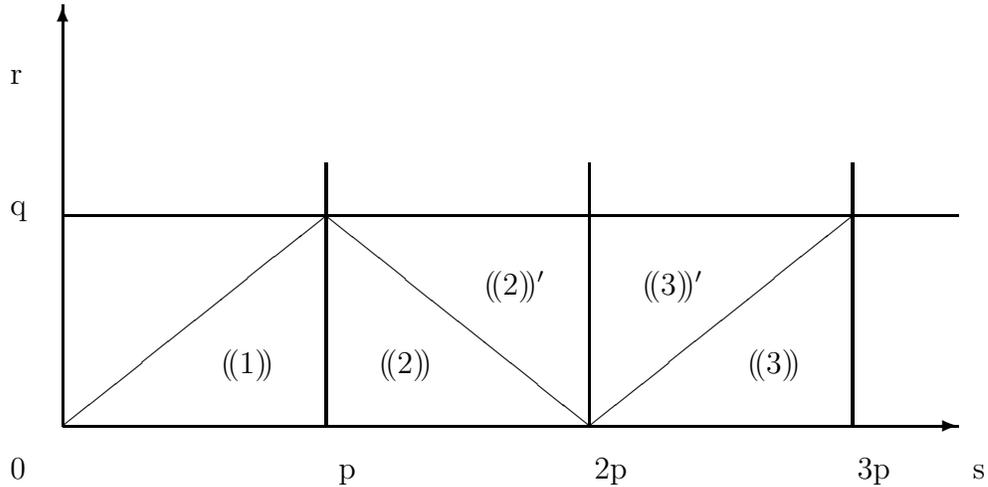

\subsection{Boundary operators}

\subsubsection{Boundary operators and touching of loops}

  In \cite{MMS} it was proposed that the scaling operators
 which do not occur in the BRST cohomology of Liouville
 theory are boundary operators and  one of them,
 which is $\hsig{3}=\hsig{1}({\cal O}_1)$ in the case of
 pure gravity, was in fact proven to be a boundary operator
 for the one-matrix model and the Ising model case.
   We would like to examine the role of 
 the operators $\hsig{n(m+1)},\;n=1,2,\cdots\ne 0\;
 ({\rm mod}\;m)$ 
 as well as $\hsig{m+1}$ for general unitary minimal models.
 
\begin{figure}
\begin{center}
 \epsfxsize=12cm
 \qquad\epsfbox{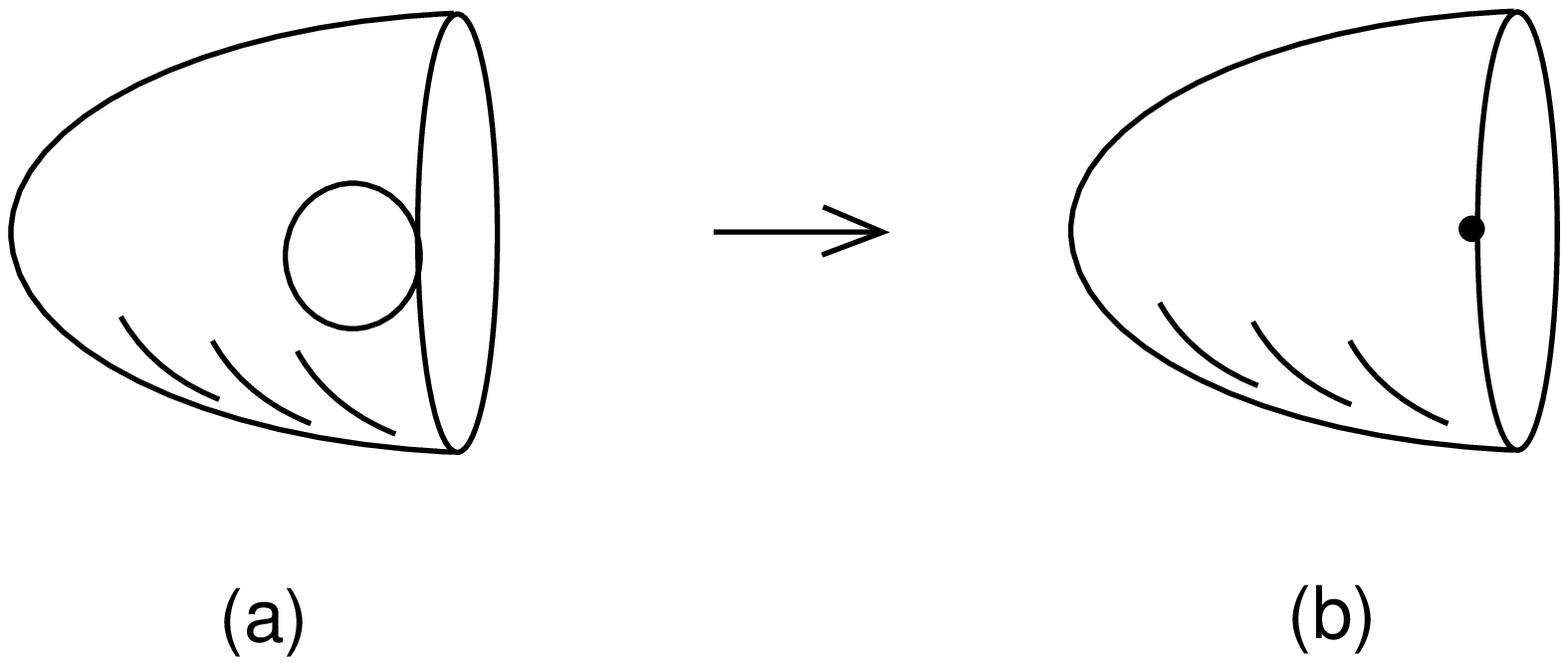}
\caption{(a): A surface with two loops touching each other 
 on a point.
 (b): When one of the loops shrinks to a microscopic loop
 the microscopic loop is equivalent to the insertion of the 
 operator denoted by the dot on the loop.}
\label{fig:bound1}
\end{center}
\begin{center}
 \epsfxsize=12cm
 \qquad\epsfbox{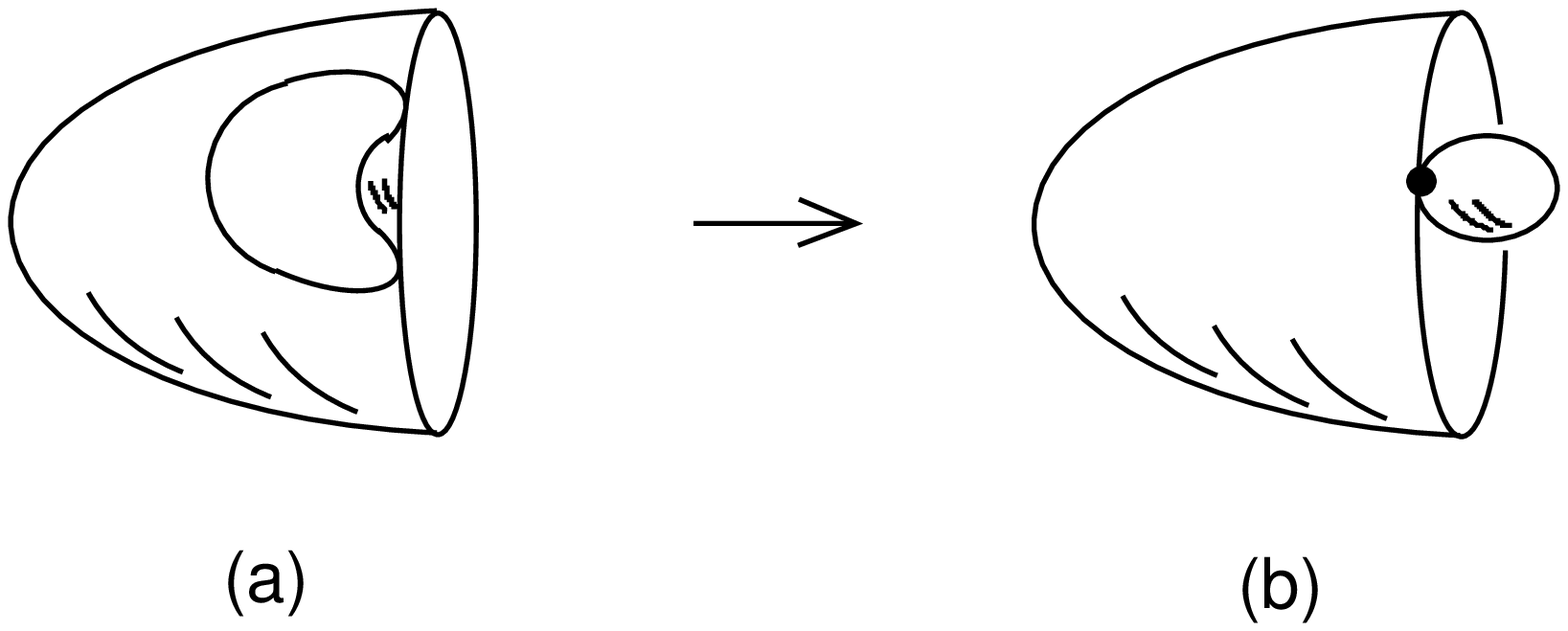}
\caption{The case of a surface with two loops touching each other
 on two points.}
\label{fig:bound2}
\end{center}
\begin{center}
 \epsfxsize=12cm
 \qquad\epsfbox{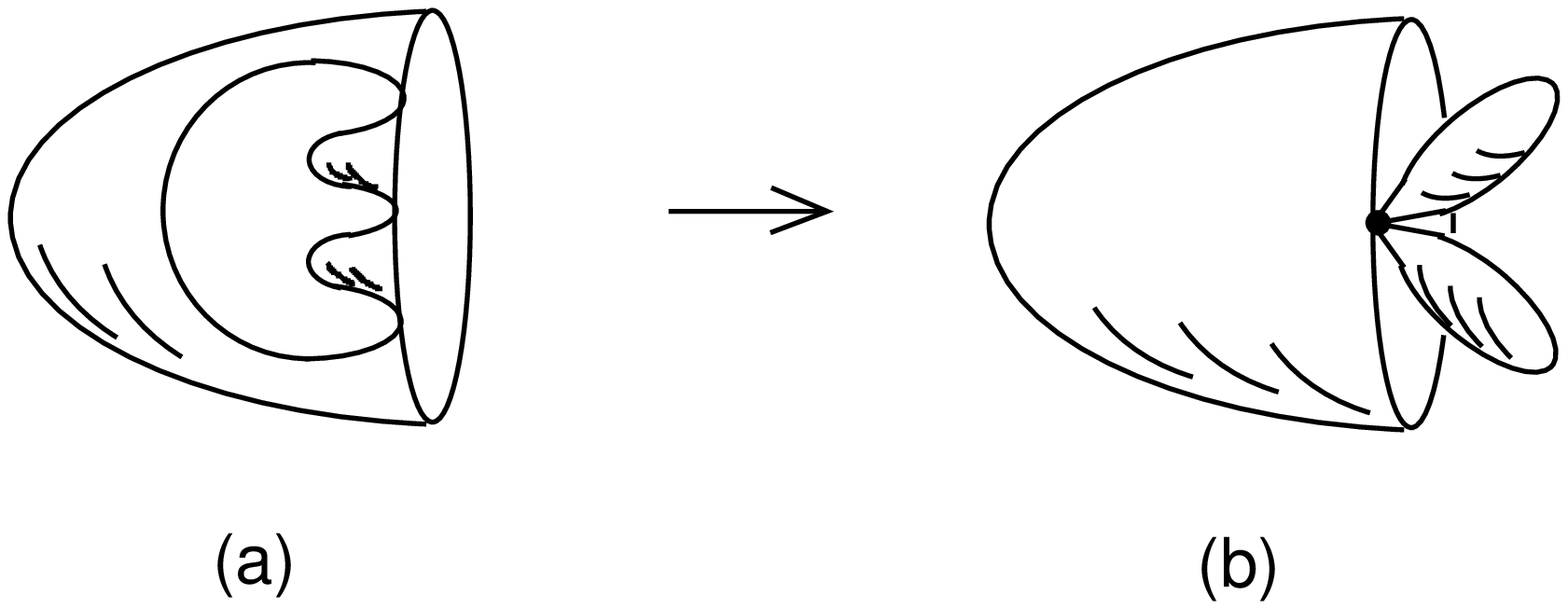}
\caption{The case of a surface with two loops touching each other
 on three points.}
\label{fig:bound3}
\end{center}
\end{figure}

  Let us denote these operators by
\beq
 \widehat{{\cal B}}_n=\hsig{n(m+1)}
 ,\;\; n=1,2,\cdots\ne 0\; ({\rm mod}\;m)
 \;.
\eeq
  In the matrix models the loop amplitudes contain the
 contribution from the configuration with loops touching
 each other.
  In two-loop case, let us consider the configuration
 in which the two loops touch  each other on n points.
  When we shrink one of the loops to a microscopic loop,
 the other loop splits into
 n loops, which are stuck each other through the
 microscopic loop 
 (see figs.\ref{fig:bound1}, \ref{fig:bound2}
 and \ref{fig:bound3}).
  Since the microscopic loop represents
 a sum of the scaling operators, the wave functions of
 some scaling operators contain the contribution from the
 surfaces with split loop.
   
  We now show that the boundary operators indeed
 represent these configurations.
  From eqs. (\ref{eq:wave-function+}) and (\ref{eq:w+-}),
 the wave function of $\widehat{{\cal B}}_n$
 and the one-loop amplitude are
\beq
 \VEV{\widehat{{\cal B}}_n w^+(\l)}
 =n(1+\frac 1m) \left(\frac M2\right)^{n(1+\frac 1m)}
 \tK{n(1+\frac 1m)}(M\l)
\;,
\eeq
\beq
 \VEV{w^+(\l)} =(1+\frac 1m) ~\l^{-1}
 \left(\frac M2\right)^{1+\frac 1m}
 \tK{1+\frac 1m}(M\l)
\;.
\eeq
 In the space of Laplace transformed coordinates,
 we have
\beq
 {\cal L}\left[\l^{-1}
 \langle \widehat{{\cal B}}_n w^+(\l)\rangle\right]
 =-\left(\frac M2\right)^{n(1+\frac 1m)}
 2 \cosh n(m+1)\theta
\;,
\eeq
\beq
 {\cal L}\left[ \langle w^+(\l)\rangle \right]
 =-\left(\frac M2\right)^{1+\frac 1m}
 2 \cosh (m+1)\theta
\;,
\eeq
 where we have used the relation
\beq
 {\cal L} \left[-\l^{-1} |\nu| \tK{\nu}(M\l)
 \right]
 = 2\cosh m\nu\theta
\;.
\eeq
  Note here that $w^+(\l)$ represents a loop with a marked
 point and $\l^{-1}w^+(\l)$ represents a loop without a
 marked point.
  Since $\cosh n(m+1)\theta$ can be expressed as
 a polynomial of $\cosh (m+1)\theta$,
\beqn
 2\cosh n(m+1)\theta
 &=&2~ T_n\Bigl(\cosh (m+1)\theta\Bigr)
 \nonumber \\
 &\equiv&
 \sum_{r=0}^{[n/2]} c^{(n)}_{r}
 \Bigl[2\cosh (m+1)\theta\Bigr]^{n-2r}
 \;,  \\
 c^{(n)}_r&=&\frac{(-1)^r n}{n-r} \left(
 {n-r \atop r}\right)
 \; \nonumber
\eeqn
 where $T_n$ is the Chebeyshev polynomial,
 we obtain the following relation:
\beq
 {\cal L}\left[ -\l^{-1}\langle \widehat{{\cal B}}_n
 w^+(\l)\rangle \right]
 = \sum_{r=0}^{[(n-1)/2]} c^{(n)}_{r}
 \left(\frac M2\right)^{2r(1+\frac 1m)}
 \left\{ {\cal L}\left[-\langle w^+(\l)\rangle \right]
 \right\}^{n-2r}
\;.
\eeq
  In the space of loop lengths, the above relation means
 that the wave function of $\widehat{{\cal B}}_n$
 is equivalent to a sum of the convolutions of disk amplitudes:
\beq
\label{eq:bound-1}
 \VEV{\widehat{{\cal B}}_n w^+(\l)}
 =-\l \sum_{r=0}^{[(n-1)/2]}c^{(n)}_{r}
 \left(\frac M2\right)^{2r(1+\frac 1m)}
 (-1)^{n-2r}\left[\odot {\cal A}_1^+\right]^{n-2r}(\l)
\;.
\eeq
 Here we introduced a notation
  ${\cal A}_1^+\equiv \langle w^+(\l)\rangle$,
 and $\left[\odot {\cal A}_1^+\right]^{s}(\l)$ denotes
 the convolution of $s$ ${\cal A}_1^+(\l)$'s, for example
\beq
 \left[\odot {\cal A}_1^+\right]^{2}(\l)
  =\int_0^{\infty}\int_0^{\infty} \d\l_1\d\l_2
 \;\delta(\l_1+\l_2-\l) 
 {\cal A}_1^+(\l_1){\cal A}_1^+(\l_2)
\;.
\eeq
   From \eq{eq:bound-1} we can conclude that
 the operator ${\widehat{\cal B}}_n$ couple to the point
 to which $s$ ($ \le n$) parts of the loop are stuck
 each other in the case of one-loop amplitude.
   When there are more than one loop, we infer
 that the operator couples to the point to which
 $s$ parts of several loops are stuck each other;
 the operator will not recognize that it is
 touching different loops this time.

 Using the following relation
\beq
\label{eq:cosh x}
 \Bigl[2\cosh x\Bigr]^{n}
 =\sum_{r=0}^{[(n-1)/2]} \left({n\atop r}\right)
 2 \cosh (n-2r)x 
,\;\;(\mbox{up to constant})
\;,
\eeq
 we also obtain
\beq
\label{eq:bound-2}
 \l\left[\odot{\cal A}_1^+\right]^{n}(\l)
 =(-1)^{n+1}\sum_{r=0}^{[(n-1)/2]} \left({n\atop r}\right)
 \left(\frac M2\right)^{2r(1+\frac 1m)}
 \VEV{\widehat{\cal B}_{n-2r} w^+(\l)}
\;.
\eeq
  Here we drop the constant term 
 in \eq{eq:cosh x} when
 we carry out the inverse Laplace transformation.
 From \eq{eq:bound-2}, we see that the boundary operator
 coupled to the point on which n parts of loops
 are touching each other  is given by
\beq
 {\cal B}_n
 =(-1)^{n+1}\sum_{r=0}^{[(n-1)/2]} \left({n\atop r}\right)
 \left(\frac M2\right)^{2r(1+\frac 1m)}
 \widehat{\cal B}_{n-2r}
\;.
\eeq

 Now let us consider the boundary operators when
 there are two loops on two-dimensional surface.
   As for ${{\cal B}}_1$, we expect that 
  $ \VEV{w^+(\l_1)w^+(\l_2){{\cal B}}_1}$
  should be proportional to
  $(\l_1+\l_2)\VEV{w^+(\l_1)w^+(\l_2)}$.
  Let us confirm this in the following.

 From the three loop correlator (\ref{eq:w+w+w+-2}),
 the expansion of loop operator (\ref{eq:w+--exp}) and
 the wave function of $\hsig{|k+2mn|}$
 (\ref{eq:wave-function+}),
 we obtain the following correlator with two loops and
 a local operator:
\beqn
\label{eq:2loop+op}
 &&\VEV{w^+(\l_1)w^+(\l_2)\hsig{|k_3+2mn_3|}}
 =\frac{-1}{m+1}\sum_{k_1,k_2}C_{k_1 k_2 k_3}
 \left(\frac M2\right)^{-\frac 1m +|\frac {k_3}m+2n_3|}
 \nonumber \\
 &&\qquad\qquad\qquad\times
 \l_1\l_2(\frac {k_3}m +2n_3)
 \tK{1-\frac{k_1}m}(M\l_1)\tK{1-\frac{k_2}m}(M\l_2)
\;.
\eeqn
  Consider the amplitude for
 ${{\cal B}}_1=\widehat{{\cal B}}_1=\hsig{m+1}=\hsig{|\frac{m-1}m-2|}$.
 Since $C_{k_1,k_2,m-1}$ is nonvanishing only for
 the case of $k_1+k_2=m$,
 we obtain the following amplitude:
\beq\label{eq:2loop-bound}
 \VEV{w^+(\l_1)w^+(\l_2){{\cal B}}_1}
 =\frac{1}{m}\sum_{k}^{m-1}
 \left(\frac M2\right) \l_1\l_2
 \tK{\frac{k}m}(M\l_1)\tK{1-\frac{k}m}(M\l_2)
\;.
 \eeq
 Comparing eqs. (\ref{eq:2loop-bound}) to (\ref{eq:w+w+-})
 we obtain the desired relation:
\beq
 \VEV{w^+(\l_1)w^+(\l_2){{\cal B}}_1}
 =(\l_1+\l_2)\VEV{w^+(\l_1)w^+(\l_2)}
\;.
\eeq

 Next, let us consider ${{\cal B}}_2$.
 Since we infer that the insertion of ${{\cal B}}_2$
 should  play the role of connecting two parts
 of loops together, we expect the following relation:
\beqn
\label{eq:2loop+B2}
  &&\VEV{w^+(\l_1)w^+(\l_2){{\cal B}}_2}
 =2\l_1\int_0^{\l_1}d\l'_1 ~
 \VEV{w^+(\l'_1)w^+(\l_2)}\VEV{w^+(\l_1-\l'_1)}
 \nonumber \\
 &&\qquad\qquad
 +2\l_2\int_0^{\l_2}d\l'_2 ~
 \VEV{w^+(\l_1)w^+(\l'_2)}\VEV{w^+(\l_2-\l'_2)}
 \nonumber \\
 &&\qquad\qquad
 +2\l_1 \l_2 \VEV{w^+(\l_1+\l_2)}
 \;.
\eeqn
  The third term in the right hand side of \eq{eq:2loop+B2}
 represents the contribution from the surfaces with
 loops $w^+(\l_1)$ and  $w^+(\l_2)$ touching
  with each other on a point.
 Let us confirm the relation (\ref{eq:2loop+B2})
 in the following.
 In this case, it is convenient to consider in the 
 space of Laplace transformed coordinates $\zeta_i$.
 In this space \eq{eq:2loop+op} reads as
\beqn
 &&\VEV{\hat{W}^+(\zeta_1)\hat{W}^+(\zeta_2)
 ~\hsig{|k_3+2mn_3|} }
 =\frac{-1}{m+1}
 \left(\frac M2\right)^{-\frac 1m -2 +|\frac {k_3}m+2n_3|}
 (\frac {k_3}m +2n_3)
 \nonumber \\
 &&\qquad\times
 \frac{\p}{\p\zeta_1}\frac{\p}{\p\zeta_2}
 \left\{
 \sum_{k_1,k_2}C_{k_1 k_2 k_3}
 ~\frac{\sinh (m-k_1)\theta_1}{\sinh m \theta_1}
 ~\frac{\sinh (m-k_2)\theta_2}{\sinh m \theta_2}
 \right\}
\;.
\eeqn
 Due to the relation
\beqn
 && \sum_{k_1,k_2}C_{k_1 k_2 k_3}
 ~\frac{\sinh (m-k_1)\theta_1}{\sinh m \theta_1}
 ~\frac{\sinh (m-k_2)\theta_2}{\sinh m \theta_2}
\nonumber \\
 &&\qquad
 =\frac{-1}{2(\cosh\theta_1-\cosh\theta_2)}
 \left(
 \frac{\sinh (m-k_3)\theta_1}{\sinh m \theta_1}
 -\frac{\sinh (m-k_3)\theta_2}{\sinh m \theta_2}
 \right)
 ,
\eeqn
 we have
\beqn
 &&\VEV{\hat{W}^+(\zeta_1)\hat{W}^+(\zeta_2)
 ~\hsig{|k_3+2mn_3|} }
 =\frac{1}{2(m+1)}
 \left(\frac M2\right)^{-\frac 1m -2 +|\frac {k_3}m+2n_3|}
 (\frac {k_3}m +2n_3)
 \nonumber \\
 &&\qquad\times
 \frac{\p}{\p\zeta_1}\frac{\p}{\p\zeta_2}
 \left\{
 \frac{1}{\cosh\theta_1-\cosh\theta_2}
 \left(
 \frac{\sinh (m-k_3)\theta_1}{\sinh m \theta_1}
 -\frac{\sinh (m-k_3)\theta_2}{\sinh m \theta_2}
 \right)
 \right\}
\;.
\nonumber \\
\eeqn
 Since we should take $k_3=2$ for
 ${\cal B}_2=-\hsig{2(m+1)}$ (for $m\ge 3$),
 we obtain the amplitude for ${\cal B}_2$
\beqn
\label{eq:2loop+B2:2}
 &&\VEV{\hat{W}^+(\zeta_1)\hat{W}^+(\zeta_2)
 {\cal B}_2 }
 \nonumber \\
 &&\;
 =\frac{-1}{m}
 \left(\frac M2\right)^{\frac 1m}
 \frac{\p}{\p\zeta_1}\frac{\p}{\p\zeta_2}
 \left\{
 \frac{1}{\cosh\theta_1-\cosh\theta_2}
 \left(
 \frac{\sinh (m-2)\theta_1}{\sinh m \theta_1}
 -\frac{\sinh (m-2)\theta_2}{\sinh m \theta_2}
 \right)
 \right\}
 .
\nonumber \\
\eeqn
  On the other hand, from the amplitudes
\beqn
 &&\VEV{\hat{W}^+(\zeta_1)\hat{W}^+(\zeta_2)}
 = \frac{\p}{\p\zeta_1}\frac{\p}{\p\zeta_2}
 \ln \frac{\cosh\theta_1-\cosh\theta_2}
 {\cosh m\theta_1-\cosh m\theta_2}
 \nonumber \\
 &&\qquad\qquad
 =\frac{\p}{\p \zeta_2}
 \left\{
 \frac{1}{\cosh\theta_1-\cosh\theta_2}
 \frac{\sinh\theta_1}{mM\sinh m\theta_1}
 -\frac{1}{\zeta_1-\zeta_2}
 \right\}
\;,
\eeqn
\beq
 \VEV{\hat{W}^+(\zeta)}=
 -\left(\frac M2\right)^{1+\frac 1m}
 2\cosh (m+1)\theta
\;,
\eeq
 we obtain the following relation
\beqn
\label{eq:2loop+B2:3}
 &&-2\frac{\p}{\p\zeta_1}
 \left\{
 \VEV{\hat{W}^+(\zeta_1)\hat{W}^+(\zeta_2)}
 \VEV{\hat{W}^+(\zeta_1)}
 \right\}
 +(1\leftrightarrow 2)
\nonumber \\
 &&-2\frac{\p}{\p\zeta_1}\frac{\p}{\p\zeta_2}
 \left\{
 \frac{\VEV{\hat{W}^+(\zeta_1)}-\VEV{\hat{W}^+(\zeta_2)} }
 {\zeta_1-\zeta_2}
 \right\}
\nonumber \\
 &&\;
 =\frac{2}{m}
 \left(\frac M2\right)^{\frac 1m}
 \frac{\p}{\p\zeta_1}\frac{\p}{\p\zeta_2}
 \left\{
 \frac{1}{\cosh\theta_1-\cosh\theta_2}
 \left(
 \frac{\sinh\theta_1 \cosh (m+1)\theta_1}{\sinh m \theta_1}
 -(1\leftrightarrow 2)
 \right)
 \right\}
 .
\nonumber \\
\eeqn
 One can easily show  that the right hand side of
 eq.~(\ref{eq:2loop+B2:3}) agrees with that of 
 eq.~(\ref{eq:2loop+B2:2}). 
 Putting eqs.~(\ref{eq:2loop+B2:3}) and
 (\ref{eq:2loop+B2:2}) together and performing 
 the inverse Laplace transformation, we obtain the desired
 relation eq.~(\ref{eq:2loop+B2}).

  We have shown that the operator ${\cal B}_2$ connects
 two parts of loops together in the case with two loops.
  We infer that similar phenomena occur in general;
 the operator ${\cal B}_n$ would 
 connect n parts of loops together in the case with any number
 of loops.
 
\subsubsection{Connection to the Schwinger-Dyson equations}

  We can observe close relationship between  the boundary
 operators and the Schwinger-Dyson equations proposed in
 \cite{IIKMNS}.
  Continuum limit of the 
 Schwinger-Dyson equations for loops in the two-
 and multi-matrix models were proposed in \cite{IIKMNS}
 under some assumptions. It was shown \cite{IIKMNS} that
 these equations contain $W_3$ constraints,
 which were derived explicitly in \cite{GN}.
  The integrability of these equations were shown in
 \cite{S-D eqs.}.
  These facts justify the proposed Schwinger-Dyson
 equations.

 Let us consider the connection of the boundary
 operators with the Schwinger-Dyson equations.
  For the $(m+1,m)$ minimal models, the following
 Schwinger-Dyson equations were proposed in
 \cite{IIKMNS}:
\beqn
\label{eq:S-D}
 &&\int_0^{\l} d\l'
 \VEV{w^{(1)}(\l')
 w^{(1)}\left(\l-\l';[{\cal H}(\sigma)]^j \right)
 w^{(1)}(\l_1)\cdots w^{(1)}(\l_n) }'
 \nonumber \\
 &&
 +g \sum_i \l_i
 \VEV{
 w^{(1)}\left(\l+\l_i;[{\cal H}(\sigma)]^j \right)
 w^{(1)}(\l_1)\cdots w^{(1)}(\l_{i-1})
 w^{(1)}(\l_{i+1})\cdots w^{(1)}(\l_n)
 }'
 \nonumber \\
 &&
 +\VEV{
 w^{(1)}\left(\l;[{\cal H}(\sigma)]^{j+1} \right)
 w^{(1)}(\l_1)\cdots w^{(1)}(\l_n)
 }'
 \approx 0 \;,
 \nonumber \\
 &&\qquad\qquad\qquad
 {\rm for}\;\; j=0,\cdots,m-2
\;,
\eeqn
 and
\beq
\label{eq:S-D:2}
 \VEV{
 w^{(1)}\left(\l;[{\cal H}(\sigma)]^{m-1} \right)
 w^{(1)}(\l_1)\cdots w^{(1)}(\l_n)
 }'
 \approx 0
 \;.
\eeq

 Here $\VEV{\cdots}'$ represent loop correlators
 that are not necessarily connected,
 and $w^{(1)}(\l)$ represents a loop operator
 corresponding to a loop created by the matrix
 $\hat{A}^{(1)}$
 in the multi-matrix model.
  The operator ${\cal H}(\sigma)$ describes an operator
 which changes  the `spin' on a loop locally from 1 to 2.
 Also $w^{(1)}\left(\l;[{\cal H}(\sigma)]^{j} \right)$
 describes a loop with $[{\cal H}(\sigma)]^{j}$ inserted.
 The symbol $\approx$ means that as a function of $\l$,
 the quantity has its support at $\l=0$.

 From \eq{eq:S-D} for $j=0$ and $n=1$, we have the relation
\beqn
 &&\l_1\VEV{ w^{(1)}\left(\l_1;{\cal H}(\sigma) \right)
 w^{(1)}(\l_2)}'
 +\l_2\VEV{ w^{(1)}(\l_1)
 w^{(1)}\left(\l_2;{\cal H}(\sigma) \right)}'
\nonumber \\
 &&
 +\l_1\int_0^{\l_1}d\l'_1
 \VEV{ w^{(1)}(\l'_1)w^{(1)}(\l_1-\l'_1)w^{(1)}(\l_2)}'
\nonumber \\
 &&
 +\l_2\int_0^{\l_2}d\l'_2
 \VEV{ w^{(1)}(\l_1)w^{(1)}(\l'_2)w^{(1)}(\l_2-\l'_2)}'
\nonumber \\
 &&
 +2 g\l_1\l_2  \VEV{ w^{(1)}(\l_1+\l_2)}'
 \approx 0
\; .
\label{eq:S-D:3}
\eeqn
 The planar part of the above relation agrees with
 \eq{eq:2loop+B2}. Note that the loop amplitudes in
 \eq{eq:2loop+B2} represent connected correlators.

  This agreement implies that ${\cal H}$ would correspond
 to $\widehat{{\cal B}}_2$.
 Taking into account the fact that $\widehat{{\cal B}}_n$
 ($n=0$ mod $m$) do not exist and \eq{eq:S-D:2},
 it is legitimate to consider that the amplitude
 (for $j=1,\cdots,m$)
\beq
 \VEV{w^+(\l_1)\cdots w^+(\l_n)\widehat{{\cal B}}_j}
\eeq
 corresponds to the connected part of the amplitude
\beq
 \sum_{i=1}^{n}
 \oint d\sigma_i
 \VEV{
  w^{(1)}(\l_1)\cdots w^{(1)}(\l_{i-1})
 w^{(1)}\left(\l_i;[{\cal H}(\sigma_i)]^{j-1} \right)
 w^{(1)}(\l_{i+1})\cdots w^{(1)}(\l_n)
 }'
\;.
\nonumber \\
\eeq
\newpage
\begin{center}
\section{Multi-loop correlators}
\label{sec:multi-loop}
\end{center}

 In this section, we generalize the discussion in 
 sect.~\ref{sec:3-loop} to the cases of higher-loop.
 First, we derive the formula of the n-resolvent correlators,
 which 
 we quoted in sect.~\ref{sec:formula for n-resolvent},
 and point out that the structure corresponding to
 the crossing symmetry of the underlying
 conformal field theory can be seen
 in the loop correlators.
 We then discuss the four-loop correlator in detail. 

\subsection{The derivation of the n-resolvent correlators}
\label{sec:n-resolvent}

  Consider in the two-matrix model  the connected  part of  the correlator
  consisting of the product of  $n$-resolvents.
\beq
\label{eq:nresolv}
   \DVEV{\prod_{i=1}^{n} \Tr\frac{1}{p_i-\hA} }
 \;\;\;.
\eeq
It should be noted that
  this expression is at most $\left( \frac{1}{N}\right)^{n-2}$
    due to the large $N$
  factorization  of the correlator  consisting of
 the product of singlet operators.
  In the second quantized notation,
  eq.~(\ref{eq:nresolv}) is expressible as

\beqn
\lefteqn{
 \langle\!\langle N|\prod_{i=1}^{n}
 :\int \d\mu_{i} \Psi^{\dagger}(\tlm_i)
 \frac{1}{p_i-\lm_i}
 \Psi(\lm_i) : |N \rangle\!\rangle  }
 \nonumber \\
 &&=
 \langle\!\langle N|
 \prod_{i=1}^{n} :a^{\dagger}_{k_i} a_{j_i} :
 |N \rangle\!\rangle
 \prod_{i=1}^{n} \int\d\mu_i
 \xi_{k_i}(\tlm_i)
 \frac{1}{p_i-\lm_i}
 \xi_{k_i}(\lm_i)
 \nonumber \\
 &&=
 \langle\!\langle N|
 \prod_{i=1}^{n} :a^{\dagger}_{k_i} a_{j_i} :
 |N \rangle\!\rangle
 \prod_{i=1}^{n} \langle k_i|
 \frac{1}{p_i-A_i}
 |l_i \rangle
 \;\; ,
\eeqn

  The normal ordering  $: ...:$ is with respect to
 the filled sea $|N \rangle\!\rangle$.
  We introduce a notation
\beqn
  \left[ \frac{1}{p-A} \right]\left( z_{i}; \Lambda_{i} , \Lambda , N\right)
  \equiv \sum_{\delta} z_{i}^{\delta} < j_i -\delta\mid \frac{1}{p-A} \mid j_i >
  \;\;\\
  \Lambda_{i}=  j_i \Lambda /N = \Lambda + \Lambda \tilde{j}_i /N \;\;\;.
\eeqn
 The evaluation of
 $ \langle\!\langle N|
 \prod_{i=1}^{n} :a^{\dagger}_{k_i} a_{j_i} :
 |N \rangle\!\rangle$
  by the Wick theorem provides $(n-1)!$ terms
   of the  following structure:
 each  term is given by the product
 of $n$-Kronecker delta's multiplied  both by a sign factor   and
  by  the product of 
  $n$-step functions to ensure that the summations over the $n$-indices
  $ \tilde{j}_{1}, \tilde{j}_{2} \cdots $  and  $\tilde{j}_{n}$ are   bounded
 either from below $(\geq 0)$ or from above$(\leq -1)$.
  We denote this product by $ \Theta( \tilde{j}_{1},
  \tilde{j}_{2}, \cdots \tilde{j}_{n}; \sigma)$.
  These $(n-1)!$ terms are in one-to-one correspondence
 with  the circular permutations of $n$ integers   $1, \cdots, n$, which
 we denote by ${\cal S}_{n}$.   The $\sigma$ is an element of ${\cal S}_{n}$.
 For large $N$, we find
\beqn
\label{eq:step1}
 &~& \left(\frac{N}{\Lambda}\right)^{n-2}
 \DVEV{ \prod_{i=1}^{n} \Tr \frac{1}{p_{i} - \hA} }
 \\ \nonumber
 &=& \sum_{\tilde{j_{1}}, \tilde{j_{2}}, \cdots \tilde{j_{n}}}
  \sum_{\sigma \in {\cal S}_{n}}  \Theta( \tilde{j}_{1},
  \tilde{j}_{2}, \cdots \tilde{j}_{n}; \sigma)  sgn (\sigma)
  \left(  \prod_{j=1}^{n} \oint
 \frac{dz_{j}}{2\pi i} \right)
  \prod_{k=1}^{n} \frac{1}{z_{k}}
 \left( \frac{z_{\sigma(k)}}{z_{k}} \right)^{\tilde{j}_{k}}  \nonumber  \\
 &~& \times \frac{1}{(n-2)!} \left( \sum_{i=1}^{n} \tilde{j}_{i}
 \frac{\partial}{ \partial \Lambda_{i} } \right)^{n-2} 
 \prod_{i^{\prime}=1}^{n}  \frac{1}{p_{i^{\prime}} - 
 A(z_{i^{\prime}}; \Lambda_{i^{\prime}})}
  \mid_{\Lambda_{i^{\prime}}
 = \Lambda}   + {\cal{O}}\left(1/N\right) \;\;\;.
\eeqn
Note that in the large $N$ limit, we can use
 $\frac{1}{p_i - A(z_i; \Lambda_i) }$
 in  place of 
 $\left[ \frac{1}{p_i -A }\right](z_i;\Lambda_i,\Lambda,N)$ 
according to the same reason as stated in the case of three-loop correlator.
  The  $sgn(\sigma)$ denotes the signature associated with  the permutation $\sigma$.  

   Let us define
\beq
  m! D_{m}(z, z^{\prime}) \equiv
 \frac{1}{z} \sum_{\tilde{j} \geq 0 } \tilde{j}^{m}
 \left(z^{\prime}/z\right)^{\tilde{j}} 
  =  - \frac{1}{z} \sum_{\tilde{j} \leq -1 } \tilde{j}^{m}
 \left(z^{\prime}/z\right)^{\tilde{j}}  \;\;,  \;\;\; m= 0, \cdots \;\;\;.
\eeq
  In the continuum limit  we will be focusing  from now on,
 it is sufficient to use
\beq
  D_{m}(z, z^{\prime}) \approx \frac{1}{ (z- z^{\prime})^{m+1}}
  \equiv D_m(z-z') \quad .
\eeq
Let $sgn_{i}(\sigma)$ be $+1$ or $-1$, depending upon whether
  the restriction on the summation over $\tilde{j}_{i}$ is
  bounded from below or from above respectively.
  It is not difficult  to show 
\beq
 sgn(\sigma) \prod_{i=1}^{n}sgn_{i}(\sigma) = -1 \;\;\;,
\eeq
  for any $\sigma$ and $n$. 
 The summations over 
 $ \tilde{j}_{1}, \tilde{j}_{2} \cdots $  and  $\tilde{j}_{n}$ 
  can then be  performed  for all $\sigma$  at once,
  leaving  with this minus sign.

  Now we turn to the integrations over $z_{i}$ $(i= 1 \sim n)$.
   The convergence on  the geometric series  
 leads to  the  successively  ordered integrations of $z_{i}'s$ 
  for each $\sigma$.
  We assume here that only the poles at $z_i=z^*_i$ give
 rise to terms with physical significance. This is
 the case for three-loop correlators.
 By simply picking up a pole  of $z_{i}$ at
  $\frac{1}{p_{i} - A(z_{i} ; \Lambda_{i}) }$  for $i= 1 \sim n$  and
  using
\beqn
 &~&   \oint \frac{dz_{i}}{2\pi i} f\left( \cdots z_{i}, \cdots \right)
 \left( \frac{\partial}{\partial \Lambda_{i}} \right)^{\ell}
\left( \frac{1}{p_{i} -
  \left[A \right] (z_{i}; \Lambda_{i}) }\right)  \nonumber \\
 &=&  -  \frac{\partial}{\partial(a \zeta_{i})}
 \left( \frac{\partial}{\partial \Lambda_{i}} \right)^{\ell}
  \int^{z_{i}^{*}} dz_{i}  f\left( \cdots z_{i}, \cdots \right) \;\;\;,
 \;\; \ell =  0,1, \cdots \;\;\;,
\eeqn
  we find that eq.~(\ref{eq:step1}) is written as
\beqn
\label{eq:step2-1}
 \left(\frac{N}{\Lambda}\right)^{n-2}
 \DVEV{ \prod_{i=1}^{n} \Tr
 \frac{1}{p_{i} - \hA} } =
 \prod_{i=1}^{n} \left( - \frac{\partial}{\partial(a \zeta_{i})} \right)
  R^{(n)} |_{\Lambda_{i} = \Lambda} \;\;\;,
\eeqn
 where
\beqn
\label{eq:step2-2}
   R^{(n)}  &\equiv&  \sum_{i_{1}}^{n} 
\left(  \frac{\partial}{\partial \Lambda_{i_{1}} }\right)^{n-2}
   \int \cdots \int 
 \sum_{\sigma \in {\cal S}_{n} }
 D_{n-2}( [ i_{1} - \sigma(i_{1})])
 \prod_{j (\neq i_{1})} D_{0}( [j- \sigma(j)]) 
  \nonumber \\
   &+&
  \sum_{m_1+m_2 \atop =n-2} \sum_{(i_{1}, i_{2})} 
 \Bigl(  \frac{\partial}{\partial \Lambda_{i_{1}} }\Bigr)^{m_1}
 \Bigl(  \frac{\partial}{\partial \Lambda_{i_{2}} }\Bigr)^{m-2}
   \int \cdots \int 
 D_{m_1}( [ i_{1} - \sigma(i_{1})] )
 D_{m_2}( [ i_{2} - \sigma(i_{2})])
     \nonumber \\
 &&\quad \times
 \prod_{j (\neq i_{1}, i_{2})} D_{0}( [j- \sigma(j)])
 \nonumber \\ 
   &+&
 \sum_{m_1+m_2+m_3 \atop =n-2} \sum_{(i_{1}, i_{2}, i_{3})} 
 \Bigl(  \frac{\p}{\p \Lm_{i_{1}} }\Bigr)^{m_1}
 \Bigl(  \frac{\p}{\p \Lm_{i_{1}} }\Bigr)^{m_2}
 \Bigl(  \frac{\p}{\p \Lm_{i_{1}} }\Bigr)^{m_3}
   \int \cdots \int 
 D_{m_1}( [ i_{1} - \sigma(i_{1})] )
     \nonumber \\
 &&\quad \times
 D_{m_2}( [ i_{2} - \sigma(i_{2})])
 D_{m_3}( [ i_{3} - \sigma(i_{3})])
 \prod_{j (\neq i_{1}, i_{2})} D_{0}( [j- \sigma(j)])
 \nonumber \\ 
  &+& \cdots
  \nonumber \\  
  &+& \sum_{(i_{1}, i_{2}, \cdots , i_{n-2})} 
 \Bigl(  \frac{\partial}{\partial \Lambda_{i_{1}} }\Bigr)
 \Bigl(  \frac{\partial}{\partial \Lambda_{i_{2}} }\Bigr)
 \cdots
 \Bigl(  \frac{\partial}{\partial \Lambda_{i_{n-2}} }\Bigr)
   \int \cdots \int  \nonumber \\
 &&\quad \times
 \prod_{j=1}^{n-2} D_{1}( [ i_{j} - \sigma(i_{j})] )
 \prod_{j (\neq i_{1}, i_{2}, \cdots, i_{n-2})} D_{0}( [j- \sigma(j)])
   \;\;\;.
\eeqn

The integrals in the equation  above are  with respect to $z^{*}_{i}$'s and we  adopt a notation
\beqn
  [i] \equiv z_{i}^{*} \;\;\;, \;\;\;  [i-j] \equiv z_{i}^{*} - z_{j}^{*}
\;\;\;.
\eeqn
 This expression is in one to one correspondence with
 the expansion of $\Bigl(\sum_{i=1}^{n} x_i \Bigr)^{n-2}$.
 The number
 of terms appearing is equal to the  number of 
  partitions of $(n-2)$  objects into parts.

In order to put eqs.~(\ref{eq:step2-1}),  (\ref{eq:step2-2}) in a simpler  form, let us introduce 
\beqn
\label{eq:braket}
  &&\left( \begin{array}{cccc}
 m_{1}, & m_{2}, & \cdots, & m_{n}  \\
 i_{1}, & i_{2}, &  \cdots, &  i_{n}
 \end{array} \right)_{n}  \nonumber \\ 
 &&\quad \equiv - \sum_{\sigma \in {\cal S}_{n} }
 D_{m_{1}}( [ i_{1} - \sigma(i_{1})])
  D_{m_{2}} ( [ i_{2} - \sigma(i_{2})] )    
 \cdots  D_{m_{n}}  ( [ i_{k} - \sigma(i_{n})] )  \nonumber \\
&&\quad =-\sum_{\sigma\in{\cal S}_n } \frac1{[i_1-\sigma(i_1)]^{m_1+1} } \;
\frac1{[i_2-\sigma(i_2)]^{m_2+1} } \;\cdots \frac1{[i_n-\sigma(i_n)]^{m_n+1} }
\;\;.\eeqn
In particular,
\beqn 
   \left( \begin{array}{cccc}
 n-2,   & 0,       & \cdots,       &  0 \\
 i_{1}, & i_{2}, & \cdots, & 0
 \end{array} \right)_{n} 
  &\equiv& - \sum_{\sigma \in {\cal S}_{n} }
 D_{n-2}( [ i_{1} - \sigma(i_{1})])
 \prod_{j (\neq i_{1})} D_{0}( [j- \sigma(j)])  \nonumber \\
  \left( \begin{array}{ccccc}
 n-3,   & 1,       & 0,        & \cdots,   &  0 \\
 i_{1}, & i_{2}, &  i_{3}, &  \cdots,  &  0
 \end{array} \right)_{n} 
  &\equiv&    - \sum_{\sigma \in {\cal S}_{n} }
  D_{n-3}( [ i_{1} - \sigma(i_{1})] )
  D_{1}( [ i_{2} - \sigma(i_{2})])  \nonumber \\
  &~& \times \prod_{j (\neq i_{1}, i_{2})} D_{0}( [j- \sigma(j)]) \nonumber \\
  &{\rm e.t.c}&  
\eeqn

  In the \ref{app:sub-culc. in n-resolvent}, we prove  that
\beqn
\label{eq:sub-recursion}
  \left( \begin{array}{cccc}
 m_{1}, & m_{2},  & \cdots, & m_{n} \\
 i_{1},   & i_{2},    &  \cdots, & i_{n}
 \end{array} \right)_{n}   = 0 \;\;\; {\rm if} \;\;\ \sum_{\ell} m_{\ell} 
 \leq  n-3 \;\;\;  ,
\eeqn
 as well as
\beqn
\label{eq:recursion}
 &~& \left( \begin{array}{ccccccc}
 m_{1}, & m_{2}, & \cdots, & m_{k}, & 0, & \cdots,  & 0 \\
 i_{1},   & i_{2},   & \cdots, &   i_{k},& i_{k+1}, & \cdots,  & 0
 \end{array} \right)_{n}     \nonumber \\
 &=&  \sum_{\ell= 1}^{k} \frac{1}{\left[ i_{\ell}- i_{n} \right]^{2}}
  \left( \begin{array}{clccccr}
 m_{1}, & \cdots,   &  m_{\ell}-1, & \cdots,  & m_{k}, & 0,  & \cdots \\
 i_{1}, & \cdots,    &  i_{\ell},  &  \cdots, &  i_{k},& \cdots, &\cdots
 \end{array} \right)_{n-1}      \;\;\; \nonumber \\
 &{\rm if }& \;\;\; \sum_{\ell} m_{\ell} = n-2 \;\;\;.
\eeqn
 In particular, 
\beqn
\label{eq:n-2,0}
   \left( \begin{array}{ccc}
 n-2, & 0, & \cdots  \\
 i_{1}, &\cdots, & \cdots
 \end{array} \right)_{n} 
=  \frac{1}{\left[i_{1} - i_{n}\right]^{2} }
   \left( \begin{array}{ccc}
 n-3, & 0, & \cdots  \\
 i_{1}, & \cdots, & \cdots 
 \end{array} \right)_{n-1}   = \frac{1}{ {\displaystyle
 \prod_{j (\neq i_{1})}^{n}  } 
 \left[ i_{1} -j \right]^{2} }
\nonumber \\
 &&  
\eeqn
  and
\beqn
\label{eq:n-3,1}
 \left( \begin{array}{ccc}
 n-3, & 1, & \cdots  \\
 i_{1}, & i_{2} & \cdots 
 \end{array} \right)_{n} 
 &=& 
  \frac{1}{\left[ i_{1} - i_{n}\right]^{2} }
 \left( \begin{array}{ccc}
 n-4, & 1, & \cdots  \\
 i_{1}, & i_{2}, & \cdots
 \end{array} \right)_{n-1}  \nonumber \\
   ~&+&
  \frac{1}{\left[ i_{2} - i_{n}\right]^{2} }
 \left( \begin{array}{ccc}
 n-3, & 0, & \cdots  \\
 i_{1}, & i_{2}, & \cdots
 \end{array} \right)_{n-1}   \;\;\;.
\eeqn

Let us introduce graphs in which    the factor $1/[i-j]^2$  is represented
 by  a double line linking  circle $i$ and circle $j$
  to handle the  quantities defined by eq.~(\ref{eq:braket})
 more easily.
For example, for n=3,
\beq
 \left( \begin{array}{ccc}
 1 & 0 & 0  \\
 1 & 2 & 3
 \end{array} \right)_{3} 
 =\frac1{[z^*_1-z^*_2]^2}\; \frac1{[z^*_1-z^*_3]^2}
 \equiv
 \epscenterboxy {2.2cm}{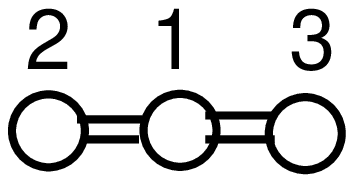}  
  .
\eeq
 Using the recursion relation eq.~(\ref{eq:recursion})
 and eq.~(\ref{eq:sub-recursion}) we have, for n=4,
\beqn
 &&\left( \begin{array}{cccc}
 2 & 0 & 0 & 0  \\
 1 & 2 & 3 & 4 
 \end{array} \right)_{4}
 = \frac{1}{[z_1^*-z_4^*]^2}
 \left( \begin{array}{ccc}
 2-1 & 0 & 0   \\
 1 & 2 & 3  
 \end{array} \right)_{3} \nonumber \\
 &&\qquad\qquad =
 \epscenterboxy{2.5cm}{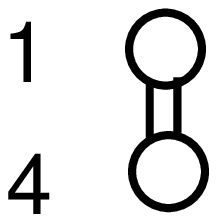}
 \times
 \epscenterboxy{2.2cm}{ai_fig/lk213.eps} 
 = \epscenterboxy{2.5cm}{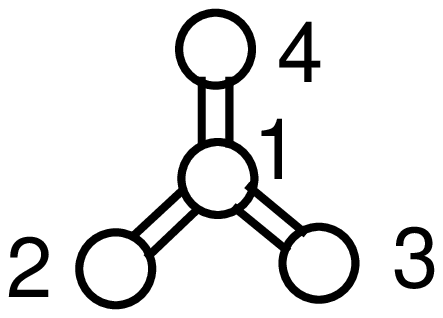} , 
\eeqn
\beqn
  &&\left( \begin{array}{cccc}
 1 & 1 & 0 & 0  \\
 1 & 2 & 3 & 4 
 \end{array} \right)_{4}
 =\frac{1}{[z_1^*-z_4^*]^2}
 \left( \begin{array}{ccc}
 1-1 & 1 & 0   \\
 1 & 2 & 3  
 \end{array} \right)_{3}
 + \frac{1}{[z_2^*-z_4^*]^2}
 \left( \begin{array}{ccc}
 1 & 1-1 & 0   \\
 1 & 2 & 3  
 \end{array} \right)_{3} \nonumber \\
 &&\qquad\qquad =
 \epscenterboxy{2.5cm}{ai_fig/lk14.eps} \times
 \epscenterboxy{2.2cm}{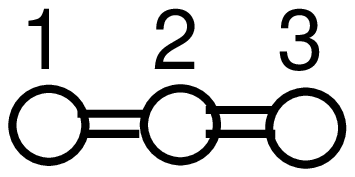} 
 + \epscenterboxy{2.5cm}{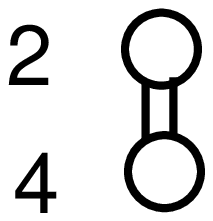} \times
 \epscenterboxy{2.2cm}{ai_fig/lk213.eps}
 \nonumber \\
 &&\qquad\qquad = 
 \epscenterboxy{2.cm}{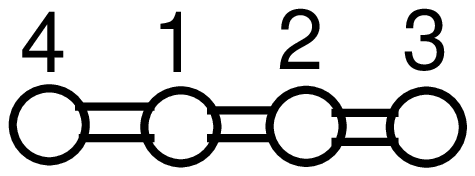}
 + \epscenterboxy{2.cm}{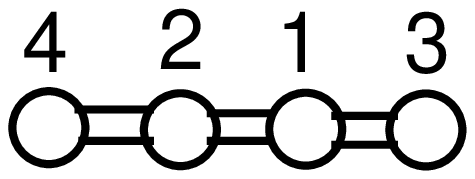}  ,
\eeqn
  and, for n=5,
\beqn
 &&\left( \begin{array}{ccccc}
 3 & 0 & 0 & 0 & 0  \\
 1 & 2 & 3 & 4 & 5
 \end{array} \right)_{5}
 =\frac{1}{[z_1^*-z_5^*]^2}
 \left( \begin{array}{cccc}
 3-1 & 0 & 0 & 0   \\
 1 & 2 & 3 & 4  
 \end{array} \right)_{4} \nonumber \\
 &&\qquad\qquad =
 \epscenterboxy{2.5cm}{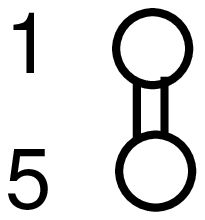}\times
 \epscenterboxy{2.5cm}{ai_fig/lk1_234.eps} 
 =\epscenterboxy{2.5cm}{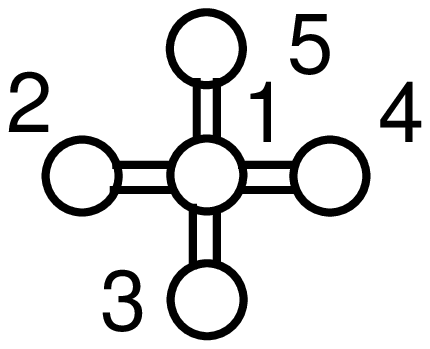} , 
\eeqn
\beqn
 &&\left( \begin{array}{ccccc}
 2 & 1 & 0 & 0 & 0  \\
 1 & 2 & 3 & 4 & 5 
 \end{array} \right)_{5}
 =\frac{1}{[z_1^*-z_5^*]^2}
 \left( \begin{array}{cccc}
 2-1 & 1 & 0 & 0   \\
 1 & 2 & 3 & 4  
 \end{array} \right)_{4}
\nonumber \\
&&\qquad\qquad
 + \frac{1}{[z_2^*-z_5^*]^2}
 \left( \begin{array}{cccc}
 2 & 1-1 & 0 & 0   \\
 1 & 2 & 3 & 4  
 \end{array} \right)_{4} \nonumber \\
 &&\qquad\qquad =
 \epscenterboxy{2.5cm}{ai_fig/lk15.eps}\times
 \Biggl\{ \epscenterboxy{2.cm}{ai_fig/lk4123.eps}
 +\epscenterboxy{2.cm}{ai_fig/lk4213.eps} 
 \Biggr\} 
 \nonumber \\
 &&\qquad\qquad \qquad+
 \epscenterboxy{2.5cm}{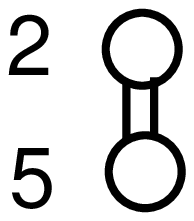}\times
 \epscenterboxy{2.5cm}{ai_fig/lk1_234.eps}
 \nonumber \\
 &&\qquad\qquad = 
 \epscenterboxy{2.5cm}{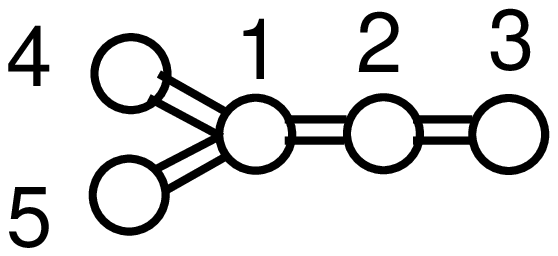}
 +\epscenterboxy{2.5cm}{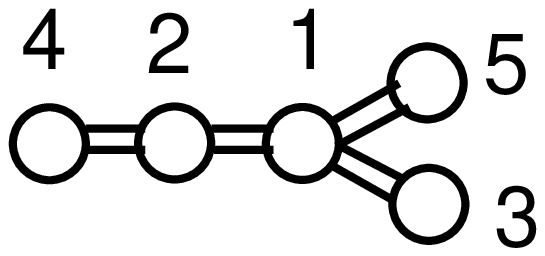}
 +\epscenterboxy{2.5cm}{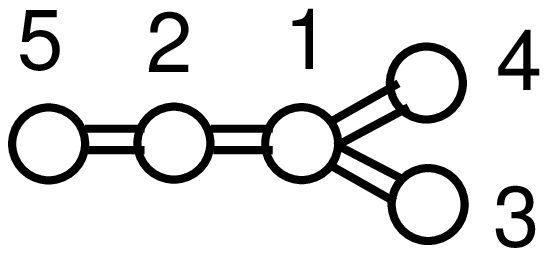}
  . 
 \nonumber \\
\eeqn
From the examples  above, it is clear that the graphs for the general
 case can be written down quite easily. 

In terms of the quantities defined by eq.~(\ref{eq:braket}), we obtain
 a formula for the n-point resolvent :
\beqn
\label{eq:result}
 \left(\frac{N}{\Lambda}\right)^{n-2}
 \DVEV{\prod_{i=1}^{n} \Tr \frac{1}{p_{i} - \hA} }   =
 \prod_{i=1}^{n} \left( - \frac{\partial}{\partial(a \zeta_{i})} \right)
 R^{(n)} |_{\Lambda_{i} = \Lambda} \;\;\;,
\eeqn
where
\beqn
\label{eq:Delta}
 \lefteqn{
 R^{(n)}(z^*_1,\cdots.z^*_n)
 } \nonumber \\
 &&
 = \sum_{i_1}^{n} \Bigl(\frac{\p}{\p\Lambda_{i_1} }\Bigr)^{n-2}
 \int \cdots \int \;
 \left( \begin{array}{cccc}
 n-2, & 0, & \cdots, & 0 \\
 i_1 , & i_2, & \cdots, & i_n
 \end{array} \right)_n   \nonumber \\
 &&
 + \sum_{m_1+m_2=n-2} \sum_{ (i_1,i_2) }
 \Bigl(\frac{\p}{\p\Lambda_{i_1} }\Bigr)^{m_1}
 \Bigl(\frac{\p}{\p\Lambda_{i_2} }\Bigr)^{m_2}
 \int \cdots \int \;
 \left( \begin{array}{ccccc}
 m_1, & m_2,& 0, & \cdots, & 0 \\
 i_1 , & i_2, & i_3, & \cdots, & i_n
 \end{array} \right)_n   \nonumber \\
 &&
 + \sum_{ {m_1+m_2+m_3\atop =n-2 }} \sum_{ (i_1,i_2,i_3) }
 \Bigl(\frac{\p}{\p\Lambda_{i_1} }\Bigr)^{m_1}
 \Bigl(\frac{\p}{\p\Lambda_{i_2} }\Bigr)^{m_2}
 \Bigl(\frac{\p}{\p\Lambda_{i_3} }\Bigr)^{m_3}
 \int \cdots \int \;
 \nonumber \\
 &&\qquad\qquad\qquad\qquad
 \qquad\qquad\qquad\qquad
 \left( \begin{array}{cccccc}
 m_1, & m_2, & m_3,& 0, & \cdots, & 0 \\
 i_1 , & i_2, & i_3, & i_4, & \cdots, & i_n
 \end{array} \right)_n 
 \nonumber \\
 &&+ \cdots \nonumber \\
 &&
 + \sum_{ (i_1,\cdots,i_{n-2}) }
 \Bigl(\frac{\p}{\p\Lambda_{i_1} } \Bigl)
 \cdots
 \Bigl(\frac{\p}{\p\Lambda_{i_{n-2}} } \Bigl)
 \int \cdots \int \;
 \left( \begin{array}{ccccc}
 1, & \cdots, & 1,& 0,  & 0 \\
 i_1 , & \cdots, & i_{n-2}, &  i_{n-1}, & i_n
 \end{array} \right)_n   
 .
\eeqn
  Here $m_{\ell} \geq 1$  and  the summation $(i_{1}, \cdots,  i_{k})$
  denotes  a set of  $k$ unequal integers from $1,2, \cdots, n$  and
 $i_{k+1}, \cdots, i_{n}$
  in the array represents  the remaining integers.

 From \eq{eq:result}, \eq{eq:Delta}, we can derive the rule
 which states how to construct the $n$-resolvent correlator
 graphically. This rule was shown in
 sect.~\ref{sec:formula for n-resolvent}.

\subsection{The selection rule in summation and crossing symmetry}
 
    In the previous section, we have derive the formula for
 the $n$-resolvent correlators, which were functions of
 $\zeta_i$ and $\mu$.
    In order to obtain the loop correlators, we have to 
 carry out the inverse Laplace transformation with respect
 to $\zeta_i$ further.
  In the processes, the following functions will
 play a fundamental role.
\beqn
\label{eq:polygon}
    P_{n}(\theta_{1}, \theta_{2}, \cdots \theta_{n}) 
  &\equiv&
   \sum_{i_{1}}^{n} 
\left(  \frac{\partial}{\partial \Lambda_{i_{1}} }\right)
   \int \cdots \int 
  \left( \begin{array}{clcr}
 n-2, & 0, & \cdots, & 0 \\
 i_{1}, &  i_{2}, &  \cdots,  &  i_{n}
 \end{array} \right) 
 \nonumber \\ 
 &=&  \sum_{i=1}^{n}
\frac{\p z^*_i}{\p \Lambda_i}
 \frac{1} { {\displaystyle
 \prod_{j (\neq i)}^{n} } [i-j] }
 \;\;\;.
\eeqn
 where
\beq
 z_i^*=\exp (2\eta_i \cosh \theta_i)
 \quad ,
\eeq
\beq
p_i - p_* =
 a \zeta_i=2 \eta_i^m \cosh m\theta_i=a M_i \cosh m\theta_i
 \quad,\quad
 \eta_i=(a M_i/2)^{1/m}
\eeq
\beq
 \Lm_{i}-\Lm_*=-(m+1) \eta_i^{2m}=-a^2 \mu_i
 =-(m+1)(aM_i/2)^2
 \quad,
\eeq
and
\beqn
  \left.\frac{\partial \theta_{i}}{ \partial \Lambda}
  \right|_{\zeta_{i}}
  = -\frac{1}{\eta} \frac{\partial \eta}{\partial \Lambda}
  \frac{\cosh m\theta_{i}}{ \sinh m \theta_{i}} \;\;\;,
 \;\;\;\;
 \frac{\p z^*_i}{\p \Lambda_i} = 2 \left(
 \frac{\partial \eta}{\partial \Lambda} \right)
  \frac{ \sinh (m-1) \theta_{i} }{ \sinh m \theta_{i} }\;\;\;.
\eeqn

 The important point here is that the above functions can
 be written as a sum of the product of $n$ factors each
 of which is a function only of the corresponding 
 $\theta_i$ and $\mu$, so that we can obtain the
 inverse-Laplace
 images of $P_n(\theta_1,\cdots,\theta_n)$ immediately. 
  We conjecture that $n$-resolvent correlator can be
 expressed in terms of $P_{n'}\;(n'\le n)$.
  In fact, as we will show later explicitly, 
 it is true for the four- and the five-resolvent
 correlator.
 
    A key manipulation we  will use is   the  partial fraction
\beq
\label{eq:pf}
  \frac{1}{[i-j][i-k]} = \frac{1}{[i-k][k-j]} + \frac{1}{[i-j][j-k]} \;\;\;.
\eeq
  One can associate a line from $i$ to $j$ with $\frac{1}{[i-j]}$.
 The following identity
  is responsible for expressing  $P_{n}$ as
 a sum of the product of  $n$ factors each of which depends
  only on  $\theta_{i}$:
\beqn
   I_{k} ( \alpha, \beta ;m ) &\equiv&
 \lefteqn{
 \frac{1}{\cosh\alpha-\cosh\beta}
 \left(\;\frac{\sinh (m-k)\alpha}{\sinh m\alpha}
 -\frac{\sinh (m-k)\beta}{\sinh m\beta} \;\right) \mbox{                 }
 } \nonumber\\
 &&=-2\sum_{j=1}^{m-k}\sum_{i=1}^{k}
 \;\frac{\sinh (m-j-i+1)\alpha}{\sinh m\alpha}
 \;\frac{\sinh (m-j-k+i)\beta}{\sinh m\beta}
 \;.
\nonumber \\
&&
\label{eq:ident}
\eeqn

Let us first work out  the cases $n=2,3,4$  to get a feeling.  For $n=2$,
\beqn
 &~&  P_{2}(\theta_{1}, \theta_{2})
 =  \frac{\frac{\p z^*_1}{\p \Lambda_1} -\frac{\p z^*_2}{\p \Lambda_2}}
 {[1-2]} = 2 \left(
 \frac{ \partial \eta}{\partial \Lambda} \right) \frac{1}{2\eta}  I_{k_{1}= 1}
 \left( \theta_{1}, \theta_{2} ; m \right)  \nonumber \\
 &=& 2 \left( \frac{ \partial \eta}{\partial \Lambda}\right)
\left( \frac{-1}{\eta} \right) \sum_{j=1}^{m-k_{1}}\sum_{i_{1}=1}^{k_{1}}
\frac{\sinh (m-j_{1}-i_{1}+1)\theta_{1} }{\sinh m\theta_{1}}
 \frac{\sinh (m-j_{1}+i_{1}- k_{1})\theta_{2}}{\sinh m\theta_{2}} 
 \nonumber \\
 &=& 2 \left( \frac{ \partial \eta}{\partial \Lambda} \right)
\left(\frac{-1}{\eta} \right)
\sum_{j_{1}=1}^{m-1}
 \;\frac{\sinh (m-j_{1})\theta_{1} }{\sinh m\theta_{1}}
 \;\frac{\sinh (m-j_{1})\theta_{2}}{\sinh m\theta_{2}} \;\;,
\eeqn
   where $i_{1}=k_{1}=1$.
  For $n=3$, we use eq.~(\ref{eq:pf}) for the  term
 containing $\frac{\p z^*_2}{\p \Lambda_2}$ to create a  link
  $[1-3]$, which is originally absent.
 This relates $P_{3}$ to $P_{2}$.    We find
\beqn 
&~& P_{3} (\theta_{1}, \theta_{2},\theta_{3})  \nonumber \\
&=&  2 \left( \frac{ \partial \eta}{\partial \Lambda} \right)
 \frac{ I_{1}\left( \theta_{2}, \theta_{1} ; m \right)- 
 I_{1}\left( \theta_{2}, \theta_{3} ; m \right) }  {[1-3]} 
  \nonumber \\
 &=& 2 \left( \frac{ \partial \eta}{\partial \Lambda}\right)
 \left(\frac{-1}{\eta}\right)^{2}
 \left(\sum_{j_{1}=1}^{m-k_{1}}\sum_{i_{1}=1}^{k_{1}=1} \right)
 \left(\sum_{j_{2}=1}^{m-k_{2}}\sum_{i_{2}=1}^{k_{2}} \right)  \\
&~& \frac{\sinh (m-j_{2}-i_{2}+1)\theta_{1} }{\sinh m\theta_{1}}
\frac{\sinh (m-j_{1}+i_{1}- k_{1})\theta_{2}}{\sinh m\theta_{2}} 
 \frac{\sinh (m-j_{2}+i_{2}- k_{2})\theta_{3}}{\sinh m\theta_{3}} 
  \;. \nonumber
\eeqn
   Here $k_{2} = j_{1} + i_{1}-1= j_{1}$.

This can be repeated  for arbitrary $n$.
 In the case $n=4$,  we use the partial fraction
 for  the two terms containing $\frac{\p z^*_2}{\p \Lambda_2}$ and 
 $\frac{\p z^*_3}{\p \Lambda_3}$
  to create a link $[1-4]$, which is originally absent. 
 This enables us to relate the  case $n=4$
  to the case $n=3$.
  In general,  $P_{n}$  is  related to $P_{n-1}$  
by using  the partial fraction for the terms containing
 $\frac{\p z^*_2}{\p \Lambda_2} \sim \frac{\p z^*_{n-1}}{\p \Lambda_{n-1}}$
   to create a link $[1-n]$.
   We obtain
\beqn
\label{eq:answer}
 &~& P_{n}(\theta_{1}, \theta_{2}, \cdots \theta_{n})  = 
  - \left( \frac{ \partial \eta^{2}}{\partial \Lambda}\right)
 \left(\frac{-1}{\eta}\right)^{n}
 \left( \prod_{\ell=1}^{n-1}
\sum_{j_{\ell}=1}^{m-k_{\ell}}\sum_{i_{\ell}=1}^{k_{\ell}} \right)
\nonumber \\
 &~& \left( \prod_{\ell^{\prime}=1}^{n-1}
 \;\frac{\sinh (m-j_{\ell^{\prime}}+
i_{\ell^{\prime}}- k_{\ell^{\prime} })\theta_{\ell^{\prime}+1 }
 }{\sinh m\theta_{\ell^{\prime}+1 }  }  \right)
 \frac{\sinh (m-j_{n-1}-i_{n-1}+1)\theta_{1} }{\sinh m\theta_{1}} \;,
\eeqn 
  where $k_{\ell} = j_{\ell-1} + i_{\ell-1} -1,$~~ for
$\ell = 2,3, \cdots , (n-1) $.
	Eq.~(\ref{eq:answer})  expresses 
 the $P_{n}(\theta_{1}, \theta_{2}, \cdots \theta_{n})$
  as a sum of the $n$-products
  of  the factor ${\displaystyle  \frac{\sinh(m-k)\theta_{i}}
{\sinh m\theta_{i}} }$.
 Owing to this property,  one can perform
 the inverse Laplace transform immediately.
 
  Let us now  discuss  the restrictions on the summations of
 $2n-3$  integers $j_{1},$ $i_{2}, j_{2},$ $\cdots$
  $i_{n-1}, j_{n-1}$ in  eq.~(\ref{eq:answer}).
  We   write these as  a set:
\beqn
 &~&{\cal F}_{n}( j_{1}, i_{2}, j_{2} \cdots
  i_{n-1}, j_{n-1})   \nonumber \\
 &\equiv& \{ ( j_{1}, i_{2}, j_{2}, \cdots
  i_{n-1}, j_{n-1}) \mid
 1 \leq i_{\ell} \leq k_{\ell}, \;
 1 \leq j_{\ell} \leq m- k_{\ell}, \;  {\rm for}~ \ell = 
  1,2, \cdots n-1 \}  \nonumber \\
 &=& {\cal F}_{2}( i_{1}=1, j_{1};k_{1}=1)
 \prod_{\ell=2}^{n-1} \cap {\cal F}_{2}( i_{\ell}, j_{\ell};k_{\ell})
 \;, 
\eeqn
  where
\beqn
  {\cal F}_{2}( i_{\ell}, j_{\ell};k_{\ell})
 \equiv  \{ (  i_{\ell}, j_{\ell} ) \mid
 1 \leq i_{\ell} \leq k_{\ell},\;
 1 \leq j_{\ell} \leq m- k_{\ell},\;\;{\rm with}\; k_{\ell}~~{\rm fixed} \}
 \;\;.
\eeqn
  We will show that  these restrictions on the sums
  are in fact in one-to-one correspondence with
 the fusion rules of the unitary minimal models
 for the diagonal primaries. 
  Let us begin with the case $n=3$.
  Define 
\beqn
\begin{array}{lcr}
    p_{1}\equiv j_{1} +k_{1} - i_{1}\;, &
 p_{2} \equiv j_{2} +k_{2} - i_{2}\;, &
q_{3} \equiv j_{2} +i_{2} - 1\;,  \\
   a_{12} \equiv p_{1} -1\;,  &   a_{23} \equiv  p_{2} -1 \;, 
  &   a_{31} \equiv q_{3} -1\;,  \end{array} \;\;
\eeqn
 The inequalities on $i_{2}, j_{2}$  are found to be   equivalent to
  the following four inequalities:
\beqn
\label{eq:trieq}
 a_{12} +  a_{23} -  a_{31} &=& 2(k_{2}-i_{2}) \geq 0 \;\;\;. \nonumber \\
a_{12} - a_{23} +  a_{31} &=& 2(i_{2}-1) \geq 0 \;\;\;. \nonumber \\
  - a_{12} +  a_{23} +  a_{31} &=& 2(j_{2}-1) \geq 0 \;\;\;\nonumber \\
 a_{12} +  a_{23} +  a_{31} &=& 2(j_{2}+ k_{2}-2) \leq  2(m-2) \;\;.
\eeqn
 From  the  third and the fourth equation of eq.~(\ref{eq:trieq}), the
 inequality $ a_{12} \leq m-2$  follows, which is a condition for
${\cal F}_{2}( i_{1}=1, j_{1};k_{1}=1)$. 
  Defining  a set
\beqn
\label{eq:D3}
{\cal D}_{3}(a_{1}, a_{2}, a_{3}) &\equiv&
   \{ (a_{1}, a_{2}, a_{3}) \mid \sum_{i(\neq j)}^{3} a_{i} - a_{j} \geq 0\;
\; {\rm for}\;\; i=1 \sim 3\; \;\;, \nonumber \\
 &~& \sum_{i=1}^{3} a_{i}
 = {\rm even} \leq 2(m-2) \} \;\;,
\eeqn
 we state  eq.~(\ref{eq:trieq})  as
\beqn
 {\cal F}_{3}(j_{1}, i_{2}, j_{2}) = 
  {\cal D}_{3} (a_{12},a_{23},a_{31}) \;\;\;.
\eeqn
  We also write
\beqn
{\cal F}_{2}(j_{1}) \equiv {\cal F}_{2}( i_{1}=1, j_{1};k_{1}=1)
 \equiv {\cal D}_{2} (a_{12}) \;\;\;.
\eeqn
  for  the case $n=2$.

 Eq.~(\ref{eq:D3}) is nothing but the condition   that
 a triangle  be formed  which is made out of
 $a_{1},a_{2}$ and $a_{3}$  and whose  circumference is less than or equal
  to  $2(m-2)$.
  It is also the selection rule for the  three point function
  of the diagonal primaries in  $m$-th minimal unitary
 conformal field theory \cite{BPZ}.

  For the case $n=4$,  introduce
$p_{3} \equiv j_{3} +k_{3} - i_{3}\;,
  a_{34}\equiv p_{3} -1\;,
q_{4} \equiv j_{3} +i_{3} - 1\;,  a_{41}\equiv q_{4} -1 \;$.
   We find
\beqn
  {\cal F}_{2}( i_{3}, j_{3};k_{3}) = 
  {\cal D}_{3} (a_{31},a_{34},a_{41}) \;\;\;\,
\eeqn
  The restrictions on the sum  in the case  $n=4$
  can be understood as gluing the two triangles:
\beqn
\label{eq:F4}
{\cal F}_{4}( j_{1}, i_{2}, j_{2}, i_{3}, j_{3}) &=&
 {\cal D}_{3} (a_{12},a_{23},a_{31})   \cap {\cal D}_{3} (a_{34},a_{41},a_{31})
  \nonumber \\
  &\equiv& {\cal D}_{4} (a_{12},a_{23},a_{34},a_{41}; a_{31})  \;\;.
\eeqn 
 The allowed integers on $a_{31}$  are  naturally interpreted as  permissible
  quantum numbers flowing through an intermediate channel.
 As one can imagine, eq.~(\ref{eq:F4}) is not the only way to represent
 the restriction: one can also represent it as
\beqn
 &~&{\cal F}_{4}( j_{1}, i_{2}, j_{2}, i_{3}, j_{3})=
 {\cal D}_{3} (a_{12},a_{24},a_{41})
   \cap {\cal D}_{3} (a_{23},a_{34},a_{24}) \nonumber \\
 &\equiv& {\cal D}_{4} (a_{12},a_{23},a_{34},a_{41}; a_{24})  \;\;,
\eeqn
  which embodies the  crossing symmetric property of the amplitude.

 The restrictions in the  general case $n$ are understood as  attaching
  a  triangle   to  the case  $(n-1)$.
   To see this, define 
\beqn
\begin{array}{lr}
  p_{\ell} = j_{\ell} + k_{\ell} - i_{\ell} \;,  &
  q_{\ell} = j_{\ell-1} +i_{\ell-1} -1 \;,  \\
 a_{\ell, 1} = q_{\ell}-1\;,  & a_{\ell, \ell+1} = p_{\ell}-1\;,
\end{array}  \;\;\; {\rm for}\;\;\;{\rm ~~\ell = 1,2, \cdots n} \;\;\;. 
\eeqn
  Using $ 1 \leq i_{n-1} \leq k_{n-1},$~~
$1 \leq j_{n-1} \leq m - k_{n-1}$,   we derive
\beqn
 a_{n-1,n} +  a_{n,1} -  a_{n-1,1} &=& 2 (j_{n-1}-1) \geq 0 \;,
  \nonumber \\
a_{n-1,n} - a_{n,1} +  a_{n-1,1} &=& 2(k_{n-1} - i_{n-1}) \geq 0 \;,
 \nonumber \\
  - a_{n-1,n} +  a_{n,1} +  a_{n-1,1} &=& 2(i_{n-1}-1) \geq 0 \;,
 \nonumber \\
 a_{n-1,n} +  a_{n,1} +  a_{n-1,1}
&=& 2(j_{n-1}+ k_{n-1} -2) \leq  2(m-2) \;.
\eeqn
The restriction on $i_{n-1}$ and $j_{n-1}$  are, therefore,
  ${\cal D}_{3}(a_{n-1,n}, a_{n,1}, a_{n-1,1})$, which
 is what we wanted to see.
   All in all, we  find 
\beqn
\label{eq:Dn}
 &~&{\cal F}_{n}( j_{1}, i_{2}, j_{2}, \cdots
  i_{n-1}, j_{n-1})   \nonumber \\
 &=&
  {\cal D}_{3}(a_{n-1,n}, a_{n,1}, a_{n-1,1}) \cap
 {\cal F}_{n-1}( j_{1}, i_{2}, j_{2}, \cdots
  i_{n-2}, j_{n-2})   \nonumber \\
 &=& {\cal D}_{3}(a_{n-1,n}, a_{n,1}, a_{n-1,1}) \cap
 {\cal D}_{n-1}(a_{1,2},a_{2,3}, \cdots a_{n-2,n-1}, a_{n-1,1};
  a_{3,1}, a_{4,1}, \cdots a_{n-2,1})  \nonumber \\
 &\equiv& {\cal D}_{n}(a_{1,2},a_{2,3}, \cdots a_{n-1,n}, a_{n,1};
  a_{3,1}, a_{4,1}, \cdots a_{n-1,1}) 
\eeqn 

 From now on,  a shortened notation ${\cal D}_{n}(a_{1,2},a_{2,3},
 \cdots a_{n-1,n}, a_{n,1})$
 is understood  to represent
 $ {\cal D}_{n}(a_{1,2},a_{2,3}, \cdots a_{n-1,n}, a_{n,1};
  a_{3,1}, a_{4,1}, \cdots a_{n-1,1}) $.

  Putting eq~(\ref{eq:answer}) and eq.~(\ref{eq:Dn}) together, we obtain a
  formula
\beqn
\label{eq:P-answer}
 &~& P_{n}(\theta_{1}, \theta_{2}, \cdots \theta_{n})  \nonumber \\
 &=& 
  - \left( \frac{ \partial \eta^{2}}{\partial \Lambda}\right)
 \left(\frac{-1}{\eta}\right)^{n}
 \sum_{ {\cal D}_{n}}  \left( \prod_{j = 2}^{n}
 \;\frac{\sinh (m- k_{j} -1)\theta_{j}
 }{\sinh m\theta_{j}  }  \right)
 \frac{\sinh (m-k_{1}-1)\theta_{1} }{\sinh m\theta_{1}}
 \;\;\;,
 \nonumber \\
\eeqn 
 where ${\cal D}_n$ means  ${\cal D}_n(k_1-1,\cdots,k_n-1)$. 
  Once again,  the  fact that
   the different divisions of ${\cal D}_{n}$ into $n-2$ triangles
   are   embodied  by   this  single expression  is precisely
  the statement of  the old duality.
 
  The object
  $P_{n}(\theta_{1}, \theta_{2}, \cdots \theta_{n})$
 is equipped with  $\theta_{j}$ and $k_j$  for $j=1,2,\cdots, n$
  and    any ${\cal D}_{3}(k_1-1,k_2-1,k_3-1)$ obeys the rule
 of  the triangle  specified above.
 It is, therefore,  natural to visualize    this as 
  a vertex which  connects $n$ external legs corresponding  to $n$ loops.
  The vertex can be regarded as a dual graph of an n-gon that
 corresponds ${\cal D}_n(k_1-1,\cdots,k_n-1)$.

   Due to the formula (\ref{eq:inverse-Laplace}),
 we can obtain  the inverse-Laplace
 image of \eq{eq:P-answer} with respect to  $\zeta_{i}$ $(i=1, \sim n)$
 immediately:
\beqn
 &&{\cal L}^{-1} \left[ P_n(\theta_1,\cdots,\theta_n) \right]
 =
 -\Bigl( \frac{\partial\eta^2}{\partial\Lambda} \Bigr)
 \Bigl(\frac{-1}{\eta}\Bigr)^n \Bigl(\frac{M}{2}\Bigr)^n
 \;\sum_{ {\cal D}_n }
 \Bigl[ \prod_i \tK{1-\frac{k_i}{m}}(M\l_i) \Bigl]
 \nonumber \\
 &&\;\;\;=
 (-1)^n \frac{1}{m(m+1)} 
 \Bigl(\frac{aM}{2}\Bigr)^{-2+(2-n)\frac 1m}
 \Bigl(\frac{M}{2}\Bigr)^{n}
 \;\sum_{ {\cal D}_n }
 \Bigl[ \prod_i \tK{1-\frac{k_i}{m}}(M\l_i) \Bigl] 
 \;\;.
\label{eq:L-1P}
\eeqn

\subsection{Four-loop correlators}
\label{sec:four-loop}

 In this subsection and in the next one,
 we show how to perform the inverse Laplace
 transformation of the resolvent correlators to get loop
 correlators in terms of loop lengths in the case of $n=4$, $5$. 
It is necessary to put 
\beqn
  R^{(n)}(\theta_1,\cdots,\theta_n)
  \equiv R^{(n)}(z^*_1,\cdots,z^*_n) |_{\Lambda_i=\Lambda}
\eeqn
 in a manageable form to the inverse Laplace transform.
 Let us recall that $P_n(\theta_1,\cdots,\theta_n)$ can be inverse
 Laplace transformed immediately.
 If $R^{(n)}(\theta_1,\cdots,\theta_n)$ is expressed as
 a polynomial of $P_j(\theta_1,\cdots,\theta_n)$ and their derivatives
 with respect to $\Lambda$,
 the inverse Laplace transform  can be carried out immediately.
 In fact, this turns out be true for $n=4$ and $5$, which we will show
 explicitly in the following.
 
 The important point we will use in the following 
 is the fact that when one of the loops shrinks
 and the loop length goes to zero,
 the $n$-loop amplitude  must become proportional to the
 derivative of the $(n-1)$-loop amplitude with
 respect to   the cosmological constant.
  We represent this fact by  
\beq
 \VEV{w^+(\l_1)\cdots w^+(\l_n)}\to ~~\propto 
 \frac{\p}{\p \mu}
 \VEV{w^+(\l_1)\cdots w^+(\l_{n-1})}
\;\;,
\eeq
 when $n$-th loop shrinks.
 Then in the limit,
 $R^{(n)}(\theta_1,\cdots,\theta_n)$ must
 satisfy the following relation:
\beq
 {\cal L}^{-1}\left[ R^{(n)}(\theta_1,\cdots,\theta_n) \right]
 \to~~\propto 
 {\cal L}^{-1}\left[ \frac{\partial}{\partial \Lambda} 
 R^{(n)}(\theta_1,\cdots,\theta_{n-1})  \right]
\label{eq:shrink-Delta}
\eeq
 This relation restricts the possible form of
 $R^{(n)}(\theta_1,\cdots,\theta_n)$ .
   We should note here that  the inverse Laplace image of 
 $P_n(\theta_1,\cdots,\theta_n)$
 satisfy the following relation
\beq
 {\cal L}^{-1}\left[ P_n(\theta_1,\cdots,\theta_n) \right]
 \to~~\propto
 {\cal L}^{-1}\left[ P_{n-1}
 (\theta_1,\cdots,\theta_{n-1}) \right]
 \quad.
\label{eq:shrink-P} 
\eeq
  This fact follows from substituting 
\beq
 \tK{1-\frac{k_n}m}(M\l_n) \approx
 \frac{1}{\Gamma(\frac{k_n}m)}
 \left(\frac{M\l_n}{2}\right)^{\frac {k_n}m -1}
 \;\;,\;\;
 M\l_n\ll 1
\eeq
 in \eq{eq:L-1P} and picking up only the $k_n=1$ parts.

 Because we want $R^{(n)}(\theta_1,\cdots,\theta_n)$ to be expressed as
 a polynomial of $P_j$ and their derivatives with respect to $\Lambda$, 
 we need here to introduce  a notation
\beqn
\label{eq:notationS}
&~& \left[ {\cal S} P_{1, \cdots, i_{1}} P_{j_{2}, \cdots, j_{2}+ i_{2}-1}
  \cdots P_{n-i_{\ell}+1, \cdots, n} \right] \left( \theta_{1}, \theta_{2},
 \cdots \theta_{n} \right) \equiv   \\
 &~& \frac{1}{n!} \sum_{\sigma \in  {\cal P}_{n}}
  P_{i_{1}} (\theta_{\sigma(1)}, \cdots,
 \theta_{\sigma(i_{1})}) P_{i_{2}}(\theta_{\sigma(j_{2})},
 \cdots, \theta_{\sigma(j_{2} +i_{2}-1)} )
  \cdots  P_{i_{\ell}}(\theta_{\sigma(n-i_{\ell} +1)},
 \cdots, \theta_{\sigma(n)} ) \;, \nonumber
\eeqn
where ${\cal P}_n$ represents   the permutations of $(1,2,\cdots,n)$. 
 To be more specific,  for example
\beqn
\label{eq:SP}
\left[{\cal S}P_{123}P_{234} \right]  (\theta_{1}, \theta_{2},
\theta_{3},\theta_{4})
  = \frac{1}{4!} \sum_{\sigma \in {\cal P}_{4}}
 P_{3}(\theta_{\sigma(1)}, \theta_{\sigma(2)},\theta_{\sigma(3)})
 P_{3}(\theta_{\sigma(2)}, \theta_{\sigma(3)},\theta_{\sigma(4)})
 \nonumber \\
\left[{\cal S}P_{1234}P_{34} \right] (\theta_{1}, \theta_{2},
\theta_{3},\theta_{4})
  = \frac{1}{4!} \sum_{\sigma \in {\cal P}_{4}}
 P_{4}(\theta_{\sigma(1)}, \theta_{\sigma(2)},\theta_{\sigma(3)}
 \theta_{\sigma(4)} )
 P_{2}(\theta_{\sigma(3)},\theta_{\sigma(4)}) \;.
\eeqn

 It is convenient to  represent $P_n(\theta_1,\cdots,\theta_n)$ by
 an n-vertex which connects n external legs. 
For example for $n=2$, 3 and 4
\beqn
  &&P_2(\theta_1,\theta_2) \equiv
  \epscenterboxy{2.5cm}{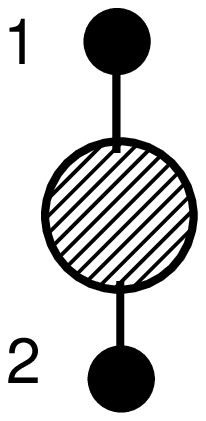} 
  \quad,\quad 
  P_3(\theta_1,\theta_2,\theta_3) \equiv
  \epscenterboxy{2.5cm}{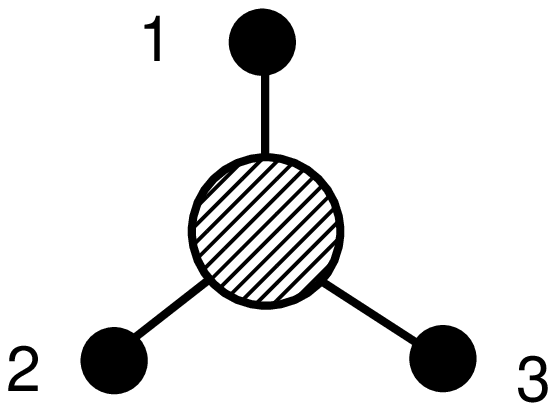}
  \nonumber \\
  &&P_4(\theta_1,\theta_2,\theta_3,\theta_4) \equiv
 \epscenterboxy{2.5cm}{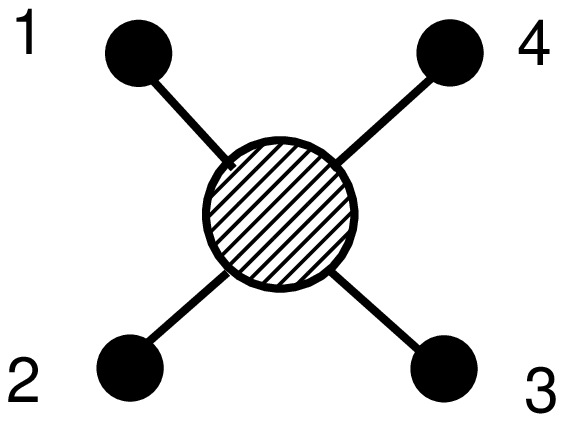}
\eeqn
The n-vertex  can be regarded as a dual graph of the n-gon (polygon)
 which corresponds to ${\cal D}_n$.
In terms of these vertices, let us express eq.~(\ref{eq:SP}) as follows.
\beqn
\left[{\cal S}P_{123}P_{234} \right] 
 (\theta_{1}, \theta_{2},\theta_{3},\theta_{4})
  &\equiv& \epscenterboxy{2.5cm}{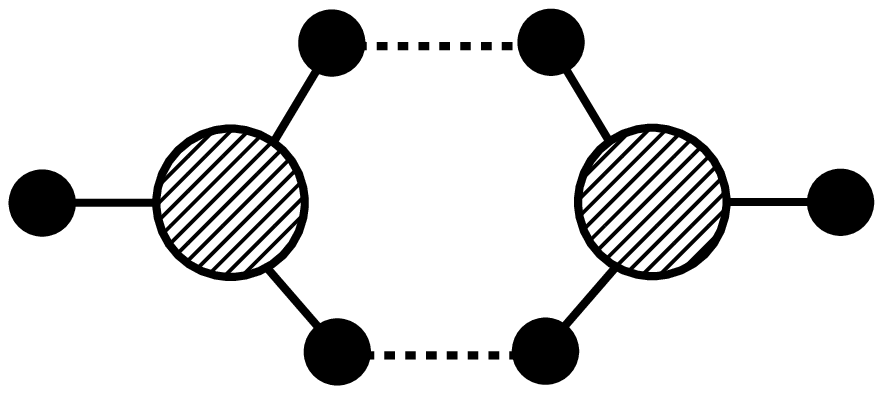} 
  \nonumber \\
 \left[{\cal S}P_{1234}P_{34} \right] (\theta_{1}, \theta_{2},\theta_{3},
 \theta_{4})  &\equiv&
 \epscenterboxy{2.5cm}{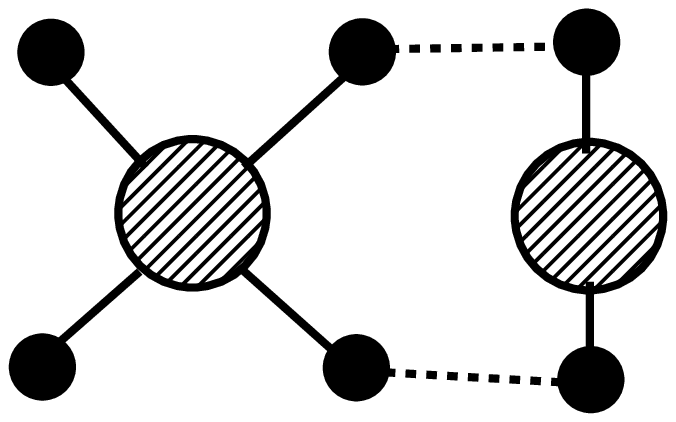}
\eeqn

The relation eq.~(\ref{eq:shrink-P}) can be represented, for example,  as 
\beq
  \epscenterboxy{2.5cm}{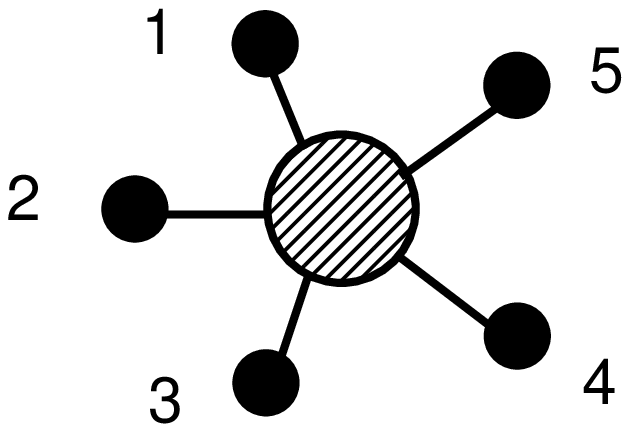}
  \qquad\to \quad\propto \qquad
  \epscenterboxy{2.5cm}{ai_fig/p1234.eps}
\eeq
 in the case of n=5.

Now we are concerned with the case of n=4 first. 
Let us recall that for n=3 
\beq
 R^{(3)}(\theta_1,\theta_2,\theta_3)
 \propto P_3(\theta_1,\theta_2,\theta_3)
\quad.\eeq
$R^{(4)}(\theta_{1}, \theta_{2}, \theta_{3}, \theta_{4})$ must
 include a
 term which becomes proportional to
 $P_3(\theta_{1}, \theta_{2}, \theta_{3})$
 in the limit $M\ell_4\to 0$,  which is
 $P_4(\theta_{1}, \theta_{2}, \theta_{3}, \theta_{4})$.
$R^{(4)}(\theta_{1}, \theta_{2},\theta_{3}, \theta_{4})$ may
 also include terms which vanish in this limit. Such terms must consist
 of the product of two  multi-vertices which have 6 external legs in total.

  By explicit computation, we find 
\beqn
  && \frac{-1}{2!} R^{(4)}(\theta_{1}, \theta_{2},
  \theta_{3}, \theta_{4}) \mid  =
  \frac{\partial}{\partial \Lambda} P_{4}
(\theta_{1}, \theta_{2},\theta_{3}, \theta_{4}) 
 -\left[ {\cal S} P_{123}P_{234} \right] (\theta_{1}, \theta_{2},\theta_{3},
 \theta_{4}) 
   \nonumber \\
 &&\qquad +
 \left[ {\cal S} P_{1234}P_{34} \right]  \left(\theta_{1}, \theta_{2},
 \theta_{3}, \theta_{4} \right)  
 \nonumber \\
 &&\qquad =
 \epscenterboxy{2.5cm}{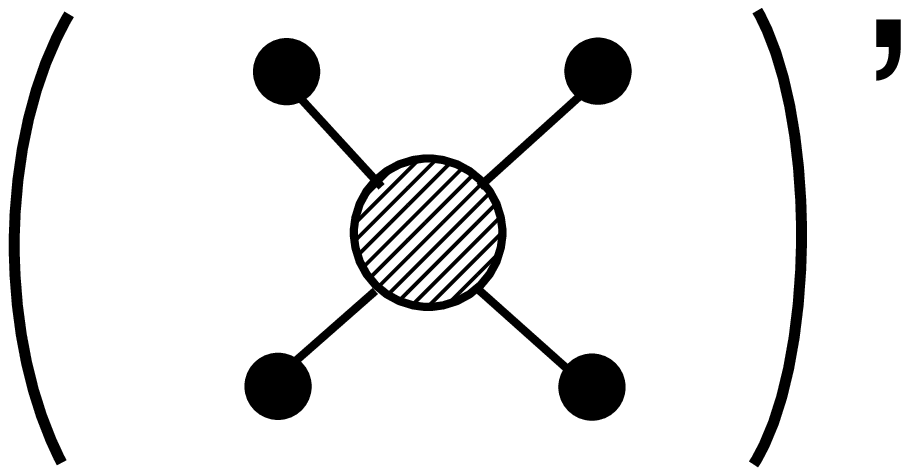}
 \nonumber \\
 &&\qquad +
 \epscenterboxy{2.5cm}{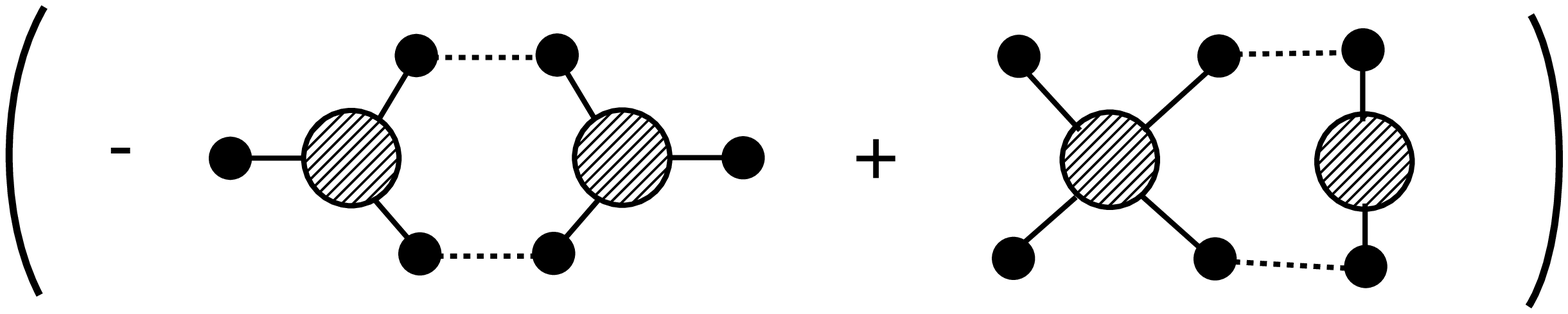}
 \quad ,
\label{eq:n=4Delta}
\eeqn
where the prime represents the differentiation with respect to $\Lambda$.


 By performing the inverse-Laplace transformation and renormalizing,
 we obtain  the complete answer for the macroscopic four loop
 correlator:
\footnote{  This was briefly reported in \cite{talk}.}
\beq
\label{eq:n=4answer}
 \VEV{w^+(\l_1)\cdots w^+(\l_4)}
 ={\cal A}_{4}^{fusion}( \ell_{1}, \cdots \ell_{4})
 +{\cal A}_{4}^{residual}( \ell_{1}, \cdots \ell_{4})
 \;\;,
\eeq
where
\beqn
 \label{eq:n=4fusionan}
 &&
 {\cal A}_{4}^{fusion}( \ell_{1}, \cdots, \ell_{4})
 \nonumber \\
 &&\;\;
 =- \frac{1}{m(m+1)} \prod_{j=1}^{4}\l_j \; 
  \frac{\p}{\p \mu}
  \left[ \Bigl(\frac M2\Bigr)^{2(1- \frac 1m)}
  \sum_{ {\cal D}_{4}}  \prod_{j=1}^{4}
  \tK{1-\frac{k_j}{m}}( M \l_{j})
  \right]
\eeqn
and
\beqn
\label{eq:n=4residual}
 \lefteqn{
 {\cal A}_{4}^{residual}( \ell_{1} \cdots \ell_{4})
 = \Bigl[\frac{1}{m(m+1)}\Bigr]^{2}
 \prod_{j=1}^{4}\l_j \;
 \Bigl(\frac M2 \Bigr)^{2(1-\frac 1m )}
 }
 \nonumber \\
 &&\left( \frac{1}{4!} 
 \sum_{\sigma \in {\cal P}_{4}} \right)
 \left\{
 - \sum_{ {\cal D}_{3}}  \sum_{ {\cal D}_{3}^{\prime} }
 \prod_{j=1}^{4} \left[ B_{j}^{123} \odot  B_{j}^{\prime 234}\right]
 \left( M \ell_{\sigma(j)} \right)
 \right.
  \nonumber \\
 &&\qquad\qquad\quad +
 \left. \sum_{ {\cal D}_{4}}  \sum_{ {\cal D}_{2}^{\prime} }
 \prod_{j=1}^{4} \left[ B_{j}^{1234} \odot  B_{j}^{\prime 34}\right]
 \left( M \ell_{\sigma(j)} \right)
 \;\right\}
 \;.
\eeqn
  Here 
\beqn
 B_{j}^{123} &=&
 (\tK{1-k_1/m}, \tK{1-k_2/m},\tK{1-k_3/m},\delta),
 \\
 B_{j}^{\prime 234} &=& 
 (\delta,\tK{1-k'_2/m},\tK{1-k'_3/m},\tK{1-k'_4/m}),
 \\
 B_{j}^{1234}&=&
 (\tK{1-k_1/m},\tK{1-k_2/m},\tK{1-k_3/m},\tK{1-k_4/m}),
 \\
 B_{j}^{\prime 34} &=&
 (\delta,\delta,\tK{1-k'_3/m},\tK{1-k'_4/m})
 \;\;.
\eeqn
   We have  introduced  
 ${\cal D}_{3} \equiv {\cal D}_{3}(k_1-1,k_2-1,k_3-1)$,
 $ {\cal D}_{3}^{\prime} 
 \equiv {\cal D}_{3}(k'_1-1,k'_2-1,k'_3-1) $,
  $ {\cal D}_{4}
\equiv  {\cal D}_{4}(k_1-1,k_2-1,k_3-1,k_4-1)$
 and
  $ {\cal D}_{2}^{\prime}  \equiv  {\cal D}_{2} 
(k'_1-1,k'_2-1 )$  and
 have defined  the convolution  $[A\odot B](M\ell)$
 with respect to $\l$ by
\beq
  \left[ A\odot B \right](M\l ')
 \equiv  \int_{0}^{\ell} \d\l '
 A(M\ell^{\prime}) B(M (\ell -\ell^{\prime}) )
 \;\;\; ,
\eeq
 in particular
\beq
  \left[ A\odot \delta \right](M\l)
 \equiv  \int_{0}^{\ell} \d\l
 A(M\ell^{\prime}) \delta(\l-\l^{\prime})=A(M\l)
 \;\;\;.
\eeq

 The important point to note is that
 ${\cal A}_4^{residual}$ seems to represent 
 the contribution from loops with mixed momenta, that is,
 a single loop seems to have two distinct parts
 each of which have different momentum.
   If two of the loops $\l_1$ and $\l_2$ touch
 each other on two point,
 the two-dimensional surface break into two surfaces and 
 the loops $\l_1$ and $\l_2$ also split into two
 pieces respectively. We infer that this configuration
 of two-dimensional surface may have connection with 
 ${\cal A}_4^{residual}$. 
   For $m=2$ (pure gravity), ${\cal A}_4^{residual}$
 vanishes. This fact is consistent with the above consideration
 because in pure gravity there is no loop with mixed boundary
 condition.

\subsection{Five-loop correlators}
\label{sec:five-loop}

Let us now turn to the n=5 case.
$R^{(5)}(\theta_{1},\cdots,\theta_{5})$ include a term
 which is proportional to
\beq
\label{eq:n=5type1}
  \frac{\partial^2} {\partial \Lambda^2}P_5(\theta_{1},\cdots,\theta_{5})
  =\epscenterboxy{2.5cm}{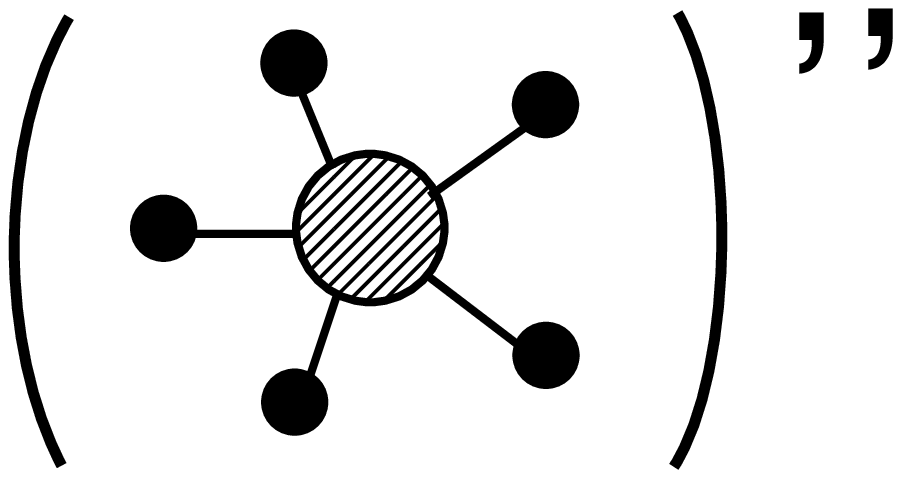}
\eeq
corresponding to the first term in eq.~(\ref{eq:n=4Delta}).
Corresponding to the second term in eq.~(\ref{eq:n=4Delta}),
 $R^{(5)}(\theta_{1},\cdots,\theta_{5})$ must include a term
 which  consists of the product of two multi-vertices with 7 external
 legs in total.
The possible form is 
\beqn
\label{eq:n=5type2}
 &&a \Bigl[{\cal S}\frac{\partial}{\partial \Lambda}(P_{12345})P_{45} \Bigr]
 +b \Bigl[{\cal S}P_{12345}\frac{\partial}{\partial \Lambda}(P_{45}) \Bigr]
 \nonumber \\
 &&\qquad+c \Bigl[{\cal S}\frac{\partial}{\partial \Lambda}
(P_{1234})P_{345} \Bigr]
 +d \Bigl[{\cal S}P_{1234}\frac{\partial}{\partial \Lambda}(P_{345}) \Bigr]
 \nonumber \\
 && =
 a \epscenterboxy{2.5cm}{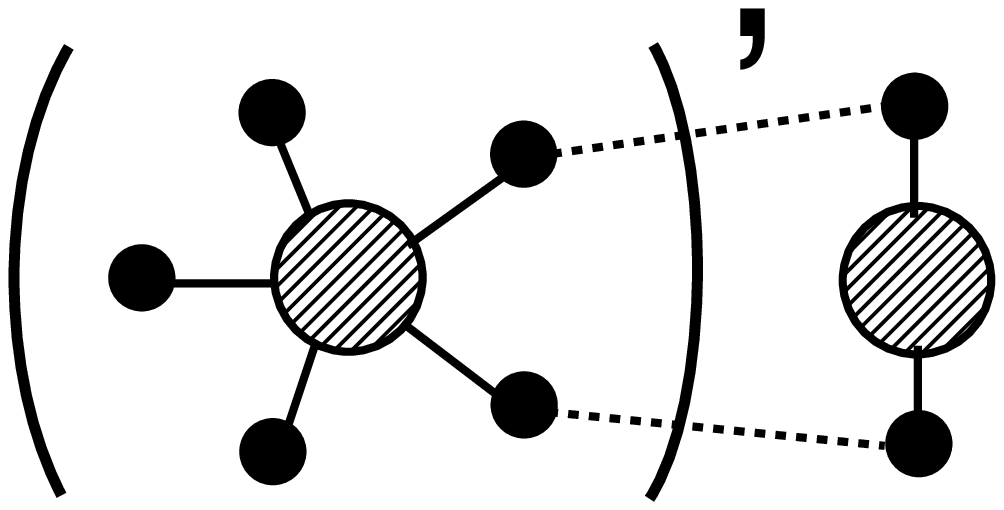} 
 +b \epscenterboxy{2.5cm}{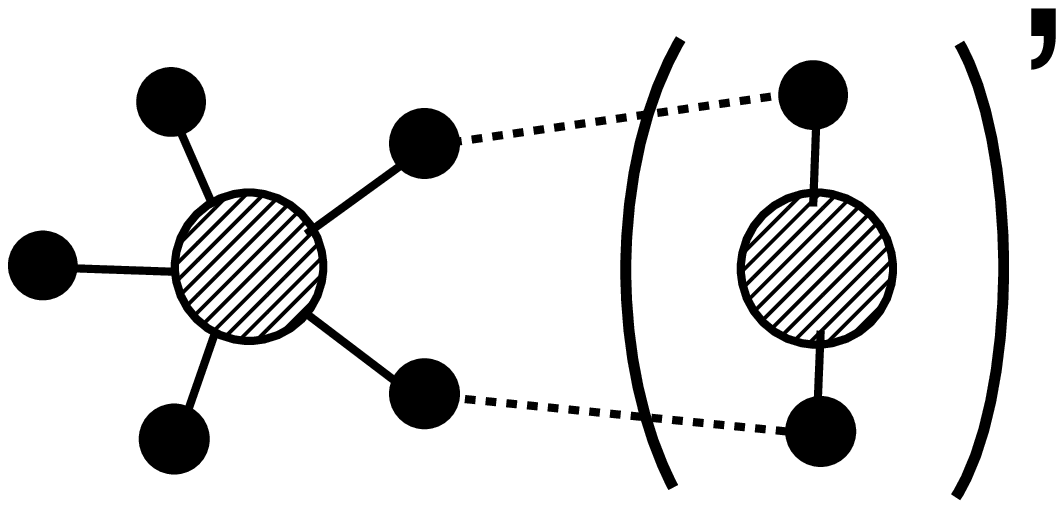} 
 \nonumber \\ 
 &&\qquad + c \epscenterboxy{2.5cm}{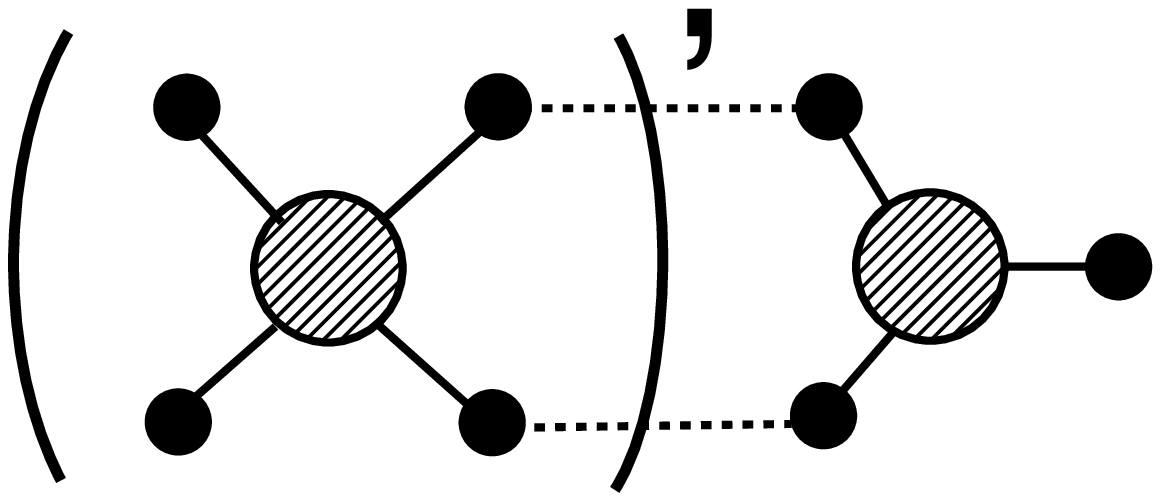}
 +d \epscenterboxy{2.5cm}{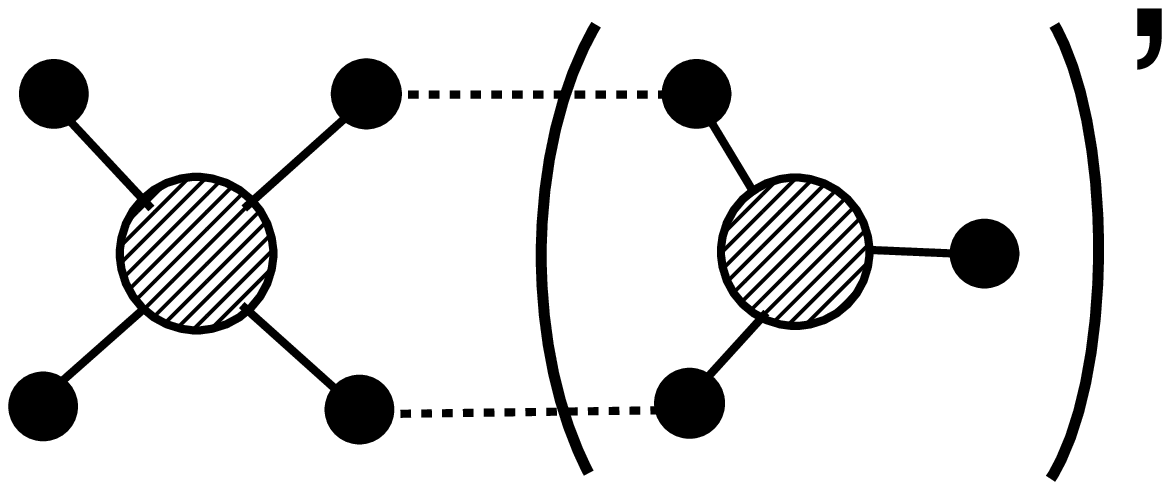} 
.
\eeqn
In the limit  $M\ell_5\to 0$, it becomes
\beqn
 (3a+c) \Bigl[{\cal S}\frac{\partial}{\partial \Lambda}(P_{1234})P_{34} \Bigr]
 +(3b+d) \Bigl[{\cal S}P_{1234}\frac{\partial}{\partial \Lambda}(P_{34}) \Bigr]
 \nonumber \\
 +2c \Bigl[{\cal S}\frac{\partial}{\partial \Lambda}(P_{123})P_{234} \Bigr]
 +2d \Bigl[{\cal S}P_{123}\frac{\partial}{\partial \Lambda}(P_{234}) \Bigr]
\quad.\eeqn
  We  require  this expression to be proportional to the $\Lambda$-derivative
 of  the second term
 in eq~.(\ref{eq:n=4Delta}).  We find  
\beq
a=(1-c)/3,\quad b=(2+c)/3,\quad d=-1-c
\quad .
\eeq
$R^{(5)}(\theta_{1},\cdots,\theta_{5})$ may include terms
 which vanish in the limit under consideration as well.
 They must  consist of the products of three multi-vertices with 9 external
 legs in total.
As one of the such terms we have 
\beqn
\label{eq:n=5type3}
 && [{\cal S}P_{1234}P_{34}P_{45}] -2[{\cal S}P_{1234}P_{23}P_{345}]
 +[{\cal S}P_{123}P_{124}P_{235}]
 \nonumber \\
 &&=
 \epscenterboxy{2.5cm}{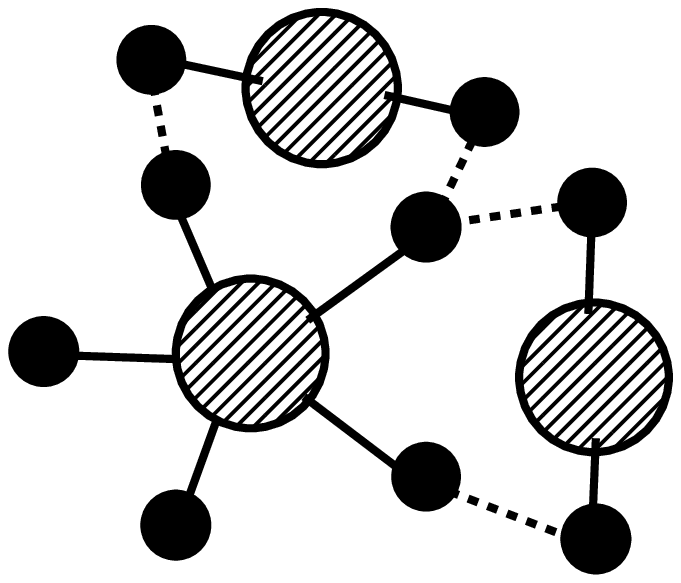} 
 -2 \epscenterboxy{2.5cm}{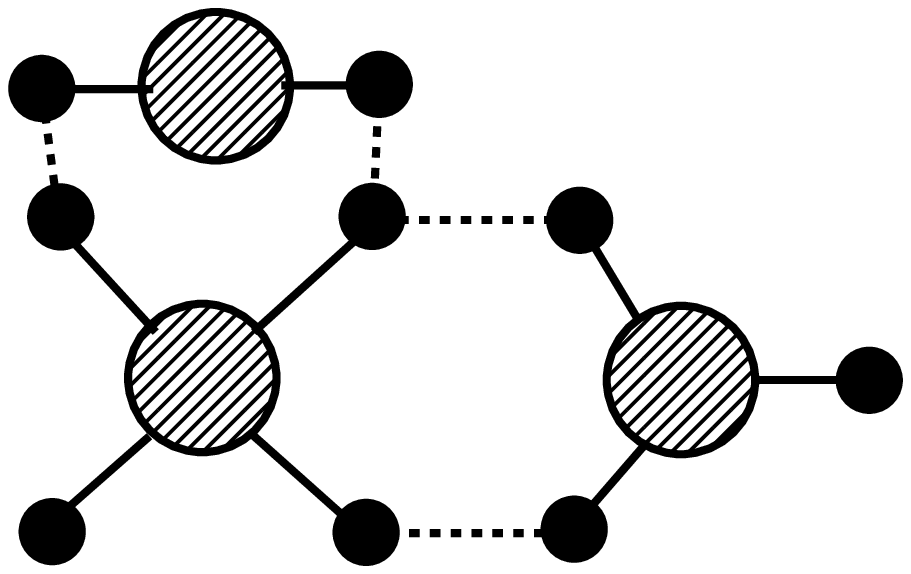} 
 \nonumber \\
 &&\qquad
 +\epscenterboxy{2.5cm}{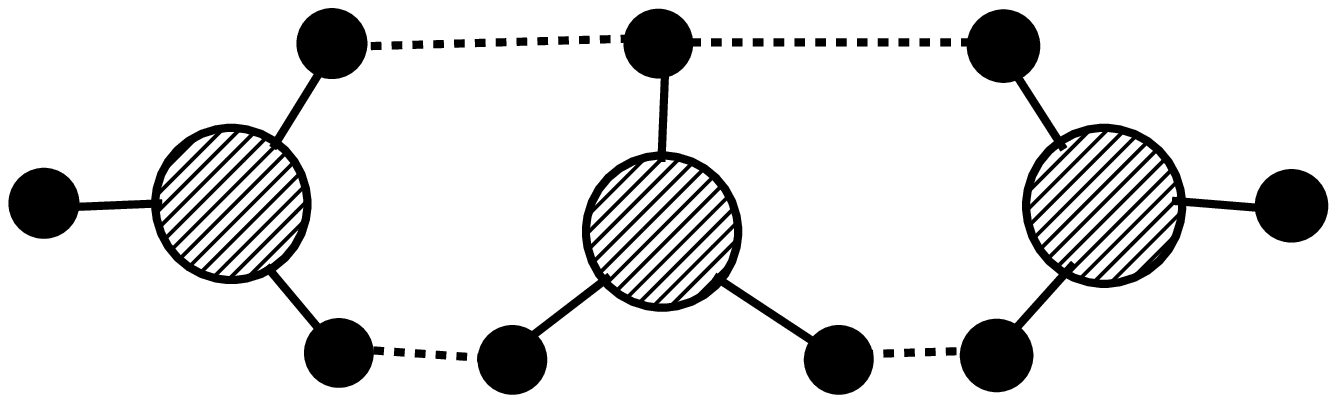} 
\quad .
\eeqn
( There are some other combinations which satisfy the conditions. )
$R^{(5)}(\theta_{1},\cdots,\theta_{5})$ must be expressed as a
 linear combination of the above three types of terms eq.~(\ref{eq:n=5type1}),
  eq.~(\ref{eq:n=5type2}) and eq.~(\ref{eq:n=5type3}) if
  the assumption under consideration is true.
By explicit calculation, we have found in fact that
 $R^{(5)}(\theta_{1},\cdots,\theta_{5})$ can be expressed
 as a linear combination of 
 eq.~(\ref{eq:n=5type1}),  eq.~(\ref{eq:n=5type2})
 and eq.~(\ref{eq:n=5type3}) :
\beqn
 \lefteqn{
 R^{(5)}(\theta_{1}, \theta_{2},  \theta_{3},
 \theta_{4}, \theta_{5}) 
 =
 \left( \frac{\partial}{\partial \Lambda} \right)^{2}
 P_{5}(\theta_{1}, \theta_{2},  \theta_{3}, \theta_{4}, \theta_{5})
  }
 \nonumber \\
 &+&\left[{\cal S}P_{12345} \left(
 2\frac{\stackrel{\rightarrow}{\partial} }
{\partial \Lambda}
 + 3\frac{\stackrel{\leftarrow}{\partial} }
{\partial \Lambda}
 \right) P_{45}  \right]
(\theta_{1}, \theta_{2},  \theta_{3}, \theta_{4}, \theta_{5})
 \nonumber \\
 &-& \left[{\cal S}  P_{1234} \left(
 \frac{ \stackrel{\rightarrow}{\partial} }{\partial \Lambda} +4
 \frac{ \stackrel{\leftarrow}{\partial} }{\partial \Lambda}
 \right) P_{345}  \right]
  (\theta_{1}, \theta_{2},  \theta_{3}, \theta_{4}, \theta_{5})  \nonumber \\
 &+& \left[ \left( 2P_{1234}P_{34}P_{45} -4P_{123}P_{23}P_{345}
 +2P_{123}P_{124}P_{235} \right) \right]
 (\theta_{1}, \theta_{2},  \theta_{3}, \theta_{4}, \theta_{5})\;.
\eeqn 

  Following the same procedure as obtaining eq.~(\ref{eq:n=4answer}),
  we find  the complete answer for
  the five loop amplitude:
\beqn
\label{eq:n=5answer}
 \VEV{w^+(\l_1)\cdots w^+(\l_5)}
 &=& {\cal A}_{5}^{fusion}(\l_{1} \cdots \l_{5})
 +{\cal A}_{5}^{residual-1}(\l_{1} \cdots \l_{5})
 \nonumber \\
 &+&{\cal A}_{5}^{residual-2}(\l_{1} \cdots \l_{5})
 \;\;,
\eeqn
where
\beqn
 \label{eq:n=5fusionan}
 \lefteqn{
 {\cal A}_{5}^{fusion}( \ell_{1}, \cdots, \ell_{5})
  }\nonumber \\
 &&
 = - \frac{1}{m(m+1)} \prod_{j=1}^{5}\l_j \; 
  \Bigl(\frac{\p}{\p \mu}\Bigr)^{2}
  \left[ \Bigl(\frac M2\Bigr)^{3(1 - \frac 1m)}
  \sum_{ {\cal D}_{5}}  \prod_{j=1}^{5}
  \tK{1-\frac{k_j}{m}}( M \l_{j})
  \right]
 \;,
\eeqn
\beqn
\label{eq:n=5residual-1an}
 &&{\cal A}_{5}^{residual-1}( \ell_{1} \cdots \ell_{5})
 \nonumber \\
 &&\;\;=
 \Bigl[\frac{1}{m(m+1)}\Bigr]^{2}
 \prod_{j=1}^{5}\l_j\;
 \left( \frac{1}{5!} \sum_{\sigma \in {\cal P}_{5}} \right)
 \nonumber \\
 &&\;\;\times
 \left\{
 \left\{ 
 \left(\frac{\p}{\p \mu} \Bigl(\frac M2\Bigr)^{3(1-\frac 1m)}
 \right)_{R}
 +3\left(\frac{\p}{\p \mu}\right)_{L}
 \Bigl(\frac M2\Bigr)^{3(1-\frac 1m)}
 \right\}
 \right. \nonumber \\
 &&\;\;\qquad
 \sum_{ {\cal D}_{5}}  \sum_{ {\cal D}_{2}^{\prime} }
 \prod_{j=1}^{5} \left[
 B_{j}^{12345} \odot  B_{j}^{\prime 45} \right]
 (M \ell_{\sigma(j)} )
 \nonumber \\
 &&\;\;
 - \left\{
 \Bigl(\frac M2\Bigr)^{2(1-\frac 1m)}
 \left(\frac{\p}{\p \mu} \Bigl(\frac M2\Bigr)^{1-\frac 1m}
 \right)_{R}
 +
 4 \left(\frac{\p}{\p \mu} \Bigl(\frac M2\Bigr)^{2(1-\frac 1m)}
 \right)_{L}
 \Bigl(\frac M2\Bigr)^{1-\frac 1m}
 \right\}
 \nonumber \\
 &&\;\;\qquad
 \left.
  \sum_{ {\cal D}_{4}}  \sum_{ {\cal D}_{3}^{\prime} }
 \prod_{j=1}^{5} \left[ B_{j}^{1234} \odot B_{j}^{\prime 345}
 \right](M \ell_{\sigma(j)} )
 \;\right\}
 \;\; 
\eeqn
\beqn
\label{eq:n=5residual-2an}
 &&{\cal A}_{5}^{residual-2}( \ell_{1} \cdots \ell_{5})
 \nonumber \\
 &&\;\;=
 \Bigl[\frac{1}{m(m+1)}\Bigr]^{3}
 \prod_{j=1}^{5}\l_j\;
 \left( \frac{1}{5!} \sum_{\sigma \in {\cal P}_{5}} \right)
 \Bigl(\frac M2 \Bigr)^{3(1-\frac 1m)}
 \nonumber \\
 &&\;\;
 \times
 \left\{
 -2\sum_{ {\cal D}_{5}}  \sum_{ {\cal D}_{2}^{\prime} }
 \sum_{ {\cal D}_{2}^{\prime \prime} }
 \prod_{j=1}^{5} \left[ B_{j}^{12345}
 \odot B_{j}^{\prime 34} \odot B_{j}^{\prime \prime 45} \right]
 (M \ell_{\sigma(j)} )  
 \right.
 \nonumber \\
 &&\;\;\quad
 +
 4\sum_{ {\cal D}_{4}}  \sum_{ {\cal D}_{2}^{\prime} }
 \sum_{ {\cal D}_{3}^{\prime \prime} }
 \prod_{j=1}^{5} \left[ B_{j}^{1234}
 \odot B_{j}^{\prime 23} \odot B_{j}^{\prime \prime 345} \right]
 (M \ell_{\sigma(j)} )
 \nonumber \\
 &&\;\;\quad
 -
 \left. 2\sum_{ {\cal D}_{3}}  \sum_{ {\cal D}_{3}^{\prime} }
 \sum_{ {\cal D}_{3}^{\prime \prime} }
 \prod_{j=1}^{5} \left[ B_{j}^{123}
 \odot B_{j}^{\prime 124} \odot B_{j}^{\prime \prime 235}
 \right] (M \ell_{\sigma(j)} )
 \right\}
 \;. 
\eeqn
  Here
 $\left( \frac{\p}{\p \mu} \right)_{L,(R)} $ means
  the derivative acting only on the left(right) part of the
 convolutions.  The rest of the notations here are similar to
   those of the $n=4$  case and will be self-explanatory.
 
  We conjecture that  $R^{(n)}$
 can be represented as a polynomial of 
$P_j,\;j\le n$ and
 their $\mu$ derivatives:  the  final answer
 would  then be obtained   by  convolutions  of various $B_{j}$'s
 and their derivatives.
If the conjecture is true in fact,  the power counting argument
  tells us that  the j-th term of $R^{(n)}$ turns out to be
  represented by a figure which
 consists of the products of j  multi-vertices with $n+2j$ external legs
 in total.
We  hope that, for higher loops, $R^{(n)}(\theta_{1},\cdots,
\theta_{n})$ can be
 put   in principle in  a form  such as eqs.~(\ref{eq:n=4answer}),
(\ref{eq:n=5answer}) in the same manner
  as we have determined  the four and five loops from the lower ones.
    From eqs.~(\ref{eq:n=4answer}) $\sim$ (\ref{eq:n=4fusionan})
 and eqs.~~(\ref{eq:n=5answer}) $\sim$ (\ref{eq:n=5fusionan}),
 we guess that the multi-loop correlators, in general,
 would include the following term corresponding 
 $P_n(\theta_{1},\cdots,\theta_{n})$ :
\beqn
 \label{eq:fusionan}
 \lefteqn{
 {\cal A}_{n}^{fusion}( \ell_{1}, \cdots, \ell_{n})
  }\nonumber \\
 &&
 = - \frac{1}{m(m+1)} \prod_{j=1}^{n}\l_j \; 
  \Bigl(\frac{\p}{\p \mu}\Bigr)^{n-3}
  \left[ \Bigl(\frac M2\Bigr)^{-(n-2)(1 - \frac 1m)}
  \sum_{ {\cal D}_{n}}  \prod_{j=1}^{n}
  \tK{1-\frac{k_j}{m}}( M \l_{j})
  \right]
 \;.
 \nonumber \\
\eeqn

\subsection{Four-point functions from loop correlators}

 The four-loop correlators we found in
 eqs.~(\ref{eq:n=4answer}) $\sim$ (\ref{eq:n=4residual})
 do not diverge  when the loop lengths approach  zero,
 so that  we expect that the loop operator can be
 replaced by the sum of the local operators in this case.
   Let us derive the four-point functions of the scaling
 operators,
 applying  the  expansion of loop to the four loop
 correlators.

   First, consider the part
 ${\cal A}_4^{fusion}(\l_1,\l_2,\l_3,\l_4)$ and 
 expand this in terms of the modified Bessel functions
 $I_\nu(\l_i)$.
   From eq.~(\ref{eq:n=4fusionan}) we have
\beqn
\lefteqn{
 {\cal A}_4^{fusion}(\l_1,\l_2,\l_3,\l_4) }
 \nonumber \\
 &&= 
 \frac 1{m^2(m+1)} \left(\frac M2\right)^{-4-\frac 2m}
 \sum_{{\cal D}_4} \prod_{j=1}^{4}
 \frac {M\l_i}2 ~\tK{1-\frac{k_j}m}(M\l_j)
 \nonumber \\
 &&\quad
 -\frac2{m(m+1)^2} \left(\frac M2\right)^{-2-\frac 2m}
 \frac 1M \frac{\p}{\p M}\left\{\;
 \sum_{{\cal D}_4} \prod_{j=1}^{4}
 \frac {M\l_i}2 ~\tK{1-\frac{k_j}m}(M\l_j)
 \;\right\}
\;.
\eeqn
  Since we can prove the relations
\beq
\label{eq:zK(z)-exp}
 \frac z2 \tK{1-p}(z)=\sum_{n=-\infty}^{\infty}
 (p+2n)~I_{|p+2n|}(z)
\;,
\eeq
and 
\beq
 z\frac{\p}{\p z}\left\{
 \frac z2 \tK{1-p}(z)\right\}
 =\sum_{n=-\infty}^{\infty}
 (p+2n)\Bigl(p+2n(p+n)\Bigr) ~I_{|p+2n|}(z)
\;,
\eeq
 for $0<p<1$,
 we obtain the following expression for
 ${\cal A}_4^{fusion}$:
\beqn
 &&{\cal A}_4^{fusion}(\l_1,\l_2,\l_3,\l_4) 
 =\frac 1{m(m+1)^2}\left(\frac M2\right)^{-4-\frac 1m}
 \nonumber \\
&&\qquad\qquad\times
 \sum_{{\cal D}_4}\prod_{j=1}^{4}
 \left(\sum_{n_j=-\infty}^{\infty}\right)
 \left\{ (1+\frac 1m) -\frac 12\sum_{j=1}^{4}
 \left[\frac{k_j}m+2n_j(\frac{k_j}m+n_j)\right]
 \right\}
 \nonumber \\
&&\qquad\qquad\times
 \prod_{j=1}^{4}\left\{(\frac{k_j}m+2n_j)
 ~I_{|\frac{k_j}m+2n_j|}(M\l_j)\right\}
\;.
\label{eq:n=4fusion-expansion}
\eeqn

 Comparing \eq{eq:n=4fusion-expansion} to the expansion
 of loop operators (\ref{eq:w+--exp}), we obtain
 the following
 contribution of the four-point functions from
 ${\cal A}_4^{fusion}(\l_1,\cdots,\l_4)$:
\beqn
 \VEV{ \prod_{j=1}^{4}\hsig{|k_i+2mn_i|} }^{fusion}
 &=&
 C_{k_1k_2k_3k_4}\frac 1{2m^2(m+1)^2}
 \left(\frac M2\right)^{-4-\frac 1m
 +\sum_{i=1}^{4}|\frac{k_i}{m}+2n_i|}
 \nonumber \\
&\times&
 \left\{ 2(m+1)-\sum_{j=1}^{4}
 \Bigl[k_j+2n_j(k_j+n_jm)\Bigr] \right\}
 \nonumber \\
&\times&
 \prod_{j=1}^{4}(k_j+2n_j m)
\;,
\eeqn
 where
\beq
 C_{k_1k_2k_3k_4}=\sum_{k'=1}^{m-1}
 C_{k_1k_2k'}C_{k'k_3k_4}
\;.
\eeq

  Next, let us consider
 ${\cal A}_4^{residual}(\l_1,\cdots,\l_4)$ part.
   We can prove that the convolution of two modified
 Bessel functions $\tK{p}(M\l)$ and $\tK{p'}(M\l)$ is
 expanded in terms of $I_{\nu}(M\l)$ as
\beqn
\label{eq:conv-exp}
 \l~\left[ \tK{p}\odot\tK{p'}\right](M\l)
 &=&\left(\frac 2M\right)^2
 \sum_{n=-\infty}^{\infty}
 \left\{ n(2n+p+p')~I_{|2n+p+p'|}(M\l) \right. 
 \nonumber \\
&&\qquad
 - \left. n(2n+p-p')~I_{|2n+p-p'|}(M\l) \right\}
\;.
\eeqn
 The above relation is easily derived in the space
 of Laplace transformed coordinate using the relations
\beq
 {\cal L}\left[~I_p(M\l)\right] 
 =\frac 1M \frac {e^{-pm\theta}}{\sinh m\theta},
\;\; {\rm for}\; p>-1
\;,
\eeq
and
\beqn
 {\cal L}\left[~\tK{p}(M\l)\right] 
 &=&{\cal L}\left[~I_{-p}(M\l)-I_{p}(M\l)\right]
 \nonumber \\
 &=&\frac 2M \frac{\sinh pm\theta}{\sinh m\theta},
\;\; {\rm for}\; p>-1
\;,
\eeqn
 because the convolution in $\l$-space corresponds
 to the product in $\zeta$-space.

   From eqs. (\ref{eq:zK(z)-exp}) and (\ref{eq:conv-exp})
 we obtain the following expression for the residual part
 of the four loop correlator:
\beqn
\lefteqn{
 {\cal A}_4^{residual}(\l_1,\cdots,\l_4)
 =\frac 1{m^2(m+1)^2}\left(\frac M2\right)
 ^{-4-\frac 2m}
 \frac{1}{4!}\sum_{\sigma\in{\cal P}_4}
 }\nonumber \\
&&\times
 \prod_{j=1}^4
 \left\{\sum_{k_j}\sum_{n_j=-\infty}^{\infty}\right\}
 \sum_{k'_1}\sum_{k'_2}
 (-C_{k_1k_2k_3}C_{k'_1k'_2k_4}
 +C_{k_1k_2k_3k_4}C_{k'_1k'_2})
 \nonumber \\
&&\times
 \sum_{+-}(\pm)n_1\Bigl(2n_1+(1-\frac{k_1}m)
 \pm(1-\frac{k'_1}m)\Bigr)
 ~I_{|(2n_1+(1-\frac{k_1}m) \pm(1-\frac{k'_1}m)|}
 (M\l_{\sigma(1)})
 \nonumber \\
&&\times
 \sum_{+-}(\pm)n_2\Bigl(2n_2+(1-\frac{k_2}m)
 \pm(1-\frac{k'_2}m)\Bigr)
 ~I_{|(2n_2+(1-\frac{k_2}m) \pm(1-\frac{k'_2}m)|}
 (M\l_{\sigma(2)})
 \nonumber \\
&&\times
 (2n_3+\frac{k_3}m)~I_{|(2n_3+\frac{k_3}m)|}
 (M\l_{\sigma(3)})~
 (2n_4+\frac{k_4}m)~I_{|(2n_4+\frac{k_4}m)|}
 (M\l_{\sigma(4)})
\;,
\label{eq:n=4res-exp}
\eeqn
 where $C_{k k'}=\delta_{k k'}$.

   From the above expression and the expansion of the
 loop operator (\ref{eq:w+--exp}), we can obtain
 the contribution of the four-point functions
 from ${\cal A}_4^{residual}$ part.
  The explicit expression, however, would be complicated.

 Let us comment on the role of the coefficients
 $(-C_{k_1k_2k_3}C_{k'_1k'_2k_4}
 +C_{k_1k_2k_3k_4}C_{k'_1k'_2})$ in \eq{eq:n=4res-exp}.
 At first sight, it appears that we would have $I_{\nu}(M\l_i)$
 with integer order in \eq{eq:n=4res-exp}.
 For example, in the case of 
 $k_1=k'_1$ or $k_1+k'_1=m$, we have $I_{\nu}(M\l_1)$
 with integer order.
  Terms including $I_{\nu}(M\l_i)$ with integer order as
 a factor cannot be explained from the viewpoint of
 the local operators.
 These terms, however, are cancelled
 due to the coefficients
 $(-C_{k_1k_2k_3}C_{k'_1k'_2k_4}
 +C_{k_1k_2k_3k_4}C_{k'_1k'_2})$,
 so that we do not have 
 $I_{\nu}(M\l_i)$
 with integer order in \eq{eq:n=4res-exp} after all.

\newpage
\begin{center}
\section{Summary and discussion}
\label{summary}
\end{center}

   In this paper we have investigated the correlators of
 macroscopic loops and those of local operators in the
 unitary minimal models coupled to
 two-dimensional gravity using
 the two-matrix model.
   We calculated the general multi-resolvent correlators,
 and examined one- to five-loop correlators explicitly.

  From these loop correlators we obtained the correlator of
 the scaling operators by applying the idea \cite{MSS} 
 that the macroscopic 
 loop can be replaced by a sum of local operators, to
 the case of the two-matrix model.
   We found that there exist the fusion rules for
 the three-loop correlators,
 which are similar to
 those for the three-point functions of the gravitational
 primaries. 
  From the three-loop correlators, we deduced the three-point
 functions of the scaling operators, and found
 the gravitational descendants as well as the gravitational
 primaries satisfy the fusion rules of the same kind.
  These fusion rules for the loops can be considered to
 express 
 those for all of the scaling operators in a compact form.

  At the $(m+1,m)$ critical point in two-matrix models,
 the scaling operators
 $\hsig{j},\;j=0\;(\rm mod\; m+1)$ have no counterparts
 in the BRST cohomology of Liouville theory coupled
 to the corresponding conformal matter.
   In \cite{MMS},these operators were argued to be
 boundary operators
 which couple to loops in the case of the one-matrix
 model. It was also shown explicitly that one of them,
 corresponding to $\hsig{m+1}$ in the case of 
 the unitary matter, is a 
 operator which measures the total length of the loops.

    We examined the role of the rest of these operators.
 We showed, in some examples,
 that the operator ${\cal B}_n$ couples to
 the points to which n parts of several loops are stuck
 each other. In other words, the operator ${\cal B}_n$
 connects n parts of loops together.
  We think these operators
 play an important role concerning the touching of the
 macroscopic loops. The emergence of these operators
 in matrix models can easily be understood from the viewpoint
 of macroscopic loops and their expansion in terms of local
 operators.

  In sect.~\ref{sec:multi-loop} we examined the property
 of the multi-loop correlators.
  We pointed out that the structure similar to those
 of the crossing symmetry in the underlying
 conformal field theory can 
 be seen in the loop correlators.
  This structure appears in the selection rules for
 the summations in the expression of loop
 correlators.

   We calculated explicitly four- and five-loop correlators.
 From the expression of these correlators,
 we inferred that  these include the
 contribution from the loops with boundary condition specified
 by more than one momentum.
  We guess this property can be understood as follows. 
 When two loops touch each other on two points, 
 each loop breaks into two pieces.
   Since a single loop breaks into two pieces,
 the broken pieces can have distinct momenta.
 The configuration
 probably have non-vanishing contribution
 to the amplitude in the case of the four- and five-loop
 amplitudes.
  Note that matter degrees of freedom are fixed only
 inside the loops when we calculate the loop
 correlators.
 Each loop therefore represents a superposition of loops with
 various momenta. Distinct lattice elements on a single
 loop can have distinct momenta.
\newpage
\begin{center}
\section{Acknowledgements}
\end{center}

   I wish to express my gratitude to Prof. H. Itoyama and Dr. A. Ishikawa
 for
 enjoyable collaboration in \cite{AII,AII2,AI} and helpful discussions. 
   I am grateful to Prof. K. Kikkawa for his encouragement and 
 useful discussions.
   I would like to thank Prof. N. Sakai, Prof. K. Hamada
 and Dr. H. Shirokura
 for valuable comments on the subject
   and  also thank Prof. A. Sato, Prof. H. Kunitomo and other
 members of the theoretical physics laboratories
 of Osaka University for their encouragement.

  This work is supported in part by
 Grant-in-Aid for  Scientific Research (07640403)
 and by the Grand-in-Aid for Scientific Research Fund (2690)
 from the Ministry of Education, Science and Culture, Japan.
\newpage
\appendix

\begin{center}
\section{ }
\end{center}

  Here we collect some formulae concerning the
 modified Bessel functions.
 $I_{\nu}(z)$ and $K_{\nu}(z)$ are linearly independent
 solutions of the Bessel equation
\beq
 \left[ \Bigl(z\frac{\p}{\p z} \Bigr)^2 -z^2 -\nu^2
 \right] Z_{\nu}(z)=0
\;.
\eeq
 $I_{\mu}(z)$ can be expanded as
\beq
 I_{\nu}(z)=\Bigl(\frac z2\Bigr)^{\nu}
 \sum_{n=0}^{\infty}\frac{1}{k!\Gamma(n+\nu+1)}
 \Bigl(\frac z2\Bigr)^{2n}
\;,
\eeq
 and $K_{\nu}(z)$ is defined as
\beq
 K_{\nu}(z)=\frac{\pi/2}{\sin \nu\pi}
 \left[I_{-\nu}(z)-I_{\nu}(z)\right]
\;.
\eeq
 We collect another useful formulae in the following.
\beq
-\frac{2\nu}z K_{\nu}(z)=K_{\nu-1}(z)-K_{\nu+1}(z)
\;
\eeq

\beq
 \Bigl(\frac{d}{zdz}\Bigr)^n[ z^{\nu} K_{\nu}(z)]
 =(-1)^n z^{\nu-n} K_{\nu-n}(z)
\;
\eeq

\beqn
 &&K_{\nu}(z) K_{\nu}(\zeta)
 =\frac 12 \int_{0}^{\infty}
 \frac{dt}{t} K_{\nu}\Bigl(\frac{z\zeta}{t}\Bigr)
 \exp\Bigl(-\frac t2 -\frac{z^2+\zeta^2}{2t}\Bigr)
 \nonumber \\
 &&\qquad [ Re~z>0,\;\;Re~\zeta>0,\;\;
 |{\rm arg}(z+\zeta)|<\pi/4]
\;
\eeqn

 Introducing  the Laplace transformation of
 a function $f(\l)$ of loop length $\l$ by
\beq
 {\cal L}\Bigl[ f(\l)\Bigr]
 =\int_{0}^{\infty}\d l ~e^{-\zeta \l} f(x)
\;,
\eeq
 we have the following relations on the Laplace
 transformations of the Bessel functions:
\beqn
 {\cal L}\Bigl[~I_p(M\l)\Bigr]
 &=&\frac 1M \frac{e^{-pm\theta}}{\sinh m\theta}
 ,\;\;p>-1,
\\
 {\cal L}\Bigl[\tK{p}(M\l)\Bigr]
 &=&\frac 2M \frac{\sinh |p|m\theta}{\sinh m\theta}
 \;,
\eeqn 
\beq
 {\cal L}\Bigl[-\l^{-1}|p|\tK{p}(M\l)\Bigr]
 = 2\cosh pm\theta
,
\eeq
 where $\zeta$ is parametrized as
\beq
 \zeta=M\cosh m\theta
\;,
\eeq
 and we introduced a notation
\beq
 \tK{p}(M\l)=\frac{\sin\pi|p|}{\pi/2}
 K_{p}(M\l)
\;.
\eeq

\begin{center}
\section{}
\label{app:sub-culc. in n-resolvent}
\end{center}

  In this  appendix, we  prove the recursion relations for
\beqn
  \left( \begin{array}{clcr}
 m_{1}, & m_{2}, & m_{3}, & \cdots \\
 i_{1}, & i_{2}, & i_{3}, &  \cdots
 \end{array} \right)_{n}   \;\;\;\;
  \;\; {\rm with}\;\;\;\; \sum_{\ell} m_{\ell}  \leq  n-2 \;\;\;
\eeqn
 introduced in the text.  The proof goes by  mathematical
 inductions.

   We will first prove  the simplest case
\beqn
\label{eq:simpler1}
 \left( \begin{array}{cccc}
 m, & 0, & \cdots, & 0 \\
 i_{1},  & i_{2}   & \cdots, & i_{n} 
 \end{array} \right)_{n}   =  
 \frac{1}{ {\displaystyle \prod_{j (\neq i_{1})}^{n} [i_1-j]^{2} } } 
 \quad,\qquad {\rm for}\quad m=n-2 
\eeqn
and
\beqn
\label{eq:simpler2}
 \left( \begin{array}{cccc}
 m, & 0, & \cdots, & 0 \\
 i_{1},  & i_{2}   & \cdots, & i_{n} 
 \end{array} \right)_{n}   
 =   0
 \quad,\qquad {\rm for}\quad m\le n-3 \quad. 
\eeqn

  Assume that eq.~(\ref{eq:simpler1})  and eq.~(\ref{eq:simpler2}) are true at $n$.
  Without loss of generality, $i_{1}$ can be taken to be $1$.
Let us consider the left hand side of eq.~(\ref{eq:simpler1}) or eq.~(\ref{eq:simpler2})
 in which $n$ is replaced by $n+1$.
To compute  them we  observe that   the elements of ${\cal S}_{n+1}$  are generated
  by associating 
  $n$  different ways of
 inserting  $[n+1]$    with each element $\sigma \in {\cal S}_{n}$.
  In the case where $[n+1]$ is inserted in between $[1]$ and $[\sigma(1)]$,
   this contribution  is equal to
\beqn
\label{eq:cont1}
  - \sum_{\sigma \in {\cal S}_{n} }
   \frac1{[1-\sigma(1)]^m} \; \frac{[1- \sigma(1)]^{m}}{ [1- (n+1)]^{m+1}
 [(n+1)- \sigma(1)] } \prod_{j=2}^{n}
 \frac1{[j - \sigma(j) ]} \;\;\;.
\eeqn
  The contributions   from  the sum of the remaining $n-1$  insertions
  are  found to be  equal to
\beqn
\label{eq:cont2}
  -  \sum_{\sigma \in {\cal S}_{n} }
  \frac1{[1-\sigma(1)]^m} \; \frac{1}{ [\sigma(1)- (n+1)]
 [1 -(n+1)] } \prod_{j=2}^{n} \frac1{[j - \sigma(j) ]} \;\;\;.
\eeqn
  Here we have used
\beqn
\label{eq:frac}
\frac{1}{[j-(n+1)] [(n+1)-m]}  =  \frac{1}{[j-m]} \left(  \frac{1}{[j-(n+1)]} -
 \frac{1}{[m-(n+1)]} \right)\;\;\;.
\eeqn

 Note also that $ {\displaystyle\prod_{j=1}^{n-1} }
 1/[\sigma^{j}(1) -\sigma^{j+1}(1) ]
 = {\displaystyle \prod_{j=2}^{n} } 1/[j - \sigma(j) ]$.
  Putting eqs.~(\ref{eq:cont1})  and (\ref{eq:cont2}) together,
  we find
\beqn
\label{eq:**}
  &~&  \left( \begin{array}{cccc}
 m, & 0, & \cdots, & 0 \\
 1, & 2, & \cdots, & n+1
 \end{array} \right)_{n+1} 
  =  \nonumber \\
   ~&-&  \sum_{\sigma \in {\cal S}_{n} }
  \frac1{[1-\sigma(1)]^m} \; 
 \frac{  \left \{ 1-  \left( \frac{[1- \sigma(1)]}{[1-(n+1)]} \right)^{m}
 \right \}  } { [\sigma(1)- (n+1)] [1 -(n+1)] }
 \prod_{j=2}^{n} \frac1{[j - \sigma (j) ]} \;\;\;.
\eeqn
  Factorizing the expression inside the bracket
 $\left \{ ~~ \cdots ~~  \right \}$, we have
\beq
\label{eq:recursion1}
 \left( \begin{array}{cccc}
 m, & 0, & \cdots, & 0 \\
 1, & 2, & \cdots, & n+1
 \end{array} \right)_{n+1} 
  =  \nonumber \\
 \sum_{l=1}^{m-1}\left( \begin{array}{cccc}
 m-l, & 0, & \cdots, & 0 \\
 1,   & 2, & \cdots, & n
 \end{array} \right)_{n}  \frac{1}{[1-(n+1)]^{1+l} } \quad.
\eeq
Then from the assumption, eq.~(\ref{eq:simpler1}) and eq.~(\ref{eq:simpler2}) are also satisfied when $n$ is replaced by $n+1$.
On the other hand for $n=3$ eq.~(\ref{eq:simpler1}) and eq.~(\ref{eq:simpler2}) are
 clearly true, so we have proven the relations.

  Now we turn to the more general case the proof of which is a straightforward
 generalization of the one given above.
  To derive the relations
\beqn
 \label{eq:relation3}
 \left( \begin{array}{cccccc}
 m_{1}, & \cdots, & m_{k}, & 0,       & \cdots, & 0 \\
 i_{1}, & \cdots  & i_{k}, & i_{k+1}, & \cdots, & i_{n}
 \end{array} \right)_{n}
 =0 \quad,\qquad {\rm for} \quad
    \sum_{\ell=1}^{k} m_{\ell}\le n-3 \quad,
\eeqn
and
\beqn
 &&\label{eq:relation4}
 \left( \begin{array}{cccccc}
 m_{1}, & \cdots, & m_{k}, & 0,          & \cdots, & 0 \\
 i_{1}, & \cdots  & i_{k}, & \ i_{k+1},  & \cdots, & i_{n}
 \end{array} \right)_{n} \nonumber \\
 &&\qquad =\sum_{j=1}^{k}
 \left( \begin{array}{cccccccc}
 m_{1}, & \cdots, & m_{j}-1, & \cdots, &m_{k}, & 0, & \cdots, & 0 \\
 i_{1}, & \cdots, & i_{j},   &\cdots,  &i_{k}, & i_{k+1}, & \cdots, & i_{n-1}
 \end{array} \right)_{n-1} \frac{1}{[j-n]^2}
 \quad, \nonumber \\
 &&\qquad\qquad\qquad\qquad\qquad\qquad\qquad\qquad\qquad
 {\rm for}\quad \sum_{\ell=1}^{k} m_{\ell}=n-2
 \quad.
\eeqn
Let us assume eq.~(\ref{eq:relation3}) at $n$ .

 We take  $ i_{\ell} = \ell,~~$ $ \ell =1 \sim k $
  without  loss of generality.  The way in which   the elements of
 ${\cal S}_{n+1}$ are generated is the same as the one given above.
  In the case where $[n+1]$ is inserted in between $[\ell]$ and
 $[\sigma \left( \ell \right)]$ $ \ell = 1 \sim k$,
  the contribution is 
\beqn
\label{eq:cont1g}
 -  \sum_{\sigma \in {\cal S}_{n} }
  \frac1{[1-\sigma(1)]^{m_1+1}} \cdots
  \frac1{[\ell-\sigma(\ell)]^{m_{\ell}}} \;
  \frac{[\ell- \sigma(\ell)]^{m_{\ell} } }
 { [\ell- (n+1)]^{m_{\ell}+1} [(n+1)- \sigma(\ell)] } \;\;\;  \nonumber \\
  \times  \frac1{ [ (\ell+1) -\sigma(\ell +1)]^{m_{\ell+1}+1}} 
 \cdots  \frac1{ [k-\sigma(k)]^{m_{k}+1}}
 \prod_{j ( \neq 1 , 2, \cdots k)}^{n-1}
 \frac1{[j - \sigma (j) ]} \;\;\;.
\eeqn
  The contributions   from  the sum of the remaining   insertions
  are 
\beqn
  &-&  \sum_{\sigma \in {\cal S}_{n} }
  \sum_{\ell ( \neq  p_{2},\cdots p_{k} )}^{n-1}
  \frac1{ [1-\sigma(1)]^{m_{1}+1 } } \cdots  \frac1{ [k-\sigma(k)]^{m_{k+1} }}
  \nonumber \\
 &\times&
   \prod_{j (\neq p_{2}, \cdots p_{k}) }^{n-1}
   \frac1{[\sigma^{j}(1) - \sigma^{j+1}(1) ]}
   \left(  \frac{1}{[ \sigma^{\ell}(1) -(n+1)]}
  - \frac{1}{[ \sigma^{\ell +1}(1) -(n+1)]} \right) \;.
 \nonumber \\
 &&
\eeqn
  Here  $p_{\ell}$  $\ell = 1 \sim k$ are such that
  $ \sigma^{p_{\ell}} (1) = \ell$.
  Using  eq. ~(\ref{eq:frac}) again,  we find that this equals
\beqn
\label{eq:cont2g}
  -  \sum_{\sigma \in {\cal S}_{n} } \sum_{\ell =1}^{k}
  \frac1{[ 1-\sigma(1)]^{m_{1}+1}}  \cdots
  \frac1{ [ \ell-\sigma(\ell)]^{m_{\ell}}}\;  \frac{1}{ [\sigma(\ell)- (n+1)]
  [\ell -(n+1)] }  \nonumber \\
  \frac1{[ (\ell +1) -\sigma(\ell + 1)]^{m_{\ell+1}+1}} 
  \cdots  \frac1{[ k -\sigma(k)]^{m_{k}+1}}
  \prod_{j (\neq 1, 2, \cdots k)}^{n-1}
  \frac1{[j - \sigma [j] ]} \;\;\;.
\eeqn
  Putting eqs.~(\ref{eq:cont1g})  and (\ref{eq:cont2g}) together,
  we find
\beqn
\label{eq:***}
 &~&  \left( \begin{array}{cccccc}
 m_{1}, & \cdots, & m_{k}, & 0,     & \cdots, & 0 \\
 1,     & \cdots  & k    , & k+1, & \cdots, & n+1
 \end{array} \right)_{n+1}    \;\;\;  \nonumber \\
 &=&   -  \sum_{\ell}^{k}  \sum_{\sigma \in {\cal S}_{n} }
 \frac1{[1-\sigma(1)]^{m_{1}+1}} \cdots 
 \frac1{[\ell-\sigma(\ell)]^{m_{\ell}}} \; 
 \frac{  \left \{ 1-  \left( \frac{[\ell - \sigma(\ell)]}{[\ell- (n+1)]}
 \right)^{m_{\ell}}
 \right \}  } { [\sigma(\ell)- n] [\ell -n] }  \nonumber \\
 &\times&  \frac1{[ (\ell +1) -\sigma(\ell +1)]^{m_{\ell +1}+1}} \cdots
 \frac1{[k- \sigma(k)]^{m_{k}+1}}
 \prod_{j ( \neq 1, 2, \cdots k)}^{n-1}
 \frac1{[j - \sigma(j)]} \;.
\eeqn
  Factorizing the expression inside the bracket, we have
\beqn
\label{eq:recursion2}
 && \left( \begin{array}{cccccc}
 m_{1}, & \cdots, & m_{k}, & 0,     & \cdots, & 0 \\
 1,     & \cdots  & k    , & k+1    & \cdots, & n+1
 \end{array} \right)_{n+1}    \;\;\;  \nonumber \\
 &&\qquad =\sum_{j=1}^{k}\sum_{l=1}^{m_j}
 \left( \begin{array}{cccccccc}
 m_{1}, & \cdots, & m_{j}-l, & \cdots, & m_{k}, & 0,     & \cdots, & 0 \\
 1,     & \cdots  & j,       & \cdots, & k    , & k+1    & \cdots, & n
 \end{array} \right)_{n}
 \frac{1}{[j-n]^{1+l} }
 \quad . \nonumber \\
\eeqn
Then from the assumption eq.~(\ref{eq:relation3}) at $n$, eq.~(\ref{eq:relation3})
 in which $n$ is replaced by $n+1$ is also satisfied.
On the other hand for $n=3$  eq.~(\ref{eq:relation3}) is clearly satisfied, so
 we have proven eq.~(\ref{eq:relation3}) and eq.~(\ref{eq:relation4}) .


\end{document}